%
%
\input harvmac
\noblackbox

\let\includefigures=\iftrue
\input epsf
\includefigures
\message{If you do not have epsf.tex (to include figures),}
\message{change the option at the top of the tex file.}
\def\figin{\epsfcheck\figin}\def\figins{\epsfcheck\figins}
\def\epsfcheck{\ifx\epsfbox\UnDeFiNeD
\message{(NO epsf.tex, FIGURES WILL BE IGNORED)}
\gdef\figin##1{\vskip2in}\gdef\figins##1{\hskip.5in}
\else\message{(FIGURES WILL BE INCLUDED)}%
\gdef\figin##1{##1}\gdef\figins##1{##1}\fi}
\def\DefWarn#1{}
\def\figinsert{\goodbreak\midinsert}
\def\ifig#1#2#3{\DefWarn#1\xdef#1{fig.~\the\figno}
\writedef{#1\leftbracket fig.\noexpand~\the\figno}%
\figinsert\figin{\centerline{#3}}\medskip\centerline{\vbox{\baselineskip12pt
\advance\hsize by -1truein\noindent\footnotefont{\bf Fig.~\the\figno:} #2}}
\bigskip\endinsert\global\advance\figno by1}
\else
\def\ifig#1#2#3{\xdef#1{fig.~\the\figno}
\writedef{#1\leftbracket fig.\noexpand~\the\figno}%
\global\advance\figno by1}
\fi
%
\newcount\yearltd\yearltd=\year\advance\yearltd by 0


\def\inbar{\,\vrule height1.5ex width.4pt depth0pt}
\def\IC{{\relax\hbox{$\inbar\kern-.3em{\rm C}$}}}

\def\IZ{\relax\ifmmode\mathchoice
{\hbox{\cmss Z\kern-.4em Z}}{\hbox{\cmss Z\kern-.4em Z}}
{\lower.9pt\hbox{\cmsss Z\kern-.4em Z}}
{\lower1.2pt\hbox{\cmsss Z\kern-.4em Z}}\else{\cmss Z\kern-.4em
Z}\fi}
\def\IR{\relax{\rm I\kern-.18em R}}
\def\fig#1#2#3{
\par\begingroup\parindent=0pt\leftskip=1cm\rightskip=1cm\parindent=0pt
\baselineskip=11pt
\global\advance\figno by 1
\midinsert
\epsfxsize=#3
\centerline{\epsfbox{#2}}
\vskip 12pt
\centerline{{\bf Figure \the\figno} #1}\par
\endinsert\endgroup\par}
\def\figlabel#1{\xdef#1{\the\figno}}

\def\pmb#1{\setbox0=\hbox{#1}%
\kern-.025em\copy0\kern-\wd0
\kern.05em\copy0\kern-\wd0
\kern-.025em\raise.0433em\box0 }
\font\cmss=cmss10
\font\cmsss=cmss10 at 7pt
\def\half{{1\over 2}}
\def\rlx{\relax\leavevmode}
\def\Cop{\relax\,\hbox{$\kern-.3em{\rm C}$}}
\def\Rop{\relax{\rm I\kern-.18em R}}
\def\Nop{\relax{\rm I\kern-.18em N}}
\def\Pop{\relax{\rm I\kern-.18em P}}

\def\Zop{\rlx\leavevmode\ifmmode\mathchoice{\hbox{\cmss Z\kern-.4em Z}}
{\hbox{\cmss Z\kern-.4em Z}}{\lower.9pt\hbox{\cmsss Z\kern-.36em Z}}
{\lower1.2pt\hbox{\cmsss Z\kern-.36em Z}}\else{\cmss Z\kern-.4em
Z}\fi}


\def\ie{{\it i.e.}}
\def\eg{{\it e.g.}}

\def\cn{{\cal N}}

\def\co{{\cal O}} 
\def\hO{{\hat O}}

\def\tj{{\tilde \jmath}}

\def\ttr{{\widetilde{\rm tr}}}
\def\psu{\Phi^{(su)}}


\lref\HIV{
K.~Hori, A.~Iqbal and C.~Vafa,
``D-branes and mirror symmetry,''
arXiv:hep-th/0005247.}   

\lref\HoriAX{
K.~Hori and A.~Kapustin,
``Duality of the fermionic 2d black hole and $N = 2$ Liouville theory as  
mirror symmetry,''
JHEP {\bf 0108}, 045 (2001)
[arXiv:hep-th/0104202].
}

\lref\forger{
K.~Foerger and S.~Stieberger,
``Higher derivative couplings and heterotic-type I duality 
in eight  dimensions,''
Nucl.\ Phys.\ B {\bf 559}, 277 (1999)
[arXiv:hep-th/9901020].
}

\lref\berva{
N.~Berkovits and C.~Vafa,
``$\cn=4$ topological strings,''
Nucl.\ Phys.\ B {\bf 433}, 123 (1995)
[arXiv:hep-th/9407190].
}

\lref\WittenFB{
E.~Witten,
``Chern-Simons gauge theory as a string theory,''
Prog.\ Math.\ {\bf 133}, 637 (1995)
[arXiv:hep-th/9207094].
}

\lref\ovcigar{
H.~Ooguri and C.~Vafa,
``Two-Dimensional black hole and singularities of CY manifolds,''
Nucl.\ Phys.\ B {\bf 463}, 55 (1996)
[arXiv:hep-th/9511164].
}

\lref\ovlargen{
H.~Ooguri and C.~Vafa,
``Worldsheet derivation of a large $N$ duality,''
Nucl.\ Phys.\ B {\bf 641}, 3 (2002)
[arXiv:hep-th/0205297].
}

\lref\ovloops{
H.~Ooguri and C.~Vafa,
``All loop $\cn=2$ string amplitudes,''
Nucl.\ Phys.\ B {\bf 451}, 121 (1995)
[arXiv:hep-th/9505183].
}

\lref\gopava{
R.~Gopakumar and C.~Vafa,
``Topological gravity as large $N$ topological gauge theory,''
Adv.\ Theor.\ Math.\ Phys.\  {\bf 2}, 413 (1998)
[arXiv:hep-th/9802016].
}

\lref\gopavatwo{
R.~Gopakumar and C.~Vafa,
``On the gauge theory/geometry correspondence,''
Adv.\ Theor.\ Math.\ Phys.\ {\bf 3}, 1415 (1999)
[arXiv:hep-th/9811131].
}

\lref\ovknot{
H.~Ooguri and C.~Vafa,
``Knot invariants and topological strings,''
Nucl.\ Phys.\ {\bf 577}, 419 (2000)
[arXiv:hep-th/9912123].
}

\lref\GiveonZM{
A.~Giveon, D.~Kutasov and O.~Pelc,
``Holography for non-critical superstrings,''
JHEP {\bf 9910}, 035 (1999)
[arXiv:hep-th/9907178].
}

\lref\labmar{
J.~M.~Labastida and M.~Marino,
``Polynomial invariants for torus knots and topological strings,''
Commun.\ Math.\ Phys.\ {\bf 217}, 423 (2001)
[arXiv:hep-th/0004196].
}

\lref\ramsar{
P.~Ramadevi and T. Sarkar,
``On link invariants and topological string amplitudes,''
Nucl.\ Phys.\ {\bf 600}, 487 (2001)
[arXiv:hep-th/0009188].
}

\lref\gkone{
A.~Giveon and D.~Kutasov,
``Little string theory in a double scaling limit,''
JHEP {\bf 9910}, 034 (1999)
[arXiv:hep-th/9909110].
}

\lref\gktwo{
A.~Giveon and D.~Kutasov,
``Comments on double scaled little string theory,''
JHEP {\bf 0001}, 023 (2000)
[arXiv:hep-th/9911039].
}

\lref\newgk{
A.~Giveon and D.~Kutasov,
``Notes on $AdS_3$,''
Nucl.\ Phys.\ B {\bf 621}, 303 (2002)
[arXiv:hep-th/0106004].
}

\lref\agnt{
I.~Antoniadis, E.~Gava, K.~S.~Narain, and T.~R.~Taylor,
``$\cn=2$ type II heterotic duality and higher derivative F-terms,''
Nucl.\ Phys.\ {\bf B455}, 109 (1995)
[arXiv:hep-th/9507115].
}

\lref\ghovaf{
D.~Ghoshal and C.~Vafa,
``$c=1$ string as the topological theory of the conifold,''
Nucl.\ Phys.\ {\bf B453}, 121 (1995)
[arXiv:hep-th/9506122].
}

\lref\kekeli{
K. Li}

\lref\maxim{
M. Kontsevich}

\lref\wittenhiggs{
E.~Witten,
``On the conformal field theory of the Higgs branch,''
JHEP {\bf 9707}, 003 (1997)
[arXiv:hep-th/9707093].
}

\lref\ahaber{
O.~Aharony and M.~Berkooz,
``IR dynamics of $d = 2$, $\cn = (4,4)$ 
gauge theories and DLCQ of 'little  string theories',''
JHEP {\bf 9910}, 030 (1999)
[arXiv:hep-th/9909101].
}

\lref\marcus{
N.~Marcus,
``The $\cn=2$ open string,''
Nucl.\ Phys.\ B {\bf 387}, 263 (1992)
[arXiv:hep-th/9207024].
}

\lref\fuku{
M.~Fukuma, H.~Kawai and R.~Nakayama,
``Infinite dimensional Grassmannian structure of two-dimensional 
quantum gravity,''
Commun.\ Math.\ Phys.\  {\bf 143}, 371 (1992).
}

\lref\VafaWI{
C.~Vafa,
``Superstrings and topological strings at large $N$,''
J.\ Math.\ Phys.\  {\bf 42}, 2798 (2001)
[arXiv:hep-th/0008142].
}

\lref\newbov{
N.~Berkovits, H.~Ooguri and C.~Vafa,
``On the worldsheet derivation of large $N$ dualities for the superstring,''
arXiv:hep-th/0310118.
}

\lref\AtiyahZZ{
M.~Atiyah, J.~M.~Maldacena and C.~Vafa,
``An M-theory flop as a large $N$ duality,''
J.\ Math.\ Phys.\  {\bf 42}, 3209 (2001)
[arXiv:hep-th/0011256].
}

\lref\chs{
C.~G.~Callan, J.~A.~Harvey and A.~Strominger,
``Supersymmetric string solitons,''
arXiv:hep-th/9112030.
}

\lref\abks{
O.~Aharony, M.~Berkooz, D.~Kutasov and N.~Seiberg,
``Linear dilatons, NS5-branes and holography,''
JHEP {\bf 9810}, 004 (1998)
[arXiv:hep-th/9808149].}

\lref\BuchbinderKN{
I.~L.~Buchbinder and E.~A.~Ivanov,
``Exact $N = 4$ supersymmetric low-energy effective action in $N = 4$  
super Yang-Mills theory,''
arXiv:hep-th/0211067.
}

\lref\KlebanovKV{
I.~R.~Klebanov and A.~Hashimoto,
``Nonperturbative solution of matrix models modified by trace squared terms,''
Nucl.\ Phys.\ B {\bf 434}, 264 (1995)
[arXiv:hep-th/9409064].
}

\lref\david{
D.~Kutasov,
``Orbifolds and solitons,''
Phys.\ Lett.\ B {\bf 383}, 48 (1996)
[arXiv:hep-th/9512145].
}

\lref\dinesei{
M.~Dine and N.~Seiberg,
``Comments on higher derivative operators in some SUSY field theories,''
Phys.\ Lett.\ B {\bf 409}, 239 (1997)
[arXiv:hep-th/9705057].
}

\lref\QiuZF{
Z.~a.~Qiu,
``Nonlocal current algebra and $N=2$ superconformal field theory in
two-dimensions,''
Phys.\ Lett.\ B {\bf 188}, 207 (1987).
}

\lref\aspinwall{
P.~S.~Aspinwall,
``Enhanced gauge symmetries and K3 surfaces,''
Phys.\ Lett.\ B {\bf 357}, 329 (1995)
[arXiv:hep-th/9507012].
}

\lref\kiritsis{
E.~Kiritsis,
``Duality and instantons in string theory,''
arXiv:hep-th/9906018.
}

\lref\KiritsisZI{
E.~Kiritsis, N.~A.~Obers and B.~Pioline,
``Heterotic/type II triality and instantons on K3,''
JHEP {\bf 0001}, 029 (2000)
[arXiv:hep-th/0001083].
}

\lref\chstwo{
C.~G.~Callan, J.~A.~Harvey and A.~Strominger,
``World sheet approach to heterotic instantons and solitons,''
Nucl.\ Phys.\ B {\bf 359}, 611 (1991).
}

\lref\rocek{
U.~Lindstrom, F.~Gonzalez-Rey, M.~Rocek and R.~von Unge,
``On $N = 2$ low energy effective actions,''
Phys.\ Lett.\ B {\bf 388}, 581 (1996)
[arXiv:hep-th/9607089].
}

\lref\mans{
M.~Henningson,
``Extended superspace, higher derivatives and $SL(2,Z)$ duality,''
Nucl.\ Phys.\ B {\bf 458}, 445 (1996)
[arXiv:hep-th/9507135].
}

\lref\StromingerCZ{
A.~Strominger,
``Massless black holes and conifolds in string theory,''
Nucl.\ Phys.\ B {\bf 451}, 96 (1995)
[arXiv:hep-th/9504090].
}

\lref\lowe{
D.~A.~Lowe and R.~von Unge,
``Constraints on higher derivative operators in maximally supersymmetric
gauge theory,''
JHEP {\bf 9811}, 014 (1998)
[arXiv:hep-th/9811017].
}

\lref\grisaru{
B.~de Wit, M.~T.~Grisaru and M.~Rocek,
``Nonholomorphic corrections to the one-loop $N=2$ Super Yang-Mills action,''
Phys.\ Lett.\ B {\bf 374}, 297 (1996)
[arXiv:hep-th/9601115].
}

\lref\kulik{
F.~Gonzalez-Rey, B.~Kulik, I.~Y.~Park and M.~Rocek,
``Self-dual effective action of $N = 4$ super-Yang-Mills,''
Nucl.\ Phys.\ B {\bf 544}, 218 (1999)
[arXiv:hep-th/9810152].
}

\lref\kuzenko{
E.~I.~Buchbinder, I.~L.~Buchbinder and S.~M.~Kuzenko,
``Non-holomorphic effective potential in $N = 4$ $SU(n)$ SYM,''
Phys.\ Lett.\ B {\bf 446}, 216 (1999)
[arXiv:hep-th/9810239].
}

\lref\buch{
I.~L.~Buchbinder and E.~A.~Ivanov,
``Complete $N = 4$ structure of low-energy effective action 
in $N = 4$ super  Yang-Mills theories,''
Phys.\ Lett.\ B {\bf 524}, 208 (2002)
[arXiv:hep-th/0111062].
}

\lref\banin{
A.~T.~Banin, I.~L.~Buchbinder and N.~G.~Pletnev,
``One-loop effective action for $N = 4$ SYM theory in the hypermultiplet  
sector: Leading low-energy approximation and beyond,''
arXiv:hep-th/0304046.
}

\lref\ellis{
J.~R.~Ellis, P.~Jetzer and L.~Mizrachi,
``One loop string corrections to the effective field theory,''
Nucl.\ Phys.\ B {\bf 303}, 1 (1988).
}

\lref\abe{
M.~Abe, H.~Kubota and N.~Sakai,
``Loop corrections to the $E_8 \times E_8$ 
heterotic string effective Lagrangian,''
Nucl.\ Phys.\ B {\bf 306}, 405 (1988);
Phys.\ Lett.\ B {\bf 200}, 461 (1988)
[Addendum-ibid.\ B {\bf 203}, 474 (1988)].
}

\lref\lerche{
W.~Lerche,
``Elliptic index and superstring effective actions,''
Nucl.\ Phys.\ B {\bf 308}, 102 (1988).
}

\lref\adscft{
J.~M.~Maldacena,
``The large $N$ limit of superconformal field theories and supergravity,''
Adv.\ Theor.\ Math.\ Phys.\  {\bf 2}, 231 (1998)
[Int.\ J.\ Theor.\ Phys.\  {\bf 38}, 1113 (1999)]
[arXiv:hep-th/9711200];
S.~S.~Gubser, I.~R.~Klebanov and A.~M.~Polyakov,
``Gauge theory correlators from non-critical string theory,''
Phys.\ Lett.\ B {\bf 428}, 105 (1998)
[arXiv:hep-th/9802109];
E.~Witten,
``Anti-de Sitter space and holography,''
Adv.\ Theor.\ Math.\ Phys.\  {\bf 2}, 253 (1998)
[arXiv:hep-th/9802150];
O.~Aharony, S.~S.~Gubser, J.~M.~Maldacena, H.~Ooguri and Y.~Oz,
``Large $N$ field theories, string theory and gravity,''
Phys.\ Rept.\  {\bf 323}, 183 (2000)
[arXiv:hep-th/9905111].
}

\lref\ZamolodchikovBD{
A.~B.~Zamolodchikov and V.~A.~Fateev,
``Operator algebra and correlation functions in the two-dimensional 
Wess-Zumino $SU(2) \times SU(2)$ chiral model,''
Sov.\ J.\ Nucl.\ Phys.\  {\bf 43}, 657 (1986)
[Yad.\ Fiz.\  {\bf 43}, 1031 (1986)].
}

\lref\TeschnerFT{
J.~Teschner,
``On structure constants and fusion rules in the $SL(2,C)/SU(2)$ WZNW  model,''
Nucl.\ Phys.\ B {\bf 546}, 390 (1999)
[arXiv:hep-th/9712256].
}

\lref\TeschnerUG{
J.~Teschner,
``Operator product expansion and factorization in the $H_3^+$ WZNW model,''
Nucl.\ Phys.\ B {\bf 571}, 555 (2000)
[arXiv:hep-th/9906215].
}

\lref\FZZ{V.~A.~Fateev, A.~B.~Zamolodchikov and Al.~B.~Zamolodchikov, 
unpublished.}

\lref\BalasubramanianNH{
V.~Balasubramanian, M.~Berkooz, A.~Naqvi and M.~J.~Strassler,
``Giant gravitons in conformal field theory,''
JHEP {\bf 0204}, 034 (2002)
[arXiv:hep-th/0107119].
}
\lref\CorleyZK{
S.~Corley, A.~Jevicki and S.~Ramgoolam,
``Exact correlators of giant gravitons from dual $N = 4$ SYM theory,''
Adv.\ Theor.\ Math.\ Phys.\  {\bf 5}, 809 (2002)
[arXiv:hep-th/0111222];
S.~Corley and S.~Ramgoolam,
``Finite factorization equations and sum rules for BPS correlators in  
$N = 4$ SYM theory,''
Nucl.\ Phys.\ B {\bf 641}, 131 (2002)
[arXiv:hep-th/0205221].
}
\lref\AharonyND{
O.~Aharony, Y.~E.~Antebi, M.~Berkooz and R.~Fishman,
``'Holey sheets': Pfaffians and subdeterminants as D-brane operators in  
large $N$ gauge theories,''
JHEP {\bf 0212}, 069 (2002)
[arXiv:hep-th/0211152].
}
\lref\BalasubramanianSN{
V.~Balasubramanian, P.~Kraus and A.~E.~Lawrence,
``Bulk vs. boundary dynamics in anti-de Sitter spacetime,''
Phys.\ Rev.\ D {\bf 59}, 046003 (1999)
[arXiv:hep-th/9805171].
}

\lref\CheungYW{
Y.~K.~Cheung, Y.~Oz and Z.~Yin,
``Families of $N = 2$ strings,''
arXiv:hep-th/0211147.
}
\lref\GluckWG{
D.~Gluck, Y.~Oz and T.~Sakai,
``The effective action and geometry of closed $N = 2$ strings,''
JHEP {\bf 0307}, 007 (2003)
[arXiv:hep-th/0304103].
}
\lref\GluckPA{
D.~Gluck, Y.~Oz and T.~Sakai,
``D-branes in $N = 2$ strings,''
JHEP {\bf 0308}, 055 (2003)
[arXiv:hep-th/0306112].
}
\lref\PolchinskiRR{
J.~Polchinski,
``String Theory. Vol. 2: Superstring theory and beyond,''
Cambridge University Press, 1998.
}
\lref\DabholkarJT{
A.~Dabholkar and J.~A.~Harvey,
``Nonrenormalization of the superstring tension,''
Phys.\ Rev.\ Lett.\  {\bf 63}, 478 (1989).
}
\lref\HarveyWM{
J.~A.~Harvey, D.~Kutasov, E.~J.~Martinec and G.~Moore,
``Localized tachyons and RG flows,''
arXiv:hep-th/0111154.
}
\lref\KutasovXB{
D.~Kutasov,
``Geometry On The Space Of Conformal Field Theories And Contact Terms,''
Phys.\ Lett.\ B {\bf 220}, 153 (1989).
}
\lref\TongIK{
D.~Tong,
``Mirror mirror on the wall: on two-dimensional black holes and Liouville  
theory,''
JHEP {\bf 0304}, 031 (2003)
[arXiv:hep-th/0303151].
}
\lref\QiuZF{
Z.~A.~Qiu,
``Nonlocal current algebra and $N=2$ superconformal field theory 
in two-dimensions,''
Phys.\ Lett.\ B {\bf 188}, 207 (1987).
}

\lref\MaldacenaHW{
J.~M.~Maldacena and H.~Ooguri,
``Strings in $AdS_3$ and $SL(2,R)$ WZW model. I,''
J.\ Math.\ Phys.\  {\bf 42}, 2929 (2001)
[arXiv:hep-th/0001053].
}
\lref\MaldacenaKM{
J.~M.~Maldacena and H.~Ooguri,
``Strings in $AdS_3$ and the $SL(2,R)$ WZW model. 
III: Correlation  functions,''
Phys.\ Rev.\ D {\bf 65}, 106006 (2002)
[arXiv:hep-th/0111180].
}

\lref\ParnachevGW{
A.~Parnachev and D.~A.~Sahakyan,
``Some remarks on D-branes in $AdS_3$,''
JHEP {\bf 0110}, 022 (2001)
[arXiv:hep-th/0109150].
}
\lref\DiFrancescoUD{
P.~Di Francesco and D.~Kutasov,
``World sheet and space-time physics in two-dimensional (Super)string theory,''
Nucl.\ Phys.\ B {\bf 375}, 119 (1992)
[arXiv:hep-th/9109005].
}
\lref\OoguriIE{
H.~Ooguri and C.~Vafa,
``N=2 heterotic strings,''
Nucl.\ Phys.\ B {\bf 367}, 83 (1991).
}

\lref\QiuZF{
Z.~A.~Qiu,
``Nonlocal current algebra and $N=2$ superconformal field theory 
in two-dimensions,''
Phys.\ Lett.\ B {\bf 188}, 207 (1987).
}

\lref\FateevMM{
V.~A.~Fateev and A.~B.~Zamolodchikov,
``Parafermionic currents in the two-dimensional conformal quantum field 
theory and selfdual critical points in $Z_N$ invariant statistical
systems,''
Sov.\ Phys.\ JETP {\bf 62}, 215 (1985)
[Zh.\ Eksp.\ Teor.\ Fiz.\  {\bf 89}, 380 (1985)].
}

\lref\BischoffBN{
J.~Bischoff and O.~Lechtenfeld,
``Path-integral quantization of the (2,2) string,''
Int.\ J.\ Mod.\ Phys.\ A {\bf 12}, 4933 (1997)
[arXiv:hep-th/9612218].
}

\lref\WittenZD{
E.~Witten,
``Ground ring of two-dimensional string theory,''
Nucl.\ Phys.\ B {\bf 373}, 187 (1992)
[arXiv:hep-th/9108004].
}
\lref\KounnasUD{
C.~Kounnas, M.~Porrati and B.~Rostand,
``On $\cn=4$ extended superliouville theory,''
Phys.\ Lett.\ B {\bf 258}, 61 (1991).
}
\lref\NappiKV{
C.~R.~Nappi and E.~Witten,
``A closed, expanding universe in string theory,''
Phys.\ Lett.\ B {\bf 293}, 309 (1992)
[arXiv:hep-th/9206078].
}

\lref\ElitzurRT{
S.~Elitzur, A.~Giveon, D.~Kutasov and E.~Rabinovici,
``From big bang to big crunch and beyond,''
JHEP {\bf 0206}, 017 (2002)
[arXiv:hep-th/0204189].
}

\lref\FateevIK{
V.~Fateev, A.~B.~Zamolodchikov and A.~B.~Zamolodchikov,
``Boundary Liouville field theory. I: Boundary state and boundary  
two-point function,''
arXiv:hep-th/0001012.
}

\lref\DornXN{
H.~Dorn and H.~J.~Otto,
``Two and three point functions in Liouville theory,''
Nucl.\ Phys.\ B {\bf 429}, 375 (1994)
[arXiv:hep-th/9403141].
}

\lref\ZamolodchikovAA{
A.~B.~Zamolodchikov and A.~B.~Zamolodchikov,
``Structure constants and conformal bootstrap in Liouville field theory,''
Nucl.\ Phys.\ B {\bf 477}, 577 (1996)
[arXiv:hep-th/9506136].
}
\lref\AharonyTH{
O.~Aharony, M.~Berkooz, S.~Kachru, N.~Seiberg and E.~Silverstein,
``Matrix description of interacting theories in six dimensions,''
Adv.\ Theor.\ Math.\ Phys.\  {\bf 1}, 148 (1998)
[arXiv:hep-th/9707079];
%
E.~Witten,
``On the conformal field theory of the Higgs branch,''
JHEP {\bf 9707}, 003 (1997)
[arXiv:hep-th/9707093].
}
\lref\KutasovUF{
D.~Kutasov,
``Introduction to little string theory,''
{\it prepared for ICTP Spring School on Superstrings and Related
Matters, Trieste, Italy, 2-10 April 2001.}
}

\lref\AharonyKS{
O.~Aharony,
``A brief review of `little string theories',''
Class.\ Quant.\ Grav.\  {\bf 17}, 929 (2000)
[arXiv:hep-th/9911147].
}

\lref\KutasovJP{
D.~Kutasov and D.~A.~Sahakyan,
``Comments on the thermodynamics of little string theory,''
JHEP {\bf 0102}, 021 (2001)
[arXiv:hep-th/0012258].
}

\lref\BershadskyCX{
M.~Bershadsky, S.~Cecotti, H.~Ooguri and C.~Vafa,
``Kodaira-Spencer theory of gravity and exact results for quantum 
string amplitudes,''
Commun.\ Math.\ Phys.\  {\bf 165}, 311 (1994)
[arXiv:hep-th/9309140].
}

\lref\McGreevyKB{
J.~McGreevy and H.~Verlinde,
``Strings from tachyons: the $c = 1$ matrix reloaded,''
arXiv:hep-th/0304224.
}

\lref\MartinecKA{
E.~J.~Martinec,
``The annular report on non-critical string theory,''
arXiv:hep-th/0305148.
}

\lref\KlebanovKM{
I.~R.~Klebanov, J.~Maldacena and N.~Seiberg,
``D-brane decay in two-dimensional string theory,''
JHEP {\bf 0307}, 045 (2003)
[arXiv:hep-th/0305159].
}

\lref\McGreevyEP{
J.~McGreevy, J.~Teschner and H.~Verlinde,
``Classical and quantum D-branes in 2D string theory,''
arXiv:hep-th/0305194.
}

\lref\AlexandrovNN{
S.~Y.~Alexandrov, V.~A.~Kazakov and D.~Kutasov,
``Non-perturbative effects in matrix models and D-branes,''
JHEP {\bf 0309}, 057 (2003)
[arXiv:hep-th/0306177].
}

\lref\SenIV{
A.~Sen,
``Open-closed duality: Lessons from matrix model,''
arXiv:hep-th/0308068.
}

\lref\TakayanagiSM{
T.~Takayanagi and N.~Toumbas,
``A matrix model dual of type 0B string theory in two dimensions,''
JHEP {\bf 0307}, 064 (2003)
[arXiv:hep-th/0307083].
}

\lref\DouglasUP{
M.~R.~Douglas, I.~R.~Klebanov, D.~Kutasov, J.~Maldacena, E.~Martinec and 
N.~Seiberg,
``A new hat for the $c = 1$ matrix model,''
arXiv:hep-th/0307195.
}

\lref\KlebanovWG{
I.~R.~Klebanov, J.~Maldacena and N.~Seiberg,
``Unitary and complex matrix models as 1-d type 0 strings,''
arXiv:hep-th/0309168.
}

\lref\AntoniadisZE{
I.~Antoniadis, E.~Gava, K.~S.~Narain and T.~R.~Taylor,
``Topological amplitudes in string theory,''
Nucl.\ Phys.\ B {\bf 413}, 162 (1994)
[arXiv:hep-th/9307158];
``$N=2$ type II heterotic duality and higher derivative F terms,''
Nucl.\ Phys.\ B {\bf 455}, 109 (1995)
[arXiv:hep-th/9507115].
}

\lref\DijkgraafQH{
R.~Dijkgraaf,
``Intersection theory, integrable hierarchies and topological field theory,''
arXiv:hep-th/9201003.
}

\lref\OoguriWW{
H.~Ooguri and C.~Vafa,
``Selfduality and $N=2$ string magic,''
Mod.\ Phys.\ Lett.\ A {\bf 5}, 1389 (1990);
``Geometry of N=2 strings,''
Nucl.\ Phys.\ B {\bf 361}, 469 (1991).
}

\lref\MarcusWI{
N.~Marcus,
``A tour through $N=2$ strings,''
arXiv:hep-th/9211059;
O.~Lechtenfeld,
``Mathematics and physics of $N = 2$ strings,''
arXiv:hep-th/9912281.
}

\lref\HullYS{
C.~M.~Hull and P.~K.~Townsend,
``Unity of superstring dualities,''
Nucl.\ Phys.\ B {\bf 438}, 109 (1995)
[arXiv:hep-th/9410167].
}

\lref\GiveonEW{
A.~Giveon and M.~Rocek,
``On the BRST operator structure of the $N=2$ string,''
Nucl.\ Phys.\ B {\bf 400}, 145 (1993)
[arXiv:hep-th/9302049].
}

\lref\BachasMC{
C.~Bachas, C.~Fabre, E.~Kiritsis, N.~A.~Obers and P.~Vanhove,
``Heterotic/type-I duality and D-brane instantons,''
Nucl.\ Phys.\ B {\bf 509}, 33 (1998)
[arXiv:hep-th/9707126].
}

\lref\StiebergerWK{
S.~Stieberger and T.~R.~Taylor,
``Non-Abelian Born-Infeld action and type I - heterotic duality. 
II: Nonrenormalization theorems,''
Nucl.\ Phys.\ B {\bf 648}, 3 (2003)
[arXiv:hep-th/0209064].
}

\lref\JunemannHI{
K.~Junemann, O.~Lechtenfeld and A.~D.~Popov,
``Non-local symmetries of the closed $N = 2$ string,''
Nucl.\ Phys.\ B {\bf 548}, 449 (1999)
[arXiv:hep-th/9901164].
}

\lref\agk{O.~Aharony, A.~Giveon and D.~Kutasov, to appear.}

\def\my_Title#1#2{\nopagenumbers\abstractfont\hsize=\hstitle\rightline{#1}%
\vskip .5in\centerline{\titlefont #2}\abstractfont\vskip .5in\pageno=0}

\rightline{EFI-03-44}
\rightline{WIS/26/03-OCT-DPP}
\my_Title{
\rightline{\tt hep-th/0310197}}
{\vbox{\centerline{Little String Theory and}
\vskip 10pt \centerline{Heterotic/Type II Duality}}}
\centerline{\it Ofer Aharony${}^{a}$, Bartomeu Fiol${}^{a}$, 
David Kutasov${}^{b}$ and David A. Sahakyan${}^{b}$}
\medskip
\centerline{${}^a$ Department of Particle Physics, 
Weizmann Institute of Science, Rehovot 76100, Israel}
\centerline{\tt Ofer.Aharony, fiol@weizmann.ac.il}
\smallskip
\centerline{${}^b$ EFI and Department of Physics, University of Chicago}
\centerline{5640 S. Ellis Av., Chicago, IL 60637, USA }
\centerline{\tt kutasov, sahakian@theory.uchicago.edu}
\bigskip
\bigskip
\noindent
Little String Theory (LST) is a still somewhat mysterious theory that
describes the dynamics near a certain class of time-like singularities
in string theory. In this paper we discuss the topological version of
LST, which describes topological strings near these singularities. For
$5+1$ dimensional LSTs with sixteen supercharges, the topological
version may be described holographically in terms of the ${\cal N}=4$
topological string (or the ${\cal N}=2$ string) on the transverse
part of the near-horizon geometry of $NS5$-branes. We show that this
topological string can be used to efficiently compute the half-BPS $F^4$
terms in the low-energy effective action of the LST. Using the strong-weak
coupling string duality relating type IIA strings on $K3$ and  
heterotic strings
on $T^4$, the same terms may also be computed in the heterotic string
near a point of enhanced gauge symmetry. We study the $F^4$ terms in
the heterotic string and in the LST, and show that they have the same
structure, and that they agree in the cases for which we compute both
of them. We also clarify some additional issues, such as the
definition and role of normalizable modes in holographic linear
dilaton backgrounds, the precise identifications of vertex operators
in these backgrounds with states and operators in the supersymmetric
Yang-Mills theory that
arises in the low energy limit of LST, and the normalization of
two-point functions.
\vfill

\Date{}



\newsec{Introduction}

One of the more mysterious outcomes of the recent progress in the
understanding of non-perturbative aspects of string theory is the
discovery of theories which are non-local (and have some stringy
aspects) but are decoupled from gravity. These theories are known
as Little String Theories (LSTs) (for reviews see
\refs{\AharonyKS,\KutasovUF}); they appear in backgrounds of string 
theory which contain singularities and/or $NS5$-branes. In these cases
interesting dynamics near the singularity (or brane)
remains even in the decoupling
limit $g_s\to 0$. This dynamics is captured by the LST associated with 
the singularity.

The best description that we have for these theories \abks\ is via an
asymptotically linear dilaton background of string theory which is
holographically dual to them\foot{In some cases there is also a DLCQ
description \AharonyTH, but this requires taking a
large $N$ limit which is difficult to control. The holographic description
does not require taking large $N$ limits.}.
This description tells us what are the observables of the LST, and some
of its properties (such as the thermodynamic behavior, which
at high energy densities resembles that of free string theories, 
with important differences  \KutasovJP). It also allows
for the computation of some correlation functions in these theories (at
least when we go out on their moduli space \refs{\gkone,\gktwo}). 
However, we still lack a direct definition of these theories.

In many supersymmetric compactifications of string theory, there is a
sector of the theory which is protected by supersymmetry (the analog
of the chiral sector in $d=4$, ${\cal N}=1$ supersymmetric gauge
theories), and which is captured by a topological version of the full
string theory. In the case of type II string theories compactified on
Calabi-Yau three-folds, this topological string theory is the ${\cal
N}=2$ topological string   \BershadskyCX , while for type II string theories
compactified on $K3$ surfaces it is the ${\cal N}=4$ topological string,
which is equivalent to the ${\cal N}=2$ string \berva\ (see also \GiveonEW).
Since the construction of a Little String Theory involves taking some limit of
a supersymmetric type II compactification, it is natural to suggest that taking
the same limit in the topological string theory will lead to a
topological version of Little String Theory. Such a version may be
easier to understand than the full LST, and it may be easier to find a
direct definition for it (the holographic description of these
theories is just the topological string version of the holographic
description of the full LSTs). In four dimensional LSTs,
some progress in this direction has been made, particularly for the case
of the conifold \refs{\gopava\gopavatwo\ovlargen\VafaWI-\newbov}. 
The main motivation
for this work is to obtain a better understanding of more general singularities
of Calabi-Yau as well as $K3$ surfaces.

In this paper we focus on the most symmetric LSTs, which are $5+1$
dimensional theories with sixteen supercharges (${\cal N}=(1,1)$
supersymmetry in six dimensions). These theories arise
from decoupling limits of type IIA strings on ALE spaces
(non-compact $K3$ manifolds that arise by blowing up the geometry
in the vicinity of ADE singularities of compact $K3$ surfaces), or
from decoupling limits of type IIB $NS5$-branes in flat space. The 
holographic description of these theories is given by string theory 
on the near-horizon geometry of $NS5$-branes, the CHS background \chs. 
The discussion above suggests that the ${\cal N}=2$ string on the CHS
background is holographically dual to a topological version of the
corresponding LST. In this paper we investigate this suggestion.

Before taking any decoupling limits, the topological string computes
various amplitudes which are protected by supersymmetry 
in the full type II string theory \berva.
These correspond to the coefficients of specific terms in the low-energy
effective action of the theory. It is interesting to ask what do 
correlation functions in the topological version of LST compute. In general, 
LSTs are known to have operators which are defined off-shell, and the 
physical observables are Green's functions of these operators. This is 
different from critical string theory, in which the observables are 
on-shell S-matrix elements. The discussion above suggests that there 
should be some sub-class of the correlation functions of the off-shell 
observables of LST which is topological in nature, and which is computed 
by the topological LST. One way to derive this sub-class is by taking a 
limit of the topological observables of the full type II string theory, 
but this is complicated by the difference between the observables in the 
two types of theories. Another way to derive this sub-class is to follow
the terms in the effective action of the type II string theory which are 
protected by supersymmetry to the LST limit, and to find which
observables in the topological LST compute these terms. This is the
route that we will follow in this paper.

The particular term in the low-energy effective action that we will
discuss has the form $t_8 F^4$, where $F$ is an Abelian gauge field 
in the low-energy theory and $t_8$ is a specific constant 
tensor which governs the contraction of the indices of the four gauge 
fields. In weakly coupled type II string theory, the field strength in 
question corresponds to a Ramond-Ramond (RR) gauge field.
The coefficient of this $F^4$
term is believed to be protected from quantum corrections by supersymmetry, 
and to be given exactly by the tree level contribution. We will show that
it corresponds to an observable in the topological string theory, and use
this to compute it. 

We will leave the analysis of other topological amplitudes in LST to
future work. Such amplitudes are also of interest; for example, it
is known that higher-loop contributions to the partition sum of the
topological string theory are related to the coefficients of certain 
$R^4 F^{4g-4}$ terms in the effective action of the type II string theory 
\berva.

Type IIA string theory on $K3$ is believed to be
dual to the heterotic string on $T^4$, and the coefficient of 
the $t_8F^4$ term may also be 
computed using the heterotic theory. In the heterotic string, this
term receives contributions only at one-loop, and it can be 
easily computed. The result simplifies significantly in the LST decoupling 
limit, where it is given by a one-loop computation in the low-energy gauge
theory. 

\ifig\diagram{A schematic diagram of the theories discussed in this paper.
The three theories on the top-right corner are related by T-dualities and
S-dualities. The different routes towards the bottom-left corner are
commutative.}
{\epsfxsize4.5in\epsfbox{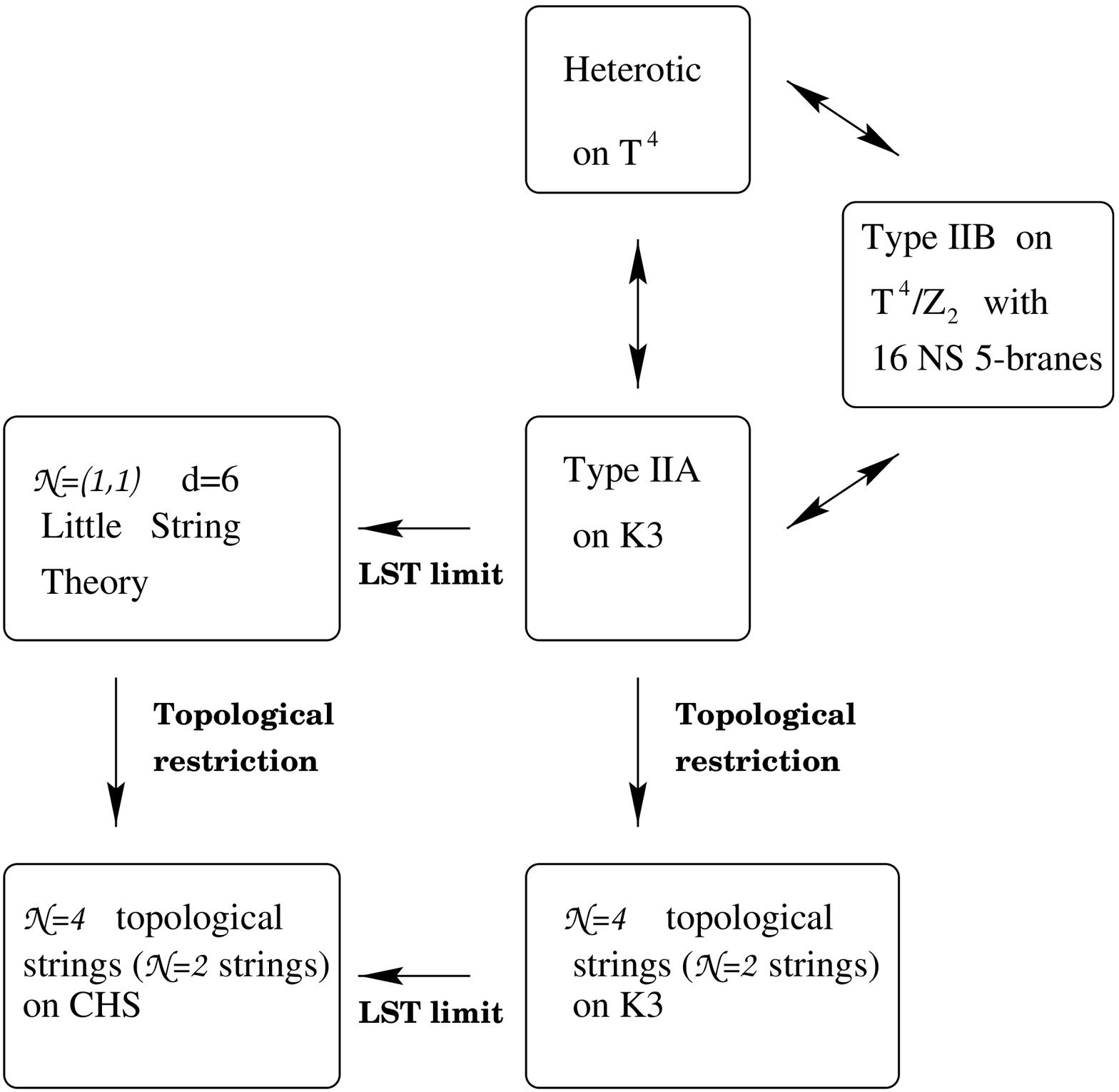}}

Heterotic/type II duality thus leads to a non-trivial prediction: a tree level
four-point function in LST should be equal to a one-loop amplitude in the 
low-energy field theory. As we review below, the two calculations are 
valid in different regions in moduli space, and the only reasons to expect 
them to agree are (1) the non-renormalization of these terms, and (2) 
heterotic/type II duality. We will show by an explicit calculation of 
both sides that they indeed agree. One can view this as a new non-trivial 
check of heterotic/type II duality and of the non-renormalization of the 
$t_8F^4$ term. Since these two circles of ideas are rather well established, 
we use these calculations to develop tools for studying LST in general, and 
particularly its topological sector.

One of our main motivations for studying these theories is the hope
that an alternative description of LST can be obtained by using 
an open-closed string large $N$ duality. There are some indications
that this should be possible. The first example of a holographic 
duality, the equivalence of certain large $N$ matrix models in one or 
less dimensions in a double scaling limit with $c\le 1$ conformal field
theories (CFTs) coupled 
to worldsheet gravity (or string theory in $D\le 2$ dimensions), is now 
understood as an open-closed string duality, with the open strings living 
on unstable $D0$-branes localized in the Liouville direction 
\refs{\McGreevyKB\MartinecKA\KlebanovKM\McGreevyEP\AlexandrovNN\SenIV
\TakayanagiSM\DouglasUP-\KlebanovWG}. These backgrounds of string
theory can be thought of as simple examples of LST \refs{\abks,\GiveonZM},
with the Liouville direction playing the role of the radial direction
away from the singularity. 

Thus, it is natural to expect more generally,
that an alternative description of LST can be obtained by studying D-branes
localized in the vicinity of the singularity (where, as we will review below,
the effective string coupling is largest). A natural first step in constructing
an open string dual of the full LST is to find one that is dual to the 
topological sector of the theory. Something like this is known to exist
for the case of the conifold, where the topological LST is dual to a
topological open string theory describing the dynamics on $N\to\infty$
D-branes localized near the conifold singularity 
\refs{\gopava,\gopavatwo,\ovlargen}. We would like to find the analog
for the case of ALE spaces. We will not discuss open-closed string 
duality in this paper (except for a few comments in the discussion),
but we hope that our results will be useful for constructing an open 
string dual for six dimensional topological LST, and perhaps eventually 
also for the full theory.

We begin in section 2 by reviewing the known results about $t_8F^4$ terms
in field theories with sixteen supercharges and in toroidal compactifications
of the heterotic string. In section 3 we review the duality between the
heterotic string on $T^4$, the type IIA string on $K3$, and type IIB
backgrounds with $NS5$-branes, and the implications of this duality for the
$t_8F^4$ terms. In section 4 we formulate type II string theory in the
near-horizon limit of ALE singularities (or coincident $NS5$-branes), and
discuss in detail its worldsheet properties. We also comment on
the analogy between our discussion of the ALE case and
previous discussions of strings on the conifold. In section 5 we
use the type II and ${\cal N}=2$ string theories on the (deformed) 
CHS background 
to compute correlation functions of the vertex operators 
which are relevant for the $t_8F^4$ terms in the low-energy
effective action. In section 6 we compare our type II and
${\cal N}=2$ string results with
the expectations from the heterotic (or low-energy field theory) analysis.
We end in section 7 with a summary of our results and a discussion of
possible future directions. Four appendices contain useful technical results.

\newsec{$F^4$ terms in theories with sixteen supercharges}

In this section we review the structure of $F^4$ terms in the effective
actions of theories with sixteen supercharges, both in field theory and
in string theory. We review the arguments for the one-loop exactness of
these terms. For the string theory case, we focus in this section on the
heterotic string, since the relation with the field theory limit there is
most straightforward. In the next section we discuss the appearance of
the $F^4$ terms in different string duals of the background considered here.

\subsec{Field theory}

In super Yang-Mills perturbation theory, there is a one-loop contribution to
$F^4$ terms in the low-energy effective action, which is given by the diagrams
with four external gauge bosons  in figure 2.
As reviewed below, for the $d=4$ ${\cal N}=4$ supersymmetric Yang-Mills 
(SYM) theory
the one-loop contribution gives the exact result, both perturbatively and
non-perturbatively. On the other hand, it is known \dinesei\ that for theories
with sixteen supercharges in $d=3$
the $F^4$ terms receive instanton corrections.

\ifig\fourleg{Diagrams contributing to $F^4$ terms in SYM theories. 
Wiggly lines denote gauge fields, solid lines are scalar fields 
and dashed lines are fermions.}
{\epsfxsize3in\epsfbox{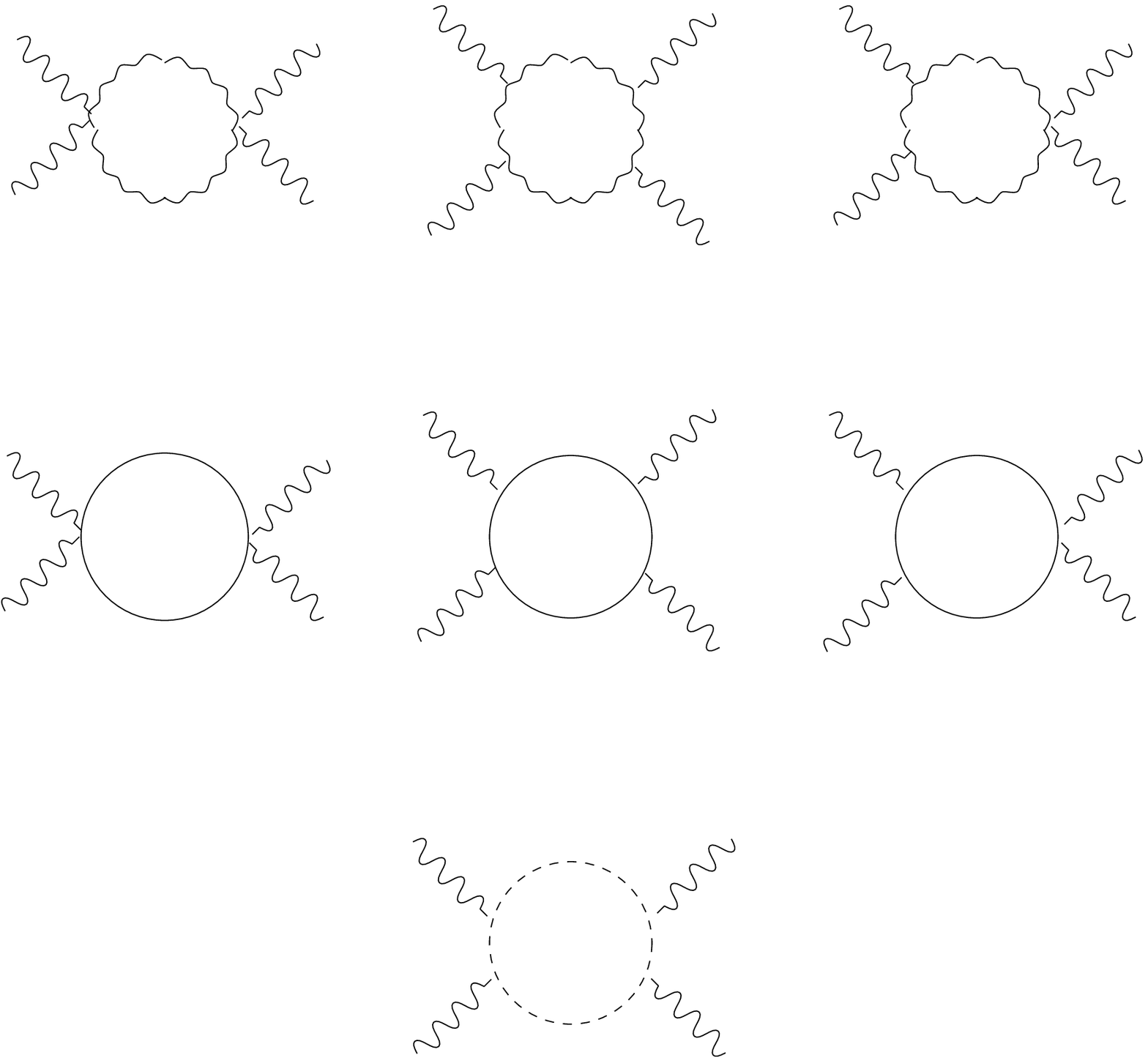}}

The non-renormalization of $F^4$ terms in $d=4$, ${\cal N}=4$ SYM has been
discussed by various authors \refs{\dinesei\mans\grisaru-\rocek}. On
general grounds, the terms for which we expect non-renormalization would be
integrals over half of ${\cal N}=4$
superspace, if such a formulation existed. Consider an $SU(2)$ gauge theory
spontaneously broken to $U(1)$. In such a case 
the low-energy effective action involves
only a $U(1)$ ${\cal N}=4$ vector multiplet. When we view the $\cn=4$
theory as an $\cn=2$ theory, the $U(1)$ gauge field strength
$F_{\mu \nu}$ appears as a component
of the ${\cal N}=2$ $U(1)$ vector multiplet $\Psi$. We consider
the $F^4$ terms in the region of moduli space where only the scalar in the
${\cal N}=2$ vector multiplet has a vacuum expectation value (VEV)\foot{The
more general case was discussed in \refs{\grisaru, \rocek , \buch} and 
reviewed in \banin.}. 
In ${\cal N}=2$ superspace language,
the most general such term in the low-energy effective action 
can be written as
\eqn\superint{
\int d^8\theta \; {\cal H}(\Psi, \Psi^\dagger, \tau, \tau ^\dagger),}
where $\tau$ is the complexified coupling constant. Invoking the scale
invariance and the $U(1)_R$ symmetry of the theory, one can argue that for
given $\tau$, ${\cal H}$ is uniquely fixed to be
\eqn\superh{
{\cal H} \sim
\hbox {ln}\left (\Psi\over \Lambda\right )\hbox {ln}
\left (\Psi^\dagger\over \Lambda\right ),}
where $\Lambda$ is a fake scale, that does not survive integration over
superspace. Furthermore, by promoting $\tau$ to a background superfield,
it also follows from scale invariance and $U(1)_R$ symmetry that ${\cal H}$
can not depend on $\tau$ at all, so the coefficient in \superh\ can be computed
at one-loop, and
there are neither perturbative
nor non-perturbative corrections to the one-loop result.

It is easy to generalize this to arbitrary gauge groups (broken to their
Abelian subgroup by a VEV for the scalar in the vector multiplet)
\refs{\lowe\kulik-\kuzenko}, where $\vec{\Psi}$ 
belongs to the Cartan subalgebra
and ${\cal H}$
is given by a sum over the positive roots of the gauge group
\eqn\superh{{\cal H} \sim \sum_{\vec{\alpha} > 0} \hbox {ln }\left
(\vec{\Psi}\cdot \vec{\alpha}\over \Lambda\right )\hbox {ln }
\left (\vec{\Psi^\dagger}\cdot \vec{\alpha}\over \Lambda\right )~.}
After integration over superspace this gives rise to $F^4$ terms.
Similar terms appear at one-loop in any dimension (in theories with sixteen
supercharges), and in the $d$-dimensional maximally supersymmetric SYM theory
they are given by
\eqn\leffrf{
{\cal L}_{eff} \sim \sum _{\vec{\alpha} > 0} {(\vec F\cdot \vec \alpha)^4\over
M_W^{8-d}(\vec \alpha)}~,}
where $M_W({\vec \alpha})$ is the mass of the W-boson associated with
the root $\vec{\alpha}$ (at a particular point in the moduli space).
The structure of the space-time indices in \leffrf\ will be described
below.

As mentioned above, for $d=4$ equation \leffrf\ is exact, while for
$d=3$ it receives instanton corrections. For $d>4$ the gauge theory is
non-renormalizable, and one has to embed it in a consistent theory, such
as string theory, in order to discuss higher loop and non-perturbative
corrections.

\subsec{String theory}

Next, we turn to  $F^4$ terms in string theories with sixteen supercharges.
For convenience, we will start by considering the field strength $F_{\mu \nu}$
to be that of the $SO(32)$ heterotic string theory in ten dimensions. The
interesting term is of the form $t_8 F^4$, where $t_8$ is a constant tensor
defined below. In ten
dimensions, the $t_8 \tr(F^4)$ term is absent at tree
level\foot{There is a 
tree-level $t_8 \tr(F^2) \tr(F^2)$ term, but this term will not be relevant in
the limit we will be interested in.}, 
but it receives a one-loop
contribution \refs{\ellis\abe\lerche-\PolchinskiRR}. 
Since it is related by
supersymmetry to the anomaly-canceling term $BF^4$,
the one-loop result is expected to be exact, both perturbatively and
non-perturbatively. The precise result for the amplitude in ten dimensions is
\eqn\terminten{
{1\over 2^8\pi^54!\alpha'}t_8^{\mu\nu\sigma\rho\alpha\beta\gamma\delta}
\hbox { Tr}_v\; 
(F_{\mu \nu}F_{\sigma \rho}F_{\alpha \beta}F_{\gamma \delta})~,}
where the trace is in the vector representation and the tensor $t_8$ is
defined as follows :
\eqn\ttensor{\eqalign{
t_8^{\mu\nu\sigma\rho\alpha\beta\gamma\delta} \equiv
-{1\over 2} \bigg \{
& (\delta ^{\mu\sigma}\delta ^{\nu\rho}-\delta ^{\mu\rho}
\delta ^{\nu\sigma})(\delta ^{\alpha\gamma}\delta ^{\beta\delta}-
\delta ^{\alpha\delta}\delta ^{\beta\gamma})+ \cr
&(\delta ^{\sigma\alpha}
\delta ^{\rho\beta}-\delta ^{\sigma\beta}\delta ^{\rho\alpha})
(\delta ^{\gamma\mu}\delta ^{\delta\nu}-\delta ^{\gamma\nu}\delta ^{\delta\mu})
+\cr
& (\delta ^{\mu\alpha}\delta ^{\nu\beta}-\delta ^{\mu\beta}
\delta ^{\nu\alpha})(\delta ^{\sigma\gamma}\delta ^{\rho\delta}-
\delta ^{\sigma\delta}\delta ^{\rho\gamma})\bigg \} + \cr
{1\over 2}\bigg \{ &
\delta^{\nu\sigma} \delta^{\rho\alpha}\delta^{\beta\gamma}
\delta^{\delta\mu}+\delta^{\nu\alpha}\delta^{\beta\sigma} \delta^{\rho\gamma}
\delta^{\delta \mu}+\delta^{\nu\alpha}\delta^{\beta\gamma}\delta^{\delta\sigma}
\delta^{\rho\mu}+\hbox{antisymmetrization}\bigg \}~.}}

When we compactify the heterotic string on $T^4$, the structure of the
$t_8F^4$ terms becomes more interesting. Now, at generic points in moduli
space the gauge group is $U(1)^{24}$. We focus on the $t_8F^4$ term
involving only the sixteen gauge bosons coming from the Cartan torus
of $SO(32)$.  The coefficient of this term is no longer fixed by
anomaly considerations, and it depends on the Narain moduli. At the
same time, there are several arguments (reviewed, for example, in
\kiritsis) that support the claim that this term does not depend on
the heterotic string coupling $g_h$, \ie\ it is one-loop exact. 
Perturbative corrections in $g_h$ can be shown to be absent, as in the
ten dimensional case (see \eg\ \BachasMC);
this was explicitly verified at two-loop order in \StiebergerWK.
Non-perturbatively, the only identifiable BPS instanton is the
heterotic fivebrane, but there are no six-cycles in $T^4$ which the
Euclidean fivebrane can wrap. Finally, the independence from the
string coupling is suggested by the decoupling between the
gravitational multiplet (to which the dilaton belongs) and the vector
multiplets, as seen in the factorization of the moduli space.

The computation of the one-loop contribution to the effective Lagrangian 
near singularities in the case at hand is very similar to the one performed
for the conifold in \AntoniadisZE. The term in the effective Lagrangian 
arising from the one-loop computation is
\eqn\leff{
{\cal L}_{eff}= l_h^2 t_8F^IF^JF^KF^L\int _{\cal F}{d^2\tau\over \tau_2^2}
\sum _{(p_L,p_R)\in \Gamma _{4,20}} p_R^Ip_R^Jp_R^Kp_R^L \; \tau_2^2 \;
{q^{{1\over 2}p_L^2}\bar q^{{1\over 2}p_R^2}\over \bar \eta (\bar
\tau)^{24}}~,} where we suppressed the space-time indices of $t_8$ and
of the gauge fields, which run over $0,\dots,5$. $\cal F$ is the
fundamental domain of the moduli space of complex structures of the
(worldsheet) torus, $l_h=1/M_h$ is the heterotic string length, and
the sum runs over the even, self-dual lattice $\Gamma_{4,20}$ corresponding
to a particular value of the Narain moduli. 
The contribution $\tau_2^2$ comes from the zero modes on
$T^4$, and the $1/\bar \eta (\bar \tau)^{24}$ is the contribution of
the right-moving bosonic lattice. To describe the points of enhanced
gauge symmetry in the moduli space, we start by recalling the mass
formula for perturbative BPS states \DabholkarJT :
\eqn\bpsmass{
{1\over 4}\alpha' M^2={1\over 2}p_L^2={1\over 2}p_R^2+(N_R-1)~.}  The
sixteen gauge bosons of the Cartan subalgebra have $p_L=p_R=0$ and
$N_R=1$, and are massless everywhere in moduli space. At the origin of
moduli space there are additional massless states with $N_R=0$ and
$p_L^2=0, p_R^2=2$. These are W-bosons corresponding to roots of
$SO(32)$. At generic points in the moduli space of $\Gamma_{4,20}$
they are massive. We will be interested in the behavior near points of
enhanced ADE gauge symmetry, where some or all of them are light. Near
such points, the expression \leff\ simplifies enormously.  The leading
contribution comes from the light W-bosons, and is dominated by
$\tau_2\rightarrow \infty$:
\eqn\asymint{
\int_{\cal F}d^2\tau e^{-\pi \tau_2 \alpha'M_W^2}=
{1\over \pi \alpha' M_W^2}\big(1+{\cal O}(\alpha' M_W^2)\; \big )~ ,}
where $M_W$ is the mass of the light $W$ boson.

All in all, near a point of enhanced gauge symmetry, the effective Lagrangian
\leff\ reduces to
\eqn\leffr{
{\cal L}_{eff}\sim \sum _{\vec{\alpha} > 0} 
{t_8(\vec F\cdot \vec \alpha)^4\over
M_W^2(\vec \alpha)}\left(1+{\cal O}(M^2_W/M^2_h)\right)~,}
where the sum runs over the roots of the gauge group corresponding to
the light gauge bosons at a particular point in moduli space. As
expected, this is the same as the one-loop result in the low-energy
field theory described above. It is easy to show that the string
calculation reduces in this limit to the field theory one-loop
calculation reviewed in \S2.1.

\newsec{Heterotic/type II duality}

In this section we review the duality between the heterotic string on $T^4$,
the type IIA string on $K3$, and configurations of $NS5$-branes in type
IIB string theory, paying special attention to the 
origin of the $t_8F^4$ terms
in these backgrounds. We will focus on the behavior near points in moduli space
where the gauge symmetry is enhanced.

\subsec{Duality for compact spaces}

There is strong evidence for a strong-weak coupling duality between type IIA
string theory on $K3$ and heterotic string theory on $T^4$ \HullYS. 
In particular, their six dimensional ${\cal N}=(1,1)$ supersymmetric effective
actions can be identified by a change of variables that implies the following
relations between the six dimensional string couplings and string scales :
\eqn\dualrel{l_h=g_{II}l_{II}~,\;\;\;\;g_h=1/g_{II}~.}
The massless content of type IIA string theory on $K3$ at generic points
in moduli space is one ${\cal N}=(1,1)$ graviton multiplet and twenty
${\cal N}=(1,1)$ vector multiplets. Since the ${\cal N}=(1,1)$ graviton
multiplet contains four graviphotons, at generic points in
moduli space the gauge group is  $U(1)^{24}$, as on the heterotic
side. Since all the gauge bosons are in the Ramond-Ramond (RR) sector, 
no perturbative
string states are charged under them, only $D$-branes.

Twenty two of the twenty four gauge bosons come from reductions of the
type IIA RR three-form potential
on two-cycles of the $K3$.   Of these twenty two, three correspond to
self-dual forms and nineteen to anti-self-dual forms. Sixteen of these
nineteen gauge bosons map to the sixteen gauge bosons that
on the heterotic side come from the Cartan subalgebra of the ten
dimensional gauge group. At the
particular point in moduli space where the $K3$ is a $T^4/\IZ_2$
orbifold, these sixteen gauge bosons arise from the twisted sectors of the
orbifold, corresponding to states localized at the fixed points, one gauge
field per fixed point.

Taking into account the normalization of the $RR$ fields in the type
IIA action, and \dualrel , the $t_8F^4$ term \leff , which was a one
loop effect on the heterotic side, must appear at {\it tree level} on
the type IIA side.  We can present some arguments for the
non-renormalization of this term directly on the type IIA side: the
identifiable BPS instantons would be Euclidean $D0$ and $D2$ branes,
but there are no one-cycles or three-cycles for these branes to wrap.
At the $T^4/\IZ_2$ orbifold point of the moduli space, the tree level
$t_8F^4$ terms in the type IIA string theory were successfully compared
to the one-loop heterotic results in \KiritsisZI.

By tuning the moduli of the $K3$, one can reach singular surfaces,
which contain some two-cycles shrunk to zero size. These singularities
of $K3$ follow an $ADE$ classification. They provide the type IIA
description of the enhanced ADE gauge symmetry that is visible
perturbatively on the heterotic side.  The W-bosons correspond in this
description to $D2$-branes wrapping the vanishing cycles.

The expression \leffr\ for the $t_8F^4$ term near a point of
enhanced gauge symmetry was derived for the heterotic string under the
condition $M_W^2\ll M_h^2 $, but without any restriction on the value of
$g_h$. For $g_h\gg 1$, the same expression must have a weakly coupled
type IIA interpretation. To see what it is,  recall
that since the fundamental string on the heterotic side is given by the
type IIA $NS5$-brane wrapping the $K3$, their tensions are related by $T_h=
T_{NS5}\; {\rm vol}(K3)\; l_{II}^4$, where ${\rm vol}(K3)$ is the dimensionless
volume of the $K3$ in type II string units. Thus,
$M_h^2\sim {\rm vol}(K3)/(l_{II}^2 g_s^2)$, where $g_s$ is the
ten dimensional IIA coupling, related
to the IIA six dimensional coupling $g_{II}$ by $g_s^2=g_{II}^2\;
{\rm vol}(K3)$. Furthermore, in the type IIA theory the W-bosons are wrapped
$D2$-branes, so $M_W\sim {\rm vol}(\hbox{two-cycle}   )/(l_{II} g_s)$,
where ${\rm vol}(\hbox{two-cycle}   )$ is the volume of the two-cycle
the $D2$-brane wraps, in string units. It follows that the
condition for the validity of \leffr\ on the type II side is 
\eqn\volcond{{\rm vol}(\hbox{two-cycles}) \ll \sqrt {{\rm vol}(\hbox {K3})}~.}
This condition is purely geometric; in particular, it is independent
of the type IIA string and Planck scales and of the type IIA string coupling. 

We see that in the limit \volcond, the structure of the $t_8F^4$
terms is only sensitive to physics near the ADE singularity on
the $K3$ (through the volumes of the shrinking two-cycles) and not
to the detailed properties of the full theory. One can isolate that
physics by studying the ``near-horizon'' region of the singularity;
we will discuss this in more detail below.

It is interesting to note that unlike the heterotic case, in the weakly
coupled type II limit $g_s\to 0$, there are actually two 
regimes that need to be analyzed separately. Denoting the string scale
of the type II theory by $M_s$ $(M_s=1/l_{II})$, the physics is qualitatively
different when $M_W\gg M_s$ and when $M_W\ll M_s$. The difference
has to do with the fact that the W-bosons correspond to wrapped $D$-branes
in the type II description. When $M_W\gg M_s$, they are very heavy, and a
perturbative string description of the physics associated with the small cycles
\volcond\ is possible. This regime will be discussed in detail in \S4, \S5.
It is there that the $t_8F^4$ interaction \leffr\ must arise at tree
level as discussed above, and we will show that it indeed does.

On the other hand, when $M_W\ll M_s$, the type II description is strongly 
coupled, and string perturbation theory breaks down, as is clear from 
the fact that in this regime the non-perturbative wrapped $D$-branes of 
mass $\simeq M_W$ are much lighter than perturbative string states.
For the purposes of studying the $t_8F^4$ term \leffr\ this region is actually
simple, since the lightest massive states are the W-bosons, and their dynamics
is given at low energies by ${\cal N}=(1,1)$ 
six dimensional SYM. Thus, the $t_8F^4$
term arises in this regime at one-loop, from the Feynman graphs discussed
in \S2, with the wrapped $D$-branes running in the loop.

So far, we have reviewed the duality between the heterotic string on $T^4$ and
the type IIA string
on $K3$. There is a further duality relating these compactifications
to a configuration of $NS5$-branes in type IIB string theory
\refs{\ovcigar,\david}. To discuss this duality, consider a
particular point in moduli space, type IIA on $T^4/\IZ_2$.
At this point in the moduli space the eighty moduli of the $K3$ CFT split
naturally into sixteen controlling the size
and shape of the $T^4/\IZ_2$, and the remaining sixty four, which
correspond to blow up modes at the sixteen orbifold singularities.

Suppose, for simplicity, that the four-torus is a product of four circles,
$T^4=(S^1)^4$. T-duality on one of the circles relates \david\ this
background to type IIB on $T^4/\IZ_2$, where the $\IZ_2$ acts in a
non-standard way. A simple way of thinking about the resulting IIB
background is as an S-dual  of type IIB with sixteen orientifold five-planes
and a $D5$-brane sitting at each of the $O5$-planes (a T-dual description
of type I string theory on $T^4$). 
Thus, each of the $\IZ_2$ fixed points carries
$(-1)$ units of $NS5$-brane charge, which is canceled by an $NS5$-brane
sitting at the fixed point. This type IIB vacuum does not have moduli that
blow up the $\IZ_2$ singularities. The original sixty four blow-up modes
of the type IIA description map to moduli describing the positions of the
sixteen $NS5$-branes on the $T^4/\IZ_2$ (four real moduli per brane).
At the orbifold point the sixteen $NS5$-branes coincide with the
fixed planes, but the duality holds also at generic points in the moduli space.
In this realization, the W-bosons are $D$-strings suspended between
different $NS5$-branes; the enhanced gauge symmetry occurs when two or
more $NS5$-branes are brought together.

\subsec{Duality for non-compact spaces}

In the previous subsection we described three dual realizations of the same
physics. By varying the moduli, we can reach points with an enhanced gauge 
group. These gauge groups follow an $ADE$ classification, but their rank
can be at most twenty four. However, if we consider
local singularities embedded in a non-compact $K3$ surface,  we can realize
configurations with arbitrary $ADE$ gauge group. Since eventually we
will be interested in all  possible LSTs with sixteen supercharges (which arise
from decoupling limits of such singularities), we discuss
this non-compact duality in some detail.

For concreteness consider type IIA on a $\IC^2/\IZ_k$ singularity.
The twisted sector includes $(k-1)$ six dimensional ${\cal N}=(1,1)$
vector multiplets, including NS-NS and RR fields, associated to the
different two-cycles which vanish at the orbifold point. The moduli
come from the NS-NS sector: the Kaluza-Klein reduction of the $B$ field on the
two-cycles gives $(k-1)$ moduli $B_i$, which are singlets under the
$SU(2)$ that rotates the complex structures; $(k-1)$
triplets of scalars $\vec \zeta _i$ account for metric
deformations. Altogether, we have $4(k-1)$ real moduli.

This is related\foot{See \HarveyWM\ for a more detailed discussion.}
by $T$-duality to $k$ $NS5$-branes of type IIB string theory arranged on a
circle of radius $R$, in the limit
\eqn\dualimit{
g_s^B \rightarrow 0,\hskip1cm R/l_{II}^B \rightarrow 0, \hskip1cm
g_s^A={g_s^Bl_{II}^B \over R} \;\; \hbox {fixed}~.}
The $(k-1)$ $B_i$ are mapped to the relative
distances between the $NS5$-branes on the circle, while the $(k-1)$
triplets correspond to their transverse positions in the remaining $\IR^3$.
The limit of enhanced $SU(k)$ gauge symmetry
in which the $B_i \to 0$ corresponds to bringing the $k$
$NS5$-branes together in the transverse $\IR^4$ (the circle can be ignored in
this limit).

We can regard this non-compact case as a limit of the compact case
described in the previous subsection. Obviously, in this limit
equation \volcond\ is satisfied, so the heterotic dual satisfies
$M_W^2 \ll M_h^2$ and equation \leffr\ should be valid.

For future reference, we recall that the type II solution
for a collection of $k$ parallel
$NS5$-branes, at different points $\vec{x}_n$ in the transverse $\IR
^4$, is given by \chs
\eqn\nsmetric{\eqalign{
&ds^2=-dt^2+dy_1^2+\dots+dy_5^2+e^{2(D-D_0)}
(dx_6^2+dx_7^2+dx_8^2+dx_9^2)
\, ;\cr
&e^{2(D-D_0)}=1+\sum _{n=1}^k
{\alpha' \over (\vec{x}-\vec{x}_n)^2}\, ;\cr
&H_{\mu \nu \lambda}=-\epsilon _{\mu \nu \lambda}^\sigma
\partial _\sigma D~.}}
Here, $D$ is the dilaton, and $D_0$ is related to the asymptotic
string coupling far from the $NS5$-branes,
$g_s=\exp(D_0)$. $H_{\mu\nu\lambda}$ is the field strength
of the NS-NS $B$ field $B_{\mu\nu}$. When all the $NS5$-branes coincide,
say at the origin, the transverse metric and dilaton near the 
branes are \chs
\eqn\chsmetric{\eqalign{
ds^2 &={k \alpha' \over r^2}(dr^2+r^2d\Omega _3^2)=d\phi^2+k\alpha' 
d\Omega _3^2~,\cr
D &=-{\phi\over\sqrt {k\alpha'}}~,
}}
where $d\Omega_3^2$ is the metric on a unit three-sphere, 
and the new radial coordinate 
$\phi$ is defined by $e^{\phi/\sqrt{k\alpha'}}=r e^{-D_0} / 
\sqrt{k\alpha'}$. Note that in
the limit \dualimit\ we take $g_s^B \rightarrow 0$. One might have
naively thought
that in this limit, the dynamics on the worldvolume of the $NS5$-branes
becomes trivial, but as we review in the next section, that is not the case.
The near-horizon limit \chsmetric\ of the $k$ coincident $NS5$-branes 
defines a non-trivial $5+1$ dimensional theory, a Little String Theory. 
Since the $t_8F^4$ terms are expected to be independent of the string
coupling, one should be able to compute them in LST. We will devote
the bulk of this paper to performing this computation and comparing it
with the heterotic result \leffr.

\newsec{Type IIA string theory on a near-singular $K3$}

As discussed in the previous section,
when the $K3$ surface develops two-cycles whose size is much smaller
than that of the whole $K3$, or in the T-dual picture, when the $NS5$-branes
approach each other to within a short distance, it is
expected that the leading contribution to the $t_8F^4$ terms comes
from the vicinity of the singularity. This is
certainly true when the mass of the W-bosons, $M_W$, is much smaller
than the string scale of the type II theory, $M_s$, since then the
low-energy field theory approximation is valid. We will see below
that it is also true for $M_W\gg M_s$, the region of interest
here\foot{This is natural, since the bulk of the $K3$ cannot
contribute terms that go like negative powers of $M_W$, such as those
in equation \leffr.}.  Thus, we are led to study the type II theory in a limit
where we decouple the bulk of the $K3$, and focus on the
``near-horizon'' geometry of the singularity.  In this section we
describe some features of this limit. We will often suppress constants and
factors of $M_s$ in the discussion below, exhibiting them only
when it seems necessary to do so.

\subsec{The decoupling limit and holography}

Near an $A_{k-1}$ ALE singularity\foot{For simplicity, we restrict in
the present discussion to A-series singularities. 
It should be possible to generalize all of our results to the D- and 
E-series.},
the $K3$ can
be described as the surface
\eqn\alea{z_1^k+z_2^2+z_3^2=0~,}
in $\IC^3$. The manifold \alea\ contains a conical singularity at 
$z_1=z_2=z_3=0$. One can think of the overall scale of $(z_1,z_2,z_3)$ 
as the radial distance away from the singularity (we will make this more
precise below).

Of course, \alea\ provides an accurate description of the geometry
only very close to the singularity. In the full geometry, it has to be
attached to the rest of the $K3$. Since the bulk of the $K3$ will not
contribute to our calculations, we will neglect it and take the target
space to be the non-compact $K3$ surface
\alea\ all the way to $z_i=\infty$.

Moreover, we would like to decouple any gravitational physics in the
bulk of the ALE space. This can be achieved by sending $g_s\to0$; by the
duality described in \S3.2, this limit in type IIA string theory 
is the same as the near-horizon limit of
$k$ $NS5$-branes in type IIB string theory, and as
described above, taking this limit 
does not affect the $t_8F^4$ terms we are interested
in. Normally,
string theory becomes free in the $g_s\to 0$ limit, 
but here this is not the case, since
non-trivial dynamics remains in the vicinity of the singularity. 
This can be seen
by embedding the ALE space \alea\ in a larger class of deformed spaces,
\eqn\aledef{z_1^k+z_2^2+z_3^2=\mu~.}
For $\mu\not=0$ the conical singularity is smoothed out, and the formerly
vanishing two-cycles get a finite volume\foot{This follows from the
form of the holomorphic two-form on the deformed ALE space \aledef,
$\Omega=dz_1\wedge dz_2/2z_3$.}
\eqn\finitev{V_{S^2}\simeq\mu^{1\over k}~.}
If one keeps the volume of the two-cycles fixed (and non-zero) as $g_s\to 0$,
the theory becomes free, as is customary in string theory. To get a non-trivial
theory, one can study the double-scaling limit \gkone
\eqn\doublesc{\eqalign{&\mu\to0, \;\;g_s\to 0,\cr
                       &M_W\simeq {\mu^{1\over k}\over g_s}={\rm fixed}~,\cr
		       }}
in which the Planck scale goes to infinity, but the mass of $D$-branes 
wrapped around the collapsing two-cycles remains finite. In effect, in 
this limit the scale $M_W$ replaces the Planck scale as the energy above
which the theory becomes strongly coupled and non-perturbative effects
become important. Since this scale is not associated with strong
gravity effects, the resulting theory is non-gravitational -- it is a
Little String Theory. The above discussion also makes it clear that
the non-trivial dynamics in the limit \doublesc\ is localized near the
singularity, as explained in \S3.

The double scaling limit \doublesc\ contains a tunable dimensionless
parameter, $M_W/M_s$. If $M_W\ll M_s$ (an extreme case of which is the
original ALE space \alea), string theory near the singularity becomes
strongly coupled well below the string scale (at $E\simeq M_W$), and
there are few useful tools for studying it except in the limit $E\to
0$, where it reduces to a (free) six dimensional $SU(k)$ $\cn=(1,1)$
supersymmetric Yang-Mills theory. This regime is of less interest to
us, since the $t_8F^4$ terms are well understood in it (see \S2 and \S3).
In the opposite limit, $M_W\gg M_s$, we expect a perturbative
description to exist for energies $E\ll M_W$, and in particular for
$E\sim M_s$.  Thus, the situation here is similar to that in
perturbative critical string theory (with the Planck scale replaced by
$M_W$). This is the regime that we will study in the rest of this paper.

A very similar story can be told in the T-dual language of
$NS5$-branes in type IIB string theory on $T^4/\IZ_2$.  The T-dual of
\alea\ involves $k$ coincident $NS5$-branes at a regular point on
$T^4/\IZ_2$.  To isolate the physics of the $NS5$-branes, one proceeds
as in the ALE case. First, focus on the geometry near the $NS5$-branes,
taking the transverse space to be $\IR^4$. This leads to the CHS
metric \nsmetric. Then, send the (asymptotic) string coupling to zero.
This gives rise to the background (see \chsmetric)
\eqn\fivebrhor{\eqalign{
ds^2&=dx^\mu dx_\mu+k\alpha'(d\sigma^2+d\Omega_3^2)~,\cr
D &=-\sigma~,\cr
}}
where $\sigma$ is related to $r=|\vec x|$ of equation 
\nsmetric\ by  $r=\sqrt{k\alpha'} g_s\exp\sigma$,
and we have suppressed the $B$-field in \nsmetric. 

The background \fivebrhor\ (with the appropriate $B$-field) 
can be described by an exact CFT,
\eqn\chsback{\IR^{5,1}\times\IR_\phi\times SU(2)_k~,}
where the real line $\IR_\phi$ is labeled by 
$\phi\equiv \sqrt{k\alpha'}\sigma$, 
the dilaton goes like 
\eqn\qtwok{D=-{Q\over2}\phi;\;\;Q\equiv {2\over\sqrt{k\alpha'}}~,}
and the linear dilaton causes the central charge of the $\phi$ CFT to be
\eqn\cphi{c_\phi=1+{3\alpha'\over 2}Q^2=1+{6\over k}~.}
The supersymmetric $SU(2)_k$ CFT consists of a level $(k-2)$ bosonic
$SU(2)$ WZW model, as well as three free fermions $\psi_a$, $a=1,2,3$,
which transform in the adjoint of an $SU(2)_2$ affine Lie algebra,
completing the total level of $SU(2)$ to $k$.  There are also free
fermions $\psi_\mu$, $\mu=0,1,\cdots,5$, and $\psi_\phi$, which are
the worldsheet superpartners of $x_\mu$ and $\phi$, respectively.  The
ADE classification of singularities of $K3$ surfaces is mapped in the
description \chsback\ to the ADE classification of $SU(2)$ modular
invariants.

Geometrically, the different components of \chsback\ can be thought of
as follows.  $\IR^{5,1}$ is the worldvolume of the $NS5$-branes,
$\IR_\phi$ parameterizes the radial direction away from the branes,
while $SU(2)_k$ describes the angular three-spheres at constant
distance from the branes. The $SU(2)_L\times SU(2)_R$ symmetry
associated with the $SU(2)_k$ CFT is identified with the $SO(4)$
rotation symmetry around the $NS5$-branes.

The background \chsback\ can be understood from the ALE point of view as well
\refs{\ovcigar,\GiveonZM}. By writing
\eqn\rescz{(z_1,z_2,z_3)=(\lambda^{2\over k}y_1,\lambda y_2,\lambda y_3)}
where $\lambda=r\exp(i\theta)\in \IC$, and $(y_1,y_2,y_3)$ take values
in a weighted complex projective space, one can argue that $r$
parameterizes the radial direction $\phi$ in \chsback, $\theta$ labels the
Cartan subgroup of $SU(2)_k$, while $(y_1,y_2,y_3)$, which satisfy
\alea\ as well, parameterize the coset $SU(2)_k/U(1)$.  Thus, the ALE
description corresponds to the parametrization
\eqn\parafz{\IR_\phi\times SU(2)_k\simeq \IR_\phi\times\left(S_k^1\times 
{SU(2)_k\over U(1)}\right)/\IZ_k~.}  
The radius of $S^1_k$ is $\sqrt{\alpha'k}$. An important fact is that
\chsback, \parafz\ are only valid for $k\ge 2$,
\ie\ for two or more coincident $NS5$-branes. For $k=1$, the ALE space
\alea\ is smooth and, in the T-dual language, a single $NS5$-brane does
not develop a throat.

The near-horizon description of the singularity \chsback\ is strongly coupled;
the string coupling $g_s= e^D$ diverges as $\phi\to-\infty$, which
corresponds to the location of the $NS5$-branes, 
or the tip of the cone in ALE. 
This behavior is in agreement with our general considerations above, since 
\chsback\ corresponds to the case $\mu=0$ in \aledef, or $M_W=0$ in \doublesc. 
Thus the theory is strongly coupled for any finite energy, 
and the weakly coupled 
description for low energy is not via string theory on \chsback, but rather
via supersymmetric Yang-Mills theory.

In order to arrive at a weakly coupled string theory description, we
would like to separate the $NS5$-branes in the transverse space, or
equivalently smooth out the tip of the cone, as in \aledef. This is
expected to eliminate the strong coupling singularity\foot{In the
$NS5$-brane picture, this is because as mentioned after equation \parafz, a
single $NS5$-brane does not have a throat along which the string
coupling can diverge.} in \chsback, and introduce the tunable
parameter, $M_W/M_s$, into the problem. To do that it is useful to
have in mind the holographic interpretation of the background
\chsback, to which we turn next.

In \abks\ it was proposed that string theory in asymptotically linear
dilaton space-times, such as \chsback, is holographically related to a
dual theory, a Little String Theory.  This relation is analogous to that of the
AdS/CFT correspondence \adscft. The analogy is most straightforward in the
$NS5$-brane picture, where the duality relates a decoupled theory on the 
world-volume of a stack of $NS5$-branes, with string theory in the near-horizon 
geometry of the branes. The $\phi$ direction in \chsback\ is expected
to play the role of the holographic direction, while the $S^3$ is
associated with the $SO(4)$ global symmetry of the LST. Observables
correspond to non-normalizable vertex operators\foot{A second class
of observables corresponds to $\delta$-function normalizable vertex 
operators. We will not discuss those here.} whose wavefunctions
diverge at the weakly coupled ``boundary'' $\phi\to\infty$.  Correlation
functions of such operators correspond to off-shell Green's functions
in the dual LST. We will be interested here in observables
corresponding to short representations of space-time supersymmetry,
which as discussed in \abks, are in one to one correspondence with
gauge-invariant operators in short representations of supersymmetry in the 
low-energy gauge theory.

The low energy limit of the LST contains $SU(k)$ ${\cal N}=(1,1)$ 
SYM theory, and we can label the operators of the LST using their 
descriptions as operators in this gauge theory. An example of the 
correspondence between LST operators and non-normalizable
vertex operators that will be useful below is:
\eqn\chiralops{
\ttr(\Phi^{i_1}\Phi^{i_2}\cdots\Phi^{i_{2j+2}})\leftrightarrow
e^{-\varphi-\bar\varphi}(\psi\bar\psi \psu_j)_{j+1;m,\bar m}e^{Q\tj \phi} 
e^{ip\cdot x}~,} 
for $j=0,1/2,1,\cdots,(k-2)/2$, 
where $\Phi^i$, $i=6,7,8,9$ are the four scalar fields in the adjoint
of $SU(k)$, which parameterize the locations of the $NS5$-branes in the
transverse directions, and on the left-hand side of
\chiralops\ one should consider only the symmetric, traceless components
in $(i_1,i_2,\cdots, i_{2j+2})$ (this is required for the operator to be
in a short representation). On the right-hand side of \chiralops,
$\varphi, \bar\varphi$ are the bosonized superconformal ghosts,
$\psu_{j;m,\bar m}$ is a primary of the bosonic $SU(2)_{k-2}$ WZW 
model\foot{We review some properties of this model in appendix B.},
and the notation $(\psi\bar\psi \psu_j)_{j+1;m,\bar m}$ means that we are
coupling the fermions in the adjoint of $SU(2)$, $\psi^a$, with the
bosonic part into a primary of total spin $(j+1)$ and $(J_3^{\rm
tot},\bar J_3^{\rm tot})=(m,\bar m)$. The values of $m$ and 
$\bar m$ depend on the precise indices appearing on the left-hand side. 
The mass shell condition provides
a relation between $\tj$, $j$ and $p_\mu$, as we review  below.

The notation $\ttr$ refers to the fact that the operator in
question has the same quantum numbers as the trace, but it is not
precisely equal to the trace. Rather, it is a combination of the
single-trace operator with multi-trace operators such as 
${\rm tr}(\Phi^{i_1}\Phi^{i_2}){\rm tr}(\Phi^{i_3}\cdots\Phi^{i_{2j+2}})$.
Such a mixing occurs quite generally in holographic dualities such as 
the AdS/CFT
correspondence, but usually one is only interested in the limit $j \ll k$
where the multi-trace contributions are negligible. 
In this
paper we are not taking the large $k$ limit, so the distinction 
is in principle important. 
When $j \sim k$ it is known in various examples that specific
combinations such as subdeterminants appear 
\refs{\BalasubramanianNH\CorleyZK-\AharonyND}. The precise
combinations of single and multi-trace operators that correspond
to single string vertex operators in our case have been determined in \agk
\foot{In different contexts
it may be more natural to choose the single string vertex operators
to correspond precisely to single-trace operators.
It is well-known that different choices of contact terms in a worldsheet CFT
lead to different parametrizations of the space of couplings \KutasovXB. We 
believe that with different choices of such contact terms we can change the 
multi-trace content of the operators \chiralops\ (which are related to
couplings as described below) and go between the theory we 
describe here and the theory in which they are single-traces.
}.
However, at the particular point in moduli
space we will be working, the multi-trace components of \chiralops\ will 
not contribute to the specific computations we will do in \S5, \S6,
and we can treat these operators as single-trace operators. 
We will normalize the expression $\ttr$ such that it is equal to the 
single-trace operator with coefficient one, plus multi-trace operators. 
Thus, for the purposes of \S5, \S6 of this paper, one can replace
$\ttr\to {\rm tr}$ in all expressions.

The mass-shell condition for
the vertex operator \chiralops\ reads
\eqn\massshellv{Q^2(\tj-j)(\tj+j+1)=p^2~,}
with the larger root $\tj$ of this equation corresponding to the
non-normalizable vertex operator.  The statement of holography is that
correlation functions of the vertex operators
\chiralops\ in the bulk theory correspond to off-shell Green's functions of the
operators $\ttr(\Phi^{i_1}\Phi^{i_2}\cdots\Phi^{i_{2j+2}})$ in the
UV completion of six dimensional super Yang-Mills theory provided by LST.

Another useful example of the correspondence is obtained by acting 
on \chiralops\ once with a chiral and once with an anti-chiral
space-time supercharge in a way which creates a two-form operator (with
polarization $\zeta^{\mu \nu}$),
\eqn\chiraltop{\eqalign{
\zeta^{\mu \nu}[\ttr(F_{\mu \nu} \Phi^{i_1} \Phi^{i_2} \cdots 
\Phi^{i_{2j+1}})& + \ttr(\lambda \gamma_{\mu \nu} {\bar \lambda} \Phi^{i_1} 
\cdots \Phi^{i_{2j}})]\leftrightarrow\cr
& \zeta^{\mu \nu}
e^{-\half (\varphi + {\bar \varphi})}
S_a\gamma_{\mu \nu}^{a {\dot{a}}}{\bar S}_{\dot{a}}
(S\bar S \psu_j)_{j+\half;m,\bar m}e^{Qj\phi}~.}}
On the left-hand side, $\lambda$ is a gaugino which transforms in the
$({\bf 4}, {\bf 2})$ representation of Spin(5,1) $\times$ Spin(4); 
$\bar\lambda$ transforms in the $({\bf \bar 4}, {\bf \bar 2})$. The RR vertex 
operator on the right-hand side is written  for $p_\mu=0$; the
general form is more complicated (and will appear below for a special
case). $(S\bar S \psu_j)_{j+\half;m,\bar m}$ corresponds to the coupling
of the spin fields constructed out of the worldsheet fermions
$\psi_1, \psi_2,\psi_3,\psi_\phi$
(and their antiholomorphic counterparts), which transform in the spin
$1/2$ representation of $SU(2)_L$ ($SU(2)_R$), with the spin
$j$ operator $\psu_j$, into an operator transforming in the spin $(j+1/2)$
representation. The spin fields
$S_a$ and ${\bar S}_{\dot{a}}$ transform as the ${\bf 4}$ and ${\bf \bar 4}$
of Spin(5,1), respectively.

To verify that the operators \chiralops, \chiraltop\ are BRST invariant,
as well as for our subsequent analysis, one needs to use the detailed
structure of the CFT corresponding to the CHS background \chsback. We
summarize some of the relevant results in appendix A.

\subsec{The moduli space of deformed ALE spaces}

As discussed in the previous subsection, to get a weakly
coupled worldsheet description of ALE spaces or $NS5$-branes,
we must deform the singularity, or equivalently separate the $NS5$-branes.
In this subsection, we discuss the relevant deformation from
the point of view of string theory in the CHS background \chsback.

From the low-energy field theory point of view, we would like to give the
scalars $\Phi^i$ (see \chiralops) VEVs of the form
\eqn\expvalphi{\vev{\Phi^i}={\rm diag}(\phi^i_1,\phi^i_2,\cdots,\phi^i_k)~,}
where $\phi^i_n$, $i=6,7,8,9$, $n=1,2,\cdots, k$, is the location of the
$n$'th $NS5$-brane in the $i$'th direction. Since the SYM potential
is proportional to ${\rm tr}[\Phi^i,\Phi^j]^2$, \expvalphi\ are indeed
flat directions of the potential. The $NS5$-brane picture makes it clear that
\expvalphi\ are flat directions even when the VEVs $\phi^i_n$ are large.
Since the low-energy gauge group is $SU(k)$ (the ``center of
mass'' $U(1)$ is not part of the decoupled interacting theory)
we will set $\sum_n\phi_n^i=0$.

How do we describe the VEVs  \expvalphi\ in the CHS background?
The vertex operators corresponding to chiral gauge-invariant combinations of
the  $\Phi$'s are given in \chiralops\foot{In the low-energy field theory
non-chiral operators also obtain vacuum expectation values in the configuration
\expvalphi. The corresponding
vertex operators in string theory are not known, but they do not seem to play
an important role for the purposes of this paper.}. 
Adding the vertex operators on
the right-hand side of \chiralops\ to the worldsheet Lagrangian corresponds, in
the low-energy field theory, to adding the operators on the left-hand side of
\chiralops\ to the space-time Lagrangian. As is well-known in the context
of the AdS/CFT correspondence \BalasubramanianSN, 
if instead we want to give expectation values to the operators
\chiralops, we have to add to the worldsheet Lagrangian the 
normalizable versions of the vertex operators  \chiralops, 
which are obtained by sending
$\tj \to - \tj-1$ in  \chiralops\foot{The precise definition and
meaning of these operators will be described in \S5.2.}.
Thus, we are led to consider worldsheet Lagrangians of the form
\eqn\defL{\CL=\CL_0+\lambda_{j;m,\bar m} G_{-\half} \bar G_{-\half}
(\psi\bar\psi \psu_j)_{j+1;m,\bar m}e^{-Q(j+1)\phi}~,}
where $\CL_0$ is the Lagrangian describing the CHS background,
$G$ is the supercurrent defined in appendix A,
and the couplings $ \lambda_{j;m,\bar m}$ are determined by the
values of the moduli \expvalphi. Note that we have shifted the operator
to the $(0,0)$ picture and that we have set the space-time
momentum $p_\mu=0$, since we are interested in describing a condensate
that is constant in space-time.

The number of couplings $\lambda_{j;m,\bar m}$ in \defL\ is in general
larger than the number of parameters determining the point in moduli
space. Indeed, the former goes like $k^3$ for large $k$, while the
latter is equal to $4(k-1)$. Thus, it must be that in order to obtain
a sensible worldsheet theory from \defL, one has to impose relations
on the $\lambda$'s, which follow from the fact that they are functions
of the $\phi^i_n$ \expvalphi.  We will see later an example of such
a relation which is understood in the worldsheet theory. In general,
the origin of such relations is not fully understood from the worldsheet
point of view.

Note also that adding the terms in \defL\ to the worldsheet Lagrangian
does not modify the background near the boundary at $\phi\to\infty$, but
as $\phi\to-\infty$, the new terms grow and regularize the
divergence coming from the strong coupling region in the CHS solution.
This is in agreement with the target space picture. Far away from the
tip of the ALE cone, or from the locations of the $NS5$-branes (but still
in the near-horizon geometry), one cannot tell whether the singularity
has been smoothed out or not. Upon approaching the singularity, one notices
that the $\{\phi^i_n\}$ have been turned on, and the singularity has been
smoothed out.

Since the general case is complicated, we  next (following \gktwo)
restrict the discussion to a subspace of the moduli space \expvalphi,
by restricting the $NS5$-branes to move in a plane. Thus, we denote
\eqn\abcdef{\eqalign{A&\equiv \Phi^6+i\Phi^7,\cr
                     B&\equiv \Phi^8+i\Phi^9,\cr
}}
and keep $\langle A\rangle=0$ while varying
\eqn\bvev{\langle B\rangle ={\rm diag} (b_1,b_2,\cdots, b_k);\;\;
\sum_{n=1}^k b_n=0~.}
It is convenient to embed 
the $SO(2)_A\times SO(2)_B$ symmetries
of rotations of the $A$, $B$ planes as follows in the $SU(2)_L
\times SU(2)_R$ symmetry of the CHS background. The generator
of  $SO(2)_A$ will be taken to be $J_3^{\rm tot}- \bar J_3^{\rm tot}$,
while that of $SO(2)_B$ is $J_3^{\rm tot}+ \bar J_3^{\rm tot}$.
The charges are normalized such that
\eqn\chargeab{\eqalign{(J_3^{\rm tot}+ \bar J_3^{\rm tot})(A)=&0~;\;\;
(J_3^{\rm tot}- \bar J_3^{\rm tot})(A)=1~;\cr
(J_3^{\rm tot}+ \bar J_3^{\rm tot})(B)=&1~;\;\;
(J_3^{\rm tot}- \bar J_3^{\rm tot})(B)=0~.\cr
}}
Then, one finds that (at zero space-time momentum),
\eqn\ops{\eqalign{\ttr(A^l B^{2j+2-l})&\leftrightarrow e^{-\varphi-\bar\varphi}
(\psi{\bar\psi}\psu_j)_{j+1;j+1,j+1-l}e^{Qj\phi}~,\cr
\ttr(A^l (B^*)^{2j+2-l})&\leftrightarrow e^{-\varphi-\bar\varphi}
(\psi{\bar\psi}\psu_j)_{j+1;-j-1+l,-j-1}e^{Qj\phi}~,}}
and in particular
\eqn\btwojtwo{\ttr(B^{2j+2})\leftrightarrow e^{-\varphi-\bar\varphi}
\psi^+\bar\psi^+\psu_{j;j,j}e^{Qj\phi}= e^{-\varphi-\bar\varphi}
e^{i(H+\bar H)} \psu_{j;j,j}e^{Qj\phi}~.}
We will also be interested later in the corresponding operators
for \chiraltop. One has
\eqn\otwojone{\co_{2j+1}^+\equiv \zeta^{\mu\nu} \ttr(F_{\mu\nu}B^{2j+1}
+{\rm fermions})\leftrightarrow \zeta^{\mu\nu}
e^{-\half(\varphi+\bar \varphi)}
S_a\gamma_{\mu \nu}^{a {\dot{a}}}{\bar S}_{\dot{a}}
e^{{i\over2}(H+H'+\bar H+\bar H')}\psu_{j;j,j}e^{Qj\phi}~.}
The field $\psi^+$ and the bosonized
fermions $H$, $H'$, $\bar H$ and $\bar H'$, appearing in
\btwojtwo\ and \otwojone,
are defined in appendix A.

In order to describe condensation of $B$, as in \bvev, we are led to
study the perturbed worldsheet Lagrangian
\eqn\pertB{\CL=\CL_0+\sum_j (\lambda_j G_{-\half} \bar G_{-\half}
\psi^+\bar\psi^+\psu_{j;j,j}e^{-Q(j+1)\phi}+{\rm c.c.})~.}
These perturbations are particularly nice, since the operators
$\psi^+\bar\psi^+\psu_{j;j,j}e^{-Q(j+1)\phi}$
are {\it chiral}. Indeed, by using the form of the
$\cn=2$ superconformal generators $G^\pm$ given in appendix A, one
can show that
\eqn\chireq{G^+(z)\psi^+\bar\psi^+\psu_{j;j,j}e^{-Q(j+1)\phi}(w)=
{\rm regular}\;{\rm as}\; z\to w~,} 
and similarly for $\bar G^+$.  Thus, one can think
of the perturbations in \pertB\ as turning on a worldsheet
superpotential. A useful way of thinking about this is in terms of the
decomposition
\parafz. We have an infinite cylinder $\IR_\phi\times S^1_k$, labeled by
$\phi$ and $Y$, where $Y$ is defined by
\eqn\defY{J_3^{\rm tot}={i\over Q}\partial Y~,}
and $Y$ is canonically normalized. Here, and in the rest of the
section, we set $\alpha'=2$. 
$SU(2)_k/U(1)$ is an $\cn=2$ minimal model, which can 
be described in terms of
a Landau-Ginzburg superfield $\chi$, with superpotential
\eqn\supmin{W=\chi^k~.}
In these variables, one can write
\eqn\prodchir{\psi^+\bar\psi^+\psu_{j;j,j}e^{-Q(j+1)\phi}=\chi^{k-2(j+1)}
e^{-Q(j+1)(\phi-iY)}=\chi^{k-2(j+1)}e^{-Q(j+1)\Phi}~,}
where in the last step we have defined a chiral superfield whose bottom
component is $\Phi=\phi-iY$ (following standard practice, we will denote
both the superfield and its bottom component by $\Phi$, and similarly for
$\chi$). The Lagrangian \pertB\ can be written as
\eqn\lpertch{\CL=\CL_0+\sum_j \left(\lambda_j \int d^2\theta
\chi^{k-2(j+1)}e^{-Q(j+1)\Phi}+{\rm c.c.} \right)~.}
%
One can use the correspondence
\btwojtwo\ to relate the $\{\lambda_j\}$ to the locations of the $k$
$NS5$-branes in the $B$-plane:
\eqn\ljb{\lambda_j \sim \langle\ttr(B^{2j+2})\rangle~.}
%

While the structure of the theory on the full moduli space labeled by
$\{\lambda_j\}$ \lpertch\ is of interest, and can probably be analyzed
using our techniques, we will further specialize to the subspace of moduli
space corresponding to $NS5$-branes which are equally spaced on a circle 
in the $B$-plane of radius $r_0$,
\eqn\bakfiv{b_n=r_0e^{2\pi i n / k}~.}
At this particular point in moduli space, ${\rm tr}\vev{B^l}=0$ for
all $l<k$, so all the possible multi-trace terms, such as ${\rm tr}\vev{B^l}
{\rm tr}\vev{B^{k-l}}$  vanish, and there is no difference between 
evaluating the VEVs of the 
operators $\ttr(B^l)$ defined in \chiralops\ and the 
ordinary single-trace operators $\tr(B^l)$. In 
this case we have, using \ljb, 
\eqn\lamres{\lambda_j=\mu\delta_{j,{k-2\over2}}~,}
with
\eqn\murrel{\mu\sim r_0^k~,}
and the deformed ALE space is described by \aledef.
The worldsheet theory \lpertch\ simplifies in this case, since
the perturbation decouples from the $\cn=2$ minimal model:
\eqn\lntwo{\CL=\CL_0+\left(\mu \int d^2\theta e^{-{1\over Q}\Phi}+{\rm c.c.}
\right)~.}
The resulting worldsheet theory is $\cn=2$ Liouville times an $\cn=2$
minimal model (with a $\IZ_k$ identification \parafz). 
The exponential superpotential
cuts off the strong coupling divergence of the string coupling 
$g_s$ as $\phi\to-\infty$.

It is known that $\cn=2$ Liouville is equivalent as a CFT to the coset
$SL(2)_k/U(1)$, which describes string propagation on a cigar of
asymptotic radius $\sqrt{2k}$ \refs{\gkone,\HoriAX,\newgk,\TongIK}. In
the cigar description of this theory, the strong coupling region
$\phi\to-\infty$ is excised altogether, and as long as $g_s$ at the
tip of the cigar $g_s^{({\rm tip})}$ (which scales as $\mu^{-1/k}$)
remains small, the theory is weakly coupled\foot{Recall that the
string coupling is largest at the tip of the cigar.}.

In fact, the space-time point of view on the deformations \defL\ sheds
interesting light on the duality between the $\cn=2$ Liouville and
cigar CFTs. In our discussion of the condensate \bakfiv\ we focused
on the expectation values of the chiral operators $\ttr(B^{2j+2})$,
but the VEV \bakfiv\ leads to non-zero expectation values of other
chiral operators as well.  
For example, the chiral operator $\ttr(BB^*- AA^*)$, 
which is a special case of the right-hand side of \chiralops, has a VEV: 
\eqn\vevbbaa{\langle \ttr(BB^*- AA^*)\rangle=k r_0^2~.}
According to the dictionary \chiralops, \defL, this corresponds to
turning on the perturbation
\eqn\bhpert{\delta\CL\sim k r_0^2 G_{-\half} \bar G_{-\half}
\psi_3\bar\psi_3e^{-Q\phi} + c.c.~,}
in the worldsheet sigma model. This perturbation is well-known to be
the leading term in the expansion of the metric of the cigar around
$\phi=\infty$, the region far from the tip. The higher order terms in
that expansion correspond to VEVs of higher order chiral operators,
involving higher powers of $(BB^*)$ and $(A A^*)$.

Thus, we see that from the space-time point of view, both the cigar and
$\cn=2$ Liouville perturbations are present in the worldsheet
Lagrangian corresponding to
\bakfiv, with related coefficients. For some purposes, one can 
focus on the superpotential terms \lpertch; for others, the cigar
picture (including \bhpert) is more useful. In general, both have to be taken
into account. The relation between the two deformations will become
clearer when we discuss relations between normalizable operators in the next 
section. 
All this ties in nicely to the worldsheet analysis of these
theories. For example, the relation between the Liouville and cigar
perturbations was determined in \newgk.

\lref\GavaGV{
E.~Gava, K.~S.~Narain and M.~H.~Sarmadi,
``Little string theories in heterotic backgrounds,''
Nucl.\ Phys.\ B {\bf 626}, 3 (2002)
[arXiv:hep-th/0112200].
}

At the special point in moduli space that we are now discussing \bakfiv,
the throat CFT becomes \ovcigar
\eqn\slsutwo{(SL(2)_k/U(1)\times  SU(2)_k/U(1))/\IZ_k~.}
The wrapped $D2$-branes discussed earlier in this section 
correspond to $D$-branes localized near the tip of the cigar, times
various $D$-branes in the $\cn=2$ minimal models  \GavaGV.
Their mass is proportional to $r_0$ and satisfies
\eqn\msgstip{M_W\simeq {M_s\over g_s^{({\rm tip})}}\gg M_s~.}
They are heavy non-perturbative objects in the limit we are studying,
as indicated in equation \msgstip.

\subsec{Vertex operators in the $(SL(2)_k/U(1)\times  SU(2)_k/U(1))/\IZ_k$ 
background}

In order to calculate correlation functions in the deformed background
\slsutwo, one has to know what the different observables of string theory
on the CHS background, discussed above, 
correspond to upon the resolution of the singularity.
This can be understood using standard CFT techniques. For example, the vertex
operator of $\ttr(B^{2j+2})$, \btwojtwo, corresponds in the deformed
background \slsutwo\ to
\eqn\glgl{e^{i(H+\bar H)} \psu_{j;j,j}e^{Qj\phi}
\leftrightarrow  
V^{(su,susy)}_{{k\over2}-j-1;-{k\over2}+j+1,-{k\over2}+j+1} 
V^{(sl,susy)}_{j;j+1,j+1}~, }
where $V^{(su,susy)}_{j;m,m}$ is a primary in the $\cn=2$ minimal model
with dimension and R-charge 
\eqn\drjmsv{\Delta=\bar\Delta={j(j+1)-m^2\over k};\;\; R=\bar R=-{2m\over k}~,}
while  $V^{(sl,susy)}_{j;m,m}$ is an $\cn=2$ primary in $SL(2)_k/U(1)$; it has
\eqn\djsl{\Delta=\bar\Delta={m^2- j(j+1) \over k};\;\; R=\bar R={2m\over k}~.}
Similarly, for the RR operators \otwojone\ one has
\eqn\ggrr{e^{{i\over2}(H+H'+\bar H+\bar H')} \psu_{j;j,j}e^{Qj\phi}
\leftrightarrow  V^{(su,susy)}_{j;j,j}(RR,+) V^{(sl,susy)}_{j;j,j}(RR,+)~,}
where   $V^{(su,susy)}_{j;m,m}(RR,\pm)$ is a RR sector operator in the
$\cn=2$ minimal model; it has
\eqn\delrrr{\Delta=\bar\Delta={1\over 8}+{j(j+1)-(m\pm\half)^2\over k};\;\;
R=\bar R=\pm\half-{2m\pm 1\over k}~,} while
$V^{(sl,susy)}_{j;m}(RR,\pm)$ is a similar object in the cigar CFT,
which has
\eqn\rrrsl{\Delta=\bar\Delta={1\over 8}-{j(j+1)-(m\pm\half)^2\over k};\;\;
R=\bar R=\pm\half+{2m\pm 1\over k}~.}
Thus, $V^{(su,susy)}_{j;j,j}(RR,+)$ has dimension 
$\Delta={1\over 8}- {1\over 4k}$
and corresponds to a RR ground state in the $\cn=2$ minimal model (which
has $c = 3 - {6\over k}$). Similarly,
$V^{(sl,susy)}_{j;j,j}(RR,+)$ has dimension $\Delta={1\over 8}+ {1\over 4k}$
and corresponds to a RR ground state of the cigar CFT.

As we will see in the next section, in our analysis we will also need
to use operators whose asymptotic form for large $\phi$ is
\eqn\asmptt{
e^{{i\over2}(H-H'+\bar H-\bar H')} \psu_{j;j,j}e^{Qj\phi} {\rm\ \ or\ \ }
e^{{i\over2}(H-H'+\bar H-\bar H')} \psu_{j;j,j}e^{Q(-j-1)\phi}~.}
In the deformed background \slsutwo, they correspond to
\eqn\drjmslv{\eqalign{
e^{{i\over2}(H-H'+\bar H-\bar H')} \psu_{j;j,j}e^{Qj\phi}
&\leftrightarrow
V^{(su,susy)}_{j;j,j}(RR,+) V^{(sl,susy)}_{j;j+1,j+1}(RR,-)~,\cr
e^{{i\over2}(H-H'+\bar H-\bar H')} \psu_{j;j,j}e^{Q(-j-1)\phi}
&\leftrightarrow
V^{(su,susy)}_{j;j,j}(RR,+) V^{(sl,susy)}_{-j-1;j+1,j+1}(RR,-)~.\cr
}}
The operator 
$V^{(sl,susy)}_{-j-1;j+1,j+1}(RR,-)$, whose precise definition will be
discussed in \S5.2 below, has the same dimension as the operator
$V^{(sl,susy)}_{j;j,j}(RR,+)$, and also corresponds to a ground state of the
cigar CFT.

It is sometimes useful to bosonize the $\cn=2$ minimal model,
by using its description as a product of a bosonic $SU(2)/U(1)$
(parafermion) theory and a compact canonically normalized
scalar field (which we will
label by $P$) \QiuZF. The NS-NS sector $\cn=2$ primaries can be written
as
\eqn\vvjjmmggns{V^{(su,susy)}_{j;m,m}=V_{j;m,m}^{(su)} e^{i\alpha_m P}~,}
where $V_{j;m,m}^{(su)}$ is a primary in the parafermion theory whose dimension
is given by
\eqn\delparajmns{\Delta(V_{j;m,m}^{(su)})=\bar\Delta(V_{j;m,m}^{(su)})=
{j(j+1)\over k}-{m^2\over k-2}~,}
and
\eqn\alphamkns{\alpha_m={2m\over\sqrt{k(k-2)}}~.}
Indeed one can check using \drjmsv, \vvjjmmggns\ and \delparajmns\ that
\eqn\checkdecparns{\Delta(V^{(su,susy)}_{j;m,m})=\Delta(V_{j;m,m}^{(su)})+\half
\alpha_m^2~.}
Note that $\alpha_m$ is related to the R-charge of the $V^{(su)}_{j;m,m}$ 
as follows:
\eqn\rchargesuns{R^{(su)}=-\alpha_m\sqrt{1-{2\over k}}=-{2m\over k}~.
}
In fact, the scalar $P$ can be thought of as a bosonized version
of the $U(1)_R$ current of the $SU(2)/U(1)$ SCFT.
In the RR sector one has
\eqn\vvjjmmgg{V^{(su,susy)}_{j;m,m}(RR,\pm)=V_{j;m,m}^{(su)} 
e^{i\alpha_m^{\pm} P}~,}
where
\eqn\alphamk{\alpha_m^{\pm}={2m\mp\half(k-2)\over\sqrt{k(k-2)}}~.}
One can check using \delrrr, \delparajmns, \alphamk, that
\eqn\checkdecpar{\Delta(V^{(su,susy)}_{j;m,m}(RR,\pm))=\Delta(V_{j;m,m}^{(su)})+
\half ({\alpha^\pm_m})^2~.}
The relation between the R-charge of the operator $V^{(su,susy)}_{j;m}(RR,\pm)$
and the $P$ charge $\alpha_m^{\pm}$ is again (as in \rchargesuns)
\eqn\rchargesu{R^{\pm(su)}=-\alpha_m^{\pm}\sqrt{1-{2\over k}}=\pm\half-
{2m\pm 1\over k}~.}

\noindent
A useful relation between the $(RR,+)$ and $(RR,-)$ operators is
\eqn\plumin{
V^{(su,susy)}_{j;m,m}(RR,+)=V^{(su,susy)}_{{k-2\over 2}-j;-{k-2\over 2}+m,
-{k-2\over 2}+m}(RR,-)~.
}
This reflection property follows from \alphamk\ and from 
the well-known  \FateevMM\ property of the parafermion theory
\eqn\paraplu{V^{(su)}_{j;m,m}=V^{(su)}_{{k-2\over 2}-j;-{k-2\over 2}+m,
-{k-2\over 2}+m}~.}

\noindent
Similarly, it is sometimes useful to separate the $\cn=2$ cigar CFT into
a bosonic $SL(2)/U(1)$ theory times a scalar field.
To generalize the above construction to $SL(2)_k/U(1)$, one simply has to take
$k\rightarrow -k$ and $j\to-j-1$ in all formulae. 
Some of the correlators of the bosonic $SL(2)/U(1)$ theory 
are described in appendix B. The analog of equation \plumin\ is now
\eqn\pluminsl{V^{(sl,susy)}_{j;m,m}(RR,+)\simeq 
V^{(sl,susy)}_{{k-2\over 2}-j;{k+2\over 2}+m,{k+2\over 2}+m}(RR,-)~.}
It can be derived by bosonizing $\cn=2$ $SL(2)_k/U(1)$ in terms of
bosonic $SL(2)/U(1)$ and a free boson and using
\refs{\MaldacenaHW,\ParnachevGW}
\eqn\paraplusl{V^{(sl)}_{j;m,m}\simeq 
V^{(sl)}_{{k-2\over 2}-j;{k+2\over 2}+m,{k+2\over 2}+m}~.}  
The expressions \pluminsl, \paraplusl\ should be understood as statements
about the normalizable states in the theory (which we will discuss in
\S5.2). For appropriate values of $j$ for which the normalizable states
exist, the left-hand side of
\pluminsl\ is equal to the right-hand side up to a $j$-dependent
multiplicative factor. 

\subsec{Analogy to conifold}

This subsection lies somewhat outside the main line of development
of the paper, and can be skipped on first reading. Its main purpose
is to point out the analogy of the foregoing discussion to another
familiar and well studied system -- the conifold, where again, the 
coefficient of a term in the effective action which is protected
by supersymmetry has different interpretations 
in different regimes.

On the geometrical side, the analogy involves replacing the ALE
singularity of $K3$, described (after deformation) by \aledef, by the
deformed conifold singularity of Calabi-Yau manifolds
\eqn\defcon{z_1^2+z_2^2+z_3^2+z_4^2=\mu~.}
Equation \defcon\ describes a conical singularity, in which a shrunken
$S^3$ has been blown up to volume $\simeq \mu$. 
A $D3$-brane wrapped around this $S^3$ gives a hypermultiplet
of mass $M_H\simeq \mu/g_s$. 

One can also realize \defcon\ in terms of a T-dual $NS5$-brane 
system, by studying two $NS5$-branes which intersect on a $3+1$-dimensional
space. 
For example, one can take one of the $NS5$-branes to span the directions
$(x^1,x^2,x^3,x^4,x^5)$, and the other to be extended in 
$(x^1,x^2,x^3,x^8,x^9)$. The intersection of the two $NS5$-branes
is the three dimensional space labeled by $(x^1,x^2,x^3)$. 
Denoting $x^4+ix^5=v$, $x^8+ix^9=w$, the deformation \defcon\
corresponds in the $NS5$-brane language to studying a $NS5$-brane 
with worldvolume $vw=\mu$. 

The analog of the $t_8F^4$ term discussed in this paper in the case
of the conifold is the kinetic $(F^2)$ term of the RR gauge field, under 
which the hypermultiplet coming from the wrapped $D3$-brane is charged. 
Its coefficient, which is related by $\cn=2$ space-time supersymmetry to the
metric on moduli space, goes as $\mu\to 0$ like $\log\mu$. This behavior
has different interpretations in different regimes. 

When the mass of the
hypermultiplet $M_H\ll M_s$, one understands the $\log\mu$ as a consequence
of integrating out at one-loop a light charged hypermultiplet in the low
energy $\cn=2$ Abelian gauge theory of the RR gauge field \StromingerCZ. 
This calculation
can also be thought of as a one-loop calculation in the S-dual heterotic
string \AntoniadisZE.

When the mass of the hypermultiplet $M_H\gg M_s$, the same behavior
arises by studying the perturbative string theory in the deformed
conifold background. Similar arguments to those reviewed earlier
in this section, lead in this case to the background
\eqn\conback{\IR^{3,1}\times{SL(2)_1\over U(1)}~,}
or, equivalently, $\IR^{3,1}\times(\cn=2\;{\rm Liouville})$
with superpotential ($Q^2=2/k=2$ in this case)
\eqn\supcon{W=\mu e^{-{1\over\sqrt2}\Phi}~.}
The string coupling at the tip of the cigar is
\eqn\gstipcon{{1\over g_s^{({\rm tip})}}={M_H\over M_s}\gg 1~,}
such that the string theory \conback\ is weakly coupled in this limit.

The metric on moduli space is given by the two-point function of the modulus
corresponding to changing $\mu$:
\eqn\metmodcon{G=\langle
e^{-{1\over\sqrt2}(\phi-iY)}e^{-{1\over\sqrt2}(\phi+iY)}
\rangle~.}
This is a bulk amplitude in the sense of Liouville theory
(see \eg\ \DiFrancescoUD\ for a discussion). It can be computed using standard
techniques, and one finds,
\eqn\glogmu{G=-\log\left|{\mu\over\Lambda}\right|~,}
where $\Lambda\gg \mu$ is a UV cutoff, and the origin of the $\log$ 
is the ``volume'' of the Liouville direction, from the wall provided by the
superpotential \supcon, or the tip of the cigar, to a cutoff $\phi_0\simeq
\log\Lambda$ far from the wall. In other words, the origin of the $\log\mu$
behavior in this limit is in the continuum of {\it perturbative} string states
living in the long throat that develops when $M_H$ is much smaller than
$M_{\rm Planck}$. Note in particular that the logarithmic contribution that
arises at one-loop for $M_H\ll M_s$, is a tree level effect for $M_H\gg M_s$.

Clearly, the situation on the conifold is similar to the ALE case discussed 
here.
Our case is more complicated, both because we are considering a larger class
of singularities, which can be thought of as analogous to generalized
conifolds,
\eqn\genconeq{z_1^k+z_2^2+z_3^2+z_4^2=\mu~,}
and because in six dimensions we are led to study four-point functions
rather than two-point functions (since we would like to compute $F^4$ terms
rather than $F^2$ terms).

\newsec {Correlation function computations}

In this section we will compute the $t_8F^4$ terms in the
low-energy effective action of $A_{k-1}$ LSTs in
the regime $M_W\gg M_s$ discussed in \S4. To do
this we will study the appropriate correlation functions of the
gauge-invariant operators  \chiraltop.

As mentioned above, we are mostly interested in performing
computations at a specific point in the moduli space of the LST where
the VEVs of the scalar fields in the low-energy $SU(k)$ gauge theory
are given by \bakfiv\ 
\eqn\vevab{\eqalign{
&\vev{A} = 0~,\cr
&\vev{B} = r_0 \, {\rm diag}(e^{2\pi i / k},
e^{4\pi i / k}, \cdots, e^{2\pi i(k-1)/k},1)~.
}}
Recall that we have set $M_s=1/\sqrt2$, and that the theory is weakly coupled
when $r_0$ is much larger than the string scale and $M_W \propto r_0$ 
obeys $M_W \gg M_s$.
We will focus on the operators
\eqn\fmu{\co^+_{2j+1} = \zeta^{\mu \nu} \ttr(F_{\mu \nu}
B^{2j+1} + {\rm fermions}\cdot B^{2j})~,} 
and their complex conjugates
\eqn\fmustar{\co^-_{2j+1} = \zeta^{\mu \nu}
\ttr(F_{\mu \nu} (B^*)^{2j+1} + {\rm fermions}\cdot
(B^*)^{2j})~.}
We will not be careful about the precise form of the fermionic
terms in these operators since they will not contribute at the leading
order in $M_s/M_W \simeq g_s$ to the correlation functions we will compute.

The form of the Ramond-Ramond vertex operators corresponding to the
operators \fmu\ with zero momentum in $\IR^{5,1}$ in the CHS theory
is given by \otwojone.
We will need the vertex operators at non-zero momentum $p_{\mu}$;
these take the form (in the CHS theory, or in the resolved theory for large
$\phi$) :
\eqn\opforms{\eqalign{
&{\hat \co}^+_{2j+1}  = \zeta^{\mu \nu}
e^{-\half \varphi -\half {\bar \varphi}}
\gamma_{\mu \nu}^{a {\dot{a}}} e^{{i\over 2} H+{i\over 2}{\bar H}} 
\psu_{j; j, j} e^{Q {\tj}\phi} e^{i p \cdot x}
\cdot \cr & \left( S_a e^{{i\over 2} H'} + {i \over{Q(j+\tj+1)}}
(\gamma^{\rho})_a^{\dot c} p_{\rho}
S_{\dot c} e^{-{i\over 2} H'} \right) \left(
{\bar S}_{\dot{a}} e^{{i\over 2} {\bar H}'} +
{i \over {Q (j+\tj+1)}} ({\gamma}^{\sigma})_{\dot{a}}^c p_{\sigma} {\bar S}_c
e^{-{i\over 2} {\bar H}'} \right)~, \cr
&{\hat \co}^-_{2j+1}  = (\zeta')^{\mu' \nu'}
e^{-\half \varphi -\half {\bar \varphi}} 
\gamma_{\mu' \nu'}^{b {\dot{b}}} e^{-{i\over 2}H - {i\over 2}{\bar H}}
\psu_{j; -j, -j} e^{Q \tj \phi} e^{i p \cdot x} 
\cdot \cr & \left( S_b e^{-{i\over 2} H'} +{i \over{Q(j+\tj+1)}} 
(\gamma^{\rho})_b^{\dot d} p_{\rho}
S_{\dot d} e^{{i\over 2} H'} \right) \left( 
{\bar S}_{\dot{b}} e^{-{i\over 2} {\bar H}'} +
{i \over {Q (j+\tj+1)}} ({\gamma}^{\sigma})_{\dot{b}}^d p_{\sigma} {\bar S}_d
e^{{i\over 2} {\bar H}'} \right)~, \cr}}
where $\tj$ is the larger of the two solutions to the mass-shell condition
\eqn\massshell{Q^2 (\tj-j)(\tj+j+1) = p^2~.}
${\hat \co}_{2j+1}^-$ (with momentum $(-p_{\mu})$) is the complex conjugate of
${\hat \co}_{2j+1}^+$ (with momentum $p_{\mu}$).  
We will begin in \S5.1 by
computing the two-point functions of these operators, in order to
verify that their low-energy behavior is the same as that  expected
for the LST operators $\co^{\pm}_{2j+1}$ (up to normalization). This will
lead us to a discussion of amputated (normalizable) versions of these
operators in \S5.2. Then, in \S5.3-\S5.5, 
we will analyze the four-point functions of these operators, 
first in the general case and
then for particular values of the $j$'s for which the computation simplifies.

\subsec{Two-point functions}

In order to verify the operator identifications described above,
we wish to compare the two-point function of the operators
\fmu, \fmustar\ $\vev{\co^+_{2j+1} \co^-_{2j+1}}$ in
the low-energy field theory to the string theory expectation value
$\vev{{\hat \co}^+_{2j+1} {\hat \co}^-_{2j+1}}$, at leading order in the
string coupling (or, equivalently, in $1/r_0$ or in 
$M_s / M_W$). We expect that the
two will be identical up to a normalization which we have not
determined above, and which will be fixed by the following computations\foot{
In fact, as shown in \agk, the string theory computation
described below receives also contributions that do not come from the low
energy gauge theory. Fortunately, the gauge theory contribution differs
from the string theory amplitude by a multiplicative
$j$-independent constant (which is
determined in \agk). Since in this paper we focus mostly on the $j$
dependence of the amplitudes, this will not affect our final results.}.

Let us start by discussing the correlator in the low-energy field theory.
The maximal power of $r_0$ arises if we replace all the $B$'s in \fmu\ by
their VEVs \vevab\ and contract the two gauge fields. The fermionic terms
give rise to lower powers of $r_0$ so they will not be relevant. Multi-trace
contributions involve factors of $\tr(\vev{B^l})$ with $l<k$, which vanish at
this point in moduli space. If
we normalize the Abelian gauge fields in the low-energy field theory to
obey $\vev{A_{\mu}(p) A_{\nu}(-p)} = \eta_{\mu \nu} / p^2$
(in Feynman gauge),
we find at leading order
\eqn\ftcorr{\eqalign{
\vev{\ttr(F_{\mu \nu} B^{2j+1})(p) & \ttr(F_{\mu' \nu'}
(B^*)^{2j+1})(-p)} =\cr
& \vev{\tr(B^{2j+1} (B^*)^{2j+1})}
{{p_{\mu} p_{\mu'} \eta_{\nu \nu'} \pm (\mu \leftrightarrow \nu, \mu'
\leftrightarrow \nu')} \over p^2}~,\cr}}
where $\vev{\tr(B^{2j+1} (B^*)^{2j+1})} = k r_0^{2(2j+1)}$.

We next turn to the two-point function in type IIB string theory. The
operators \opforms\ were written in the $(-1/2,-1/2)$-picture,
so in order to compute their two-point function on the sphere 
we need to either shift them to another picture or
add another operator in the $(-1,-1)$-picture\foot{For example, by 
differentiating
the two-point function with respect to the coupling $\mu$, \lntwo.}. 
We will use the former method.
Using the fact that the picture-changing operator is given by $(G(z)
e^{\varphi}(z) +  ghost\ terms)$ and the form of $G$ from appendix A, we
find that 
the form of ${\hat \co}_{2j+1}^+$ in the $(-3/2,-1/2)$, $(-1/2,-3/2)$ and
$(-3/2,-3/2)$ pictures is given by the following expressions\foot{One can
write these vertex operators in other, BRST-equivalent, ways.} (at large 
$\phi$):
\eqn\nopforms{\eqalign{
{\hat \co}^+_{2j+1} = & \zeta^{\mu \nu} e^{-{3\over 2}\varphi -
\half {\bar \varphi}} 
\gamma_{\mu \nu}^{a {\dot{a}}} e^{{i\over 2}H + {i\over 2}{\bar H}} 
\psu_{j; j, j} e^{Q \tj \phi} e^{i p \cdot x} 
\cdot \cr & 
\left( {\sqrt{2} i \over {Q(j+\tj+1)}} S_a 
e^{- {i\over 2}H'}\right) \left( 
{\bar S}_{\dot{a}} e^{{i\over 2} {\bar H}'} +
{i \over {Q (j+\tj+1)}} ({\gamma}^{\sigma})_{\dot{a}}^c p_{\sigma} {\bar S}_c
e^{-{i\over 2} {\bar H}'} \right)~, \cr
{\hat \co}^+_{2j+1} = & \zeta^{\mu \nu}
e^{-\half \varphi -{3\over 2} {\bar \varphi}} 
\gamma_{\mu \nu}^{a {\dot{a}}} e^{{i\over 2} H+{i\over 2}{\bar H}} 
\psu_{j; j, j} 
e^{Q {\tj}\phi} e^{i p \cdot x} 
\cdot \cr & \left( S_a e^{{i\over 2} H'} + {i \over{Q(j+\tj+1)}} 
(\gamma^{\rho})_a^{\dot c} p_{\rho}
S_{\dot c} e^{-{i\over 2} H'} \right) \left( 
{{\sqrt{2} i}\over {Q(j+\tj+1)}}
{\bar S}_{\dot{a}} e^{-{i\over 2} {\bar H}'} \right)~, \cr
{\hat \co}^+_{2j+1} = & \zeta^{\mu \nu}
e^{-{3\over 2} \varphi -{3\over 2} {\bar \varphi}} 
\gamma_{\mu \nu}^{a {\dot{a}}} e^{{i\over 2} H+{i\over 2}{\bar H}} 
\psu_{j; j, j} 
e^{Q {\tj}\phi} e^{i p \cdot x} 
\cdot \cr & 
\left( {\sqrt{2} i \over {Q(j+\tj+1)}} S_a 
e^{- {i\over 2}H'}\right) \left( 
{{\sqrt{2} i}\over {Q(j+\tj+1)}}
{\bar S}_{\dot{a}} e^{-{i\over 2} {\bar H}'} \right)~, \cr}}
with $\tj$ related to $j$ as in \massshell. The expressions for ${\hat
\co}_{2j+1}^-$ are the complex conjugates of these. Note that the situation
with Ramond-Ramond vertex operators here is different than in flat
space-time.  There, the $(-1/2,-1/2)$ picture vertex operator
corresponds to the field strength of the RR gauge field, while the
$(-1/2,-3/2)$ and $(-3/2,-1/2)$ picture vertex operators correspond to
the gauge field \PolchinskiRR. Here, the vertex operators in all
pictures correspond to the gauge-invariant operators \chiraltop.

Consider \eg\ the two-point function of ${\hat \co}_{2j+1}^+$, with
space-time momentum $p_{\mu}$ and ``Liouville momentum'' $\tj_1$,
in the $(-3/2,-3/2)$ picture (last line of \nopforms) with
${\hat \co}_{2j+1}^-$, with space-time
momentum $(-p_{\mu})$ and ``Liouville momentum'' $\tj_2=\tj_1=\tj$,
in the $(-1/2,-1/2)$ picture  \opforms. It is useful to start by
analyzing the scaling and conservation laws of this two-point function
using the description of the theory as an $\cn=2$ Liouville theory times
a minimal model, as described in the previous section \lntwo, \slsutwo.

The  correlator in question scales with  the Liouville coupling
$\mu$ as follows:
\eqn\scltwo{\langle {\hat \co}_{2j+1}^+{\hat \co}_{2j+1}^-\rangle
\sim \mu^a\bar \mu^b~.}
The scaling exponents $a$, $b$ can be determined by using the symmetries.
Momentum conservation in the $Y$ direction\foot{Using
\defY, and noting that $J_3^{\rm tot}({\hat \co}_{2j+1}) = j+{1\over 2}$.}
leads to the condition $a=b$, since the contributions of the operators
${\hat \co}_{2j+1}^{\pm}$ cancel (they have opposite $J_3^{\rm tot}$ charges).
By looking at the momentum in the $\phi$ direction, taking into account the
background charge of this scalar on the sphere, we find that 
$Q\tj_1+Q\tj_2- {1\over Q}(a+b)=-Q$, or
$a+b=2(\tj_1+\tj_2+1)/k$. At low momentum $\tj \sim j$ and we get
$a = b \simeq (2j+1) / k$, meaning that the correlator
scales as $(\mu {\bar
\mu})^{{2j+1}\over k} \simeq r_0^{2(2j+1)}$, just
as we found in the field theory \ftcorr.

Next, we look at the momentum in the $H'$ direction (a similar analysis
holds for $\bar H'$). Since $a=b$, the Liouville interactions do not
contribute to this, and ${\hat \co}_{2j+1}^+$ in the $(-3/2,-3/2)$ picture
goes like $e^{-{i\over 2}H'}$,
so we only get a contribution from the term scaling as $e^{{i\over 2}H'}$
in the operator ${\hat \co}_{2j+1}^-$.

Now, we are ready to compute the two-point function. As usual in theories
including $SL(2)/U(1)$, we fix the positions of the two operators, and the
additional zero from dividing by the volume of the conformal Killing group
of the sphere with two punctures is canceled by an infinity due to integration
over bosonic zero modes in the $SL(2)/U(1)$ CFT. This infinity is reflected
in the factor $\delta(\tj_1-\tj_2)$ in the $SL(2)$ two-point function (see 
equation (B.19) in appendix B). 
The ratio of the two infinities is a $j$ dependent
constant that is determined in \MaldacenaKM\ and in appendix C.

The ghost contribution to the two-point function is one, and using the fact
that $S_a(z) S_{\dot a}(w) \sim \eta_{a \dot{a}} / (z - w)^{3/4}$ we find
that the $\IR^{5,1}$ contribution (including both left-movers and
right-movers) is
\eqn\rsix{\eqalign{\zeta_{\mu \nu} \zeta'_{\mu' \nu'}
(\gamma^{\mu \nu})^{a \dot{a}}
(\gamma^{\mu' \nu'})^{b \dot{b}} 
\left({\sqrt{2} \over {Q^2(j+\tj+1)^2}} \right)^2 
(\gamma^{\rho})_b^{\dot{d}} p_{\rho} \eta_{a \dot{d}}
(\gamma^{\sigma})_{\dot{b}}^d p_{\sigma} \eta_{d \dot{a}}
& = \cr
\zeta_{\mu \nu} \zeta'_{\mu' \nu'}
{2\over {Q^4 (j+\tj+1)^4}} \tr(\gamma^{\mu \nu} \gamma^{\sigma}
\gamma^{\mu' \nu'} \gamma^{\rho}) p_{\sigma} p_{\rho} & = \cr
\zeta_{\mu \nu} \zeta'_{\mu' \nu'}
{8\over {Q^4 (j+\tj+1)^4}} [2 (\eta^{\mu \mu'} p^{\nu} p^{\nu'} \pm
(\mu \leftrightarrow \nu, \mu' \leftrightarrow \nu')) + p_{\rho}^2 (\eta^{\nu
\mu'} \eta^{\mu \nu'} & - \eta^{\mu \mu'} \eta^{\nu \nu'})]~.
 \cr}}
The same answer arises if we make different choices for the
pictures of the two operators.

To compute the ``throat'' contribution, we use a decomposition 
similar to \drjmslv. In the $SU(2)/U(1)$ theory we have a
two-point function which is normalized to one in the conventions that
we are using. We are left with a two-point function in the
supersymmetric $SL(2)/U(1)$ theory, which (using the decompositions of
\S4.3) is equal to a two-point
function in the bosonic $SL(2)/U(1)$ theory involving
$\vev{V^{(sl)}_{\tj;j+1,j+1} V^{(sl)}_{\tj;-j-1,-j-1}}$,
with $\tj$ related to $j$ and the momentum as in \massshell.
Two-point functions of this type were computed in \gktwo\ and
references therein, and were found to have a pole at zero
momentum\foot{Additional two-point functions, which are not supposed
to have poles in the low-energy field theory, were also found to have
poles in \gktwo. We will not discuss this issue here; in this paper we
will limit ourselves to operators for which this problem does not arise.}.
At this stage we will find it convenient to change the normalization of
the operators \opforms, \nopforms, so that, as described in appendix B,
they are proportional to ${\tilde V}_{\tj;j+1,j+1}$ rather than to
$V^{(sl)}_{\tj;j+1,j+1}$, where $\tilde V$ is the $SL(2)/U(1)$ reduction
of the $SL(2)$ operator $\tilde \Phi$ discussed in appendix B.
In this normalization,
and taking into account
the results of appendix C on the normalization of two-point functions in
string theory on $SL(2)$, we find the following low-momentum
(${\tj} \to j$) behavior\foot{
Here and below, in our $SL(2)/U(1)$ computations 
$\mu$ \lntwo\ is fixed to a particular
value, and a particular normalization is chosen for the $SL(2)/U(1)$ operators.
These choices are implicitly specified by the values of the two- and 
three-point functions in appendix B. As we discuss in the text, the overall
power of $\mu$ in each correlation function can always be reinstated
by a KPZ-type scaling analysis.}
\eqn\sltwopart{\eqalign{
\vev{{\tilde V}_{\tj;j+1,j+1} {\tilde V}_{\tj;-j-1,-j-1}}  =&
{1\over2\pi^2}\left(2j+1\over k\right)\pi\,{{
\Gamma(-2\tj-1) \Gamma(\tj-j) \Gamma(2+j+\tj)}\over {
\Gamma(2\tj+2) \Gamma(-\tj-j-1) \Gamma(j+1-\tj)}}\cr
\simeq &{1\over\pi}\left(2j+1\over k \right)^2{1\over p^2}~.
\cr}}

Putting everything together, we find two terms. The second term in
\rsix\ gives a contribution scaling as a constant at low momentum,
which is presumably non-universal and corresponds to a contact
term in space-time. The first term in \rsix, together with \sltwopart, gives a
contribution which is precisely proportional to our expected answer \ftcorr:
\eqn\ttwwoo{\langle {\hat \co}_{2j+1}^+{\hat \co}_{2j+1}^-\rangle
\simeq{4\over \pi(2j+1)^2} \zeta^{\mu \nu} (\zeta')^{\mu' \nu'}
{{p_{\mu} p_{\mu'} \eta_{\nu \nu'} \pm (\mu \leftrightarrow \nu, \mu'
\leftrightarrow \nu')} \over p^2}~.}
Comparing to \fmu, \fmustar, \ftcorr, we see that the relation between the 
gauge
theory operators $\co_{2j+1}$ and the 
string theory operators ${\hat \co}_{2j+1}$ is (for small $p^2$)
\eqn\relgastr{\co_{2j+1}^\pm\leftrightarrow
r_0^{2j+1}(2j+1) \sqrt{\pi k\over4}{\hat \co}_{2j+1}^\pm~.}

\subsec{Amputated correlation functions and normalizable operators}

Correlation functions of  non-normalizable vertex operators in
holographic backgrounds correspond to correlators of local operators
in the dual theory (in this case a LST).  Because of the
momentum dependence coming from the propagator (for example in \rsix)
these turn out to be somewhat complicated; this problem occurs
also in other correlation functions of these operators, such as their
contractions with the $t_8F^4$ vertex \leffr\ discussed above. It would be
nice if we could study some other class of operators whose correlation
functions would give us directly the amputated correlators in the dual
theory. Such amputated correlators are  related to the S-matrix in the dual 
theory, which is another reason to be interested in them.

Luckily, there is a simple way to get such amputated correlation
functions in holographic theories. Generally, the
non-normalizable modes in such theories are related to insertions of
local operators, while the normalizable modes are related to states
created by these operators. In conformal theories there is a precise
state/operator correspondence, but in general theories there is no such
correspondence, and we expect that
an operator $\co_i(p_{\mu})$ would create (on-shell) states in the
theory for (one or more) specific values of $p^2$. Indeed, the two-point
functions of operators like the ones we discuss here were found
\refs{\gkone,\gktwo} to have a series of poles which may be
interpreted as corresponding to such states. For the appropriate
values of $p^2$, we expect to have normalizable operators which
create these states, and the correlation
functions of the normalizable modes should give us directly the
S-matrix for scattering states of this type, without the ``external''
propagators discussed above. One way to define these amputated (normalizable)
operators is by looking at the limit of the non-normalizable operators as 
their momentum approaches a pole in the two-point function (which behaves 
as $1/(p^2+M^2)$), multiplied by $(p^2+M^2)$
(in order to cancel the ``external'' propagator); obviously this definition
only makes sense when the amputated operator is on-shell.

In the CHS background, one cannot meaningfully talk about normalizable
operators, since they would necessarily be supported in the strong coupling
region. This is connected with the fact, mentioned after \defL, that
naively one would have of order $k^3$ independent normalizable
operators, while we know that when we go on the moduli space
only $4(k-1)$ combinations of them make sense. To have a well-defined notion
of normalizable operators, we need to resolve the singularity, $e.g$ by the
deformation \lntwo. We next discuss the normalizable operators for the 
resolved background.

In theories like the one we are discussing here, the non-normalizable 
operators involve vertex operators of the type ${\tilde V}_{\tj;m,{\bar m}}$ 
in the bosonic $SL(2)/U(1)$ CFT (when we use the decompositions of 
\S4.3), where $\tj$ is the larger root of the mass-shell 
condition \massshell; they behave for large 
$\phi$ as $e^{Q\tj \phi}$. As we mentioned in \S4, we
expect the normalizable operators to have a similar form but with the 
dependence on $\phi$ being instead (for large $\phi$) $e^{Q \tj' \phi}$ 
where $\tj'=-\tj-1$ 
is the smaller root (instead of the larger root) of \massshell.
These operators can be defined by studying the limit of 
${\tilde V}_{\tj;m,m}$
as one approaches a pole of the two-point function. For generic $\tj$, the
expansion of ${\tilde V}_{\tj;m,m}$ at large $\phi$ has the form
(see \eg\ \newgk)
\eqn\expvjm{{\tilde V}_{\tj;m,m}\simeq a(\tj,m)e^{iQmY}\left(
e^{Q\tj \phi}+C(\tj,m)e^{-Q(\tj+1) \phi}+\cdots\right)~,} 
where $a(\tj,m)$ is an overall normalization factor\foot{This
appears because we are using the operators $\tilde V$ rather than
the operators $V^{(sl)}$ (defined in \S4.3), which have the same 
form with $a(\tilde j,m)=1$.} 
and $C(\tj,m)$ is related to the two-point function given in appendix
B (equation (B.19)). As we approach a pole of the two-point function,
for example at $\tj=|m|-1$, $C(\tj,m)$ blows up, and the second term in 
\expvjm\ becomes much
larger than the first. We can define a normalizable operator as
follows:
\eqn\normlim{{\tilde V}_{-|m|;m,m}\equiv\lim_{\tj\to|m|-1}(\tj-|m|+1)
{\tilde V}_{\tj;m,m}~.}
This operator behaves at large $\phi$ like $e^{-Q(\tj+1)
\phi}=e^{-Q|m| \phi}$. Thus, it is indeed normalizable; it creates
from the vacuum the state with the relevant quantum numbers.

Using these normalizable operators in the bosonic $SL(2)/U(1)$ CFT
and the decompositions discussed in \S4.3, we can construct normalizable
versions of the operators \opforms. Using the decompositions \ggrr,
\drjmslv, it
is easy to see that the terms in ${\hat \co}_{2j+1}^+$ scaling as
$e^{{i\over 2}H'}$ or $e^{{i\over 2}{\bar H}'}$ in \opforms\ vanish in
the limit \normlim, and we are left purely with the terms scaling as
$e^{-{i\over 2}(H'+\bar H')}$.

The correlation functions of the normalizable versions of \opforms\ constructed
using \normlim\ compute the amputated correlation functions in the LST, which
are directly related to the S-matrix.  In the next subsections we will use 
them to
compute the $t_8F^4$ term in the low-energy effective Lagrangian of LST.
Similarly, we can construct normalizable versions of the operators
\chiralops, which we already used in our constructions of the
deformed worldsheet Lagrangians in \S4.2.

In the low-energy SYM 
theory, acting on the vacuum with the operators \fmu, \fmustar\
creates various linear combinations of single-particle states of the
$(k-1)$ massless gauge bosons corresponding to the Cartan subalgebra
of $SU(k)$, with coefficients obtained by replacing the $B$'s in the
definition of the operators by their VEVs. At the specific point in moduli
space we are interested in, \vevab, the operators $\co_{2j+1}^+$ and
$\co_{k-2j-1}^-$ create precisely the same combination of massless gauge
fields (up to an overall constant). Thus, we expect that in string theory,
the normalizable versions of these  operators \opforms\ should be the
same (up to a constant).

Indeed, the reflection properties of the $SU(2)/U(1)$ and $SL(2)/U(1)$
vertex operators discussed in \S4.3 lead to precisely such a relation.
The reflection symmetries \plumin,\pluminsl\ imply that the 
non-normalizable operators (in the normalization we used above)
obey the relation
\eqn\reflprop{{\hat \co}_{2j+1}^-={{k-2j-1}\over {2j+1}}\,
{\hat \co}_{k-2j-1}^+}
near the massless pole in their correlation functions.
This follows from the equality (up to a constant) of
the corresponding normalizable operators. This provides 
one more check of the correspondence between the string theory and field 
theory analysis. At other points in the moduli space (away from the origin)
there are still 
relations between the normalizable operators corresponding to the 
$\co^+$'s and $\co^-$'s, but they are more complicated than \reflprop.

\subsec{Four-point functions -- generalities}

In the remainder of this section we will compute the four-point
function of the operators \opforms, for small energies and momenta,
in type II string theory. In the next
section we will compare these results with the heterotic (or low-energy
field theory) computations. We will discuss only the four-point
functions of the operators \opforms\ which include the scalar field
$B$ but not the scalar field $A$ in the low-energy field theory. Note
that the operators \opforms\ are already in the right picture to have a
non-vanishing four-point function.

We start  by analyzing the selection rules for obtaining a non-zero result
and its scaling with $\mu$, $\bar\mu$. Consider the four-point function
of the operators $\hat{\co}_{2j_i+1}^{\alpha_i}$ \opforms\ with
$i=1,2,3,4$, $\alpha_i = \pm$. Assume that this correlator is non-zero
and scales as
\eqn\fourptscal{\langle \prod_{i=1}^4\hat{\co}_{2j_i+1}^{\alpha_i}
\rangle\sim\mu^a\bar\mu^b~.}
First, we impose $Y$-momentum (or $J_3^{\rm tot}$ charge \defY)
conservation. The operators $\hat{\co}_{2j+1}^{\alpha}$ behave as
$e^{iQ\alpha (j+1/2) Y}$. Taking into account the $Y$-dependence of the
$\cn=2$ Liouville interaction \lntwo, we have
\eqn\yscaling{\sum_i Q \alpha_i (j_i+{1\over 2}) -{1\over Q}(a-b) =
0~,}
or
\eqn\selamb{a-b = \sum_i {2j_i+1\over k}\alpha_i ~.}
Next, we impose  $H'$ momentum conservation (for left or right-movers).
The Liouville interaction carries charge $(a-b)$. The non-normalizable vertex
operators \opforms\  have two terms, one that goes like $p_\rho^0
e^{iH'\alpha_i/2}$ and a second one that goes like $p_\rho^1
e^{-iH'\alpha_i/2}$. If we take the $p_\rho^0$ terms from all
four operators, we get the sum rule
\eqn\selection{a-b + \sum_i{\alpha_i \over 2}=0~.}
Clearly, if all $\alpha_i$ have the same sign there are no solutions
of \selamb\ and \selection, so this contribution to the amplitude vanishes.
This is consistent with the 
field theory analysis, since the contraction of four operators involving
$F_{\mu \nu}$ with the vertex \leffr\ in free field theory 
actually involves eight factors of momentum $p_\rho$. 
Therefore, 
one expects the $t_8F^4$ term we are after to arise 
from the term that goes like $p_\rho e^{-iH'\alpha_i/2}$ in 
each of the four vertex operators \opforms\ (times another 
term of the same form from the other worldsheet chirality). 
Using these terms, the $H'$ sum rule takes the form
\eqn\hprimesum{a-b = \sum_i{\alpha_i \over 2}~.}
Combining \selamb\ and \hprimesum\ we see that up to permutations
and complex conjugation, there are three distinct possibilities for the
values of $(\alpha_1,\alpha_2,\alpha_3,\alpha_4)$, and in each
case we get one condition on the $j$'s for obtaining a non-zero
correlation function :
\eqn\casesd{\eqalign{
&(\alpha_1,\alpha_2,\alpha_3,\alpha_4)=(+,+,+,+);
\;\;\;\;j_1+j_2+j_3+j_4=k-2;\;\;\;\;a-b=2,\cr
&(\alpha_1,\alpha_2,\alpha_3,\alpha_4)=(+,+,+,-);
\;\;\;\;j_1+j_2+j_3-j_4={k-2\over2};\;\;\;\;a-b=1,\cr
&(\alpha_1,\alpha_2,\alpha_3,\alpha_4)=(+,+,-,-);
\;\;\;\;j_1+j_2-j_3-j_4=0;\;\;\;\;\qquad a-b=0~.\cr
}}
Note that the different cases are related by the reflection
property \reflprop. To go from the first line to the second,
one takes $j_4\to {k-2\over2}-j_4$; to go from the second to
the third, $j_3\to {k-2\over2}-j_3$. 

Finally, we can find the total power of $r_0$ 
(or $\mu$, \murrel) associated with the
four-point functions \fourptscal, by using KPZ scaling (in $\phi$), 
which gives rise (in the low-momentum limit) to the sum rule
\eqn\sumphi{Q\sum_i j_i-{1\over Q}(a+b)=-Q~.}
Recalling that $r_0\propto |\mu|^{1\over k}$ we conclude that
the four-point function scales like $r_0^{2\sum_i j_i+2}$. 
This is in agreement with the field theory expectation. Indeed,
we expect a factor of $M_W^{-2} \propto r_0^{-2}$ from the vertex \leffr,
and a factor of $r_0^{2\sum_i j_i+4}$ from the expectation
value of $B$ in \fmu, \fmustar. In \S6 we will see that the
selection rules \casesd\ are also in agreement with the field theory
analysis.

We still need to compute the correlation function \fourptscal\
in cases when the selection rules \casesd\ are satisfied. 
In order to compute the four-point function \fourptscal\ in string
theory we have to integrate over the position of one of the four
vertex operators.
Before performing this integration, the amplitude splits into
two parts: a simple part involving the kinematic factors and the 
expectation value associated with the free fields on
$\IR^{5,1}$, and a non-trivial part associated with the 
directions transverse to the $NS5$-branes. 

The correlation function of the $Spin(5,1)$ spin fields
is given by
\eqn\corspin{\langle S_{\dot a}(0)S_{\dot b}(1)
S_{\dot c}(\infty)S_{\dot d}(z)\rangle=
\epsilon_{\dot a \dot b \dot c \dot d}
[z(1-z)]^{-{1\over4}}~.}
Using this equation and $\gamma$ matrix identities
one can show that the kinematic structure of the four-point
function \fourptscal\ at low momentum is precisely what one 
would obtain in field theory by contracting four operators
of the form \fmu,\fmustar\ against the vertex \leffr. Due to
the non-trivial kinematic factors one
must compute the four-point function at non-zero momentum, and send
the momentum to zero at the end of the calculation.

As mentioned above, the kinematic structure is complicated by the presence
of the four external leg propagators. The discussion of the previous
subsection suggests a nice way to simplify the calculation. 
By going to the poles as a function of the external momenta,
and computing the residue of the poles, as in \normlim, one finds
that the amputated $F^4$ amplitude is proportional to the four-point
function of the normalizable versions of the vertex operators \opforms,
which for large $\phi$ behave as
\eqn\ampnorm{\eqalign{
\hO^+_{2j+1}  = &a(j,j+1)\zeta^{\mu \nu}\gamma_{\mu \nu}^{{\dot{a}}a}
e^{-\half \varphi -\half {\bar \varphi}}S_{\dot a}\bar S_a
e^{{i\over 2}(H-H'+\bar H -\bar {H'})} \psu_{j; j, j}
e^{-Q(j+1)\phi}, \cr
\hO^-_{2j+1}  = &a(j,-j-1)\zeta^{\mu \nu}\gamma_{\mu \nu}^{{\dot{a}}a}
e^{-\half \varphi -\half {\bar \varphi}}S_{\dot a}\bar S_a
e^{{i\over 2}(H'-H+\bar H' -\bar H)} \psu_{j; -j, -j}
e^{-Q(j+1)\phi}~.
\cr}}
Here we have already taken the zero momentum limit; this limit
is non-singular for the amputated amplitude. In \ampnorm\ we
chose to normalize the amputated operators such that their asymptotic
form at large $\phi$ is similar to that of the non-normalizable operators
but with a different exponent of $\phi$. This 
is not the same as the naive normalization of the amputated operators,
in which we simply remove the external free field 
propagators from the operators \opforms.
If we define ``naive amputated operators'' ${\hat \co}_{2j+1}^{\pm}$ by
just removing these external propagators, 
then the two-point function of such an amputated
operator with a non-normalizable operator at $p^2=0$ would be the same as
\ttwwoo\ but without the momentum dependence, or
\eqn\ttwwoon{\langle ({\hat \co}_{2j+1}^+)_{\rm amputated} \cdot
{\hat \co}_{2j+1}^-\rangle
= {4\over \pi(2j+1)^2} \zeta^{\mu \nu} \zeta'_{\mu \nu}~.}
By following through the computation of an
amputated four-point function, one can show that
\eqn\ampfac{
\vev{\prod_{i=1}^4
(\hat{\co}_{2j_i+1}^{\alpha_i})_{\rm amputated}} = 
\vev{
\prod_{i=1}^4\left(-{1\over Q^2(2j_i+1)^2}\cdot
Q^2(2j_i+1)\cdot
\hO_{2j_i+1}^{\alpha_i} \right)},
}
where the first factor comes from the explicit factors
in \opforms, while the second is due to \normlim\ and the fact that the
mass-shell condition \massshell\ implies that near the pole
at $\tj=j$, $1/(\tj-j)=Q^2(2j+1)/p^2$. Thus, we see
that the amputated version of $\hat{\co}^\pm_{2j+1}$ is
\eqn\ampo{\left(\hat{\co}^\pm_{2j+1}\right)_{\rm amputated}=
-{1\over 2j+1} {\hat O}^\pm_{2j+1}~,}
and using \ttwwoon\ we obtain
\eqn\ttwwoonn{\langle {\hat O}_{2j+1}^+ \cdot
{\hat \co}_{2j+1}^-\rangle
= -{4\over \pi(2j+1)}
\zeta^{\mu \nu} \zeta'_{\mu \nu}~.}
In fact, one can simplify the four-point function even further, by 
using \corspin\ and the identity
\eqn\idteight{
\zeta^{\mu_1 \nu_1}_1 \gamma_{\mu_1 \nu_1}^{{\dot{a_1}}a_1}
\cdots\zeta^{\mu_4 \nu_4}_4 \gamma_{\mu_4 \nu_4}^{{\dot{a_4}}a_4}
\epsilon_{\dot a_1 \dot a_2 \dot a_3 \dot a_4}
\epsilon_{a_1 a_2 a_3 a_4}=t_8\zeta_1\zeta_2\zeta_3\zeta_4~.}
Roughly, this means that we can remove the $\gamma$ matrices and 
polarization tensors from \ampnorm, and the 
$\epsilon$ tensor from \corspin, and the resulting
amplitude will compute the object that we are interested in -- the coefficient
of $t_8 F^4$ in the vertex \leffr, written in terms of the variables
\fmu, \fmustar. In the next subsection we will make this more precise.

\subsec{Relation to ${\cal N}=2$ strings}

In the previous subsection we have seen that in order to compute the
coefficient of $t_8 F^4$ in the low-energy effective action of LST we
have to evaluate the four-point function of the normalizable operators
\ampnorm, removing the kinematic parts which refer to $\IR^{5,1}$.

An elegant reformulation of the problem which achieves precisely this
was proposed in \berva. The key
observation is that the CHS background
\parafz, its resolved version \slsutwo, 
and more generally any compact or non-compact
$K3$, is a good background for the $\cn=2$ string. In order to explain
the utility of this observation in our context, we next briefly review
some relevant aspects of $\cn=2$ string theory (see \MarcusWI\ for reviews
and references).

Ordinary type II string theory can be thought of as $\cn=1$ worldsheet
supergravity coupled to matter. In superconformal gauge, this means
that the $\cn=1$ superconformal group generated by the superconformal
generators $G$, $\bar G$, is gauged. The $\cn=2$ string is obtained by
studying $\cn=2$ worldsheet supergravity coupled to $\cn=2$ supersymmetric
matter. In
superconformal gauge, an $\cn=2$ superconformal group is gauged. The
critical central charge is equal to six in this case; the critical
dimension is four. Hence, $K3$ is a good background of this theory.

In addition to the usual bosonic 
reparametrization ghosts $(c,b)$, $\cn=2$ string
theory in superconformal gauge contains two pairs of superconformal
ghosts $(\gamma_1,\beta_1)$ and $(\gamma_2,\beta_2)$ associated with
the two superconformal generators $G^-$ and $G^+$, respectively, and a
pair of $U(1)$ ghosts $(\tilde c,\tilde b)$ associated with the $U(1)$
current in the $\cn=2$ superconformal algebra (as well as the
right-moving counterparts of all these fields).  The BRST current
takes the form
\eqn\brstcharge{J_B=cT+\gamma_1 G^-+\gamma_2 G^++\tilde c J+\cdots~,}
where the $\cdots$ stand for ghost terms that will not play a role below.
We can ``bosonize'' the superconformal ghosts in the usual way
(see \eg\ \PolchinskiRR) by defining
\eqn\phionetwo{\partial\varphi_j=\beta_j\gamma_j;\;\;\;j=1,2~,}
and adding two $(\eta,\xi)$ systems.
In \berva\ it was argued that the $\cn=2$ string (or, equivalently, the
$\cn=4$ topological string) on $K3$ computes
BPS terms in the Lagrangian of the full type II string theory on $K3
\times \IR^{5,1}$. Our $t_8F^4$ coupling is an example of such a term; hence
it should correspond to an observable in the $\cn=2$ string. We next show that
this is indeed the case.

Using formulae in appendix A, it is not difficult to verify that the following
normalizable vertex operators are in the BRST cohomology of the $\cn=2$ string 
in the CHS background $\IR_\phi\times SU(2)_k$, \chsback:
\eqn\ntwoform{\eqalign{
\hO^+_{2j+1} & = a(j,j+1)e^{-\half (\varphi_1 + \varphi_2+{\bar \varphi_1}+
{\bar \varphi_2})
+{i\over 2}(H-H'+\bar H -\bar {H'})} \psu_{j; j, j}e^{-Q(j+1)\phi}~, \cr
\hO^-_{2j+1} & = a(j,-j-1)e^{-\half (\varphi_1 + \varphi_2+{\bar \varphi_1}+
{\bar \varphi_2})
-{i\over 2}(H-H'+\bar H -\bar {H'})} \psu_{j; -j, -j} e^{-Q(j+1)\phi}~.
\cr}}
Comparing these operators to their type II string counterparts
\ampnorm\ we see that the difference is the replacement
\eqn\replii{
\zeta^{\mu \nu}\gamma_{\mu \nu}^{{\dot{a}}a}
e^{-\half \varphi -\half {\bar \varphi}}S_{\dot a}\bar S_a
\rightarrow
e^{-\half (\varphi_1 + \varphi_2+{\bar \varphi_1}+{\bar
\varphi_2})}~.}  This replacement is just what we needed above. Using
equations \corspin\ and \idteight\ it is easy to see that the
unintegrated four-point function of the operators \ampnorm\ is equal
to $t_8\zeta_1\zeta_2\zeta_3\zeta_4$ times the unintegrated four-point
function of the operators \ntwoform. Therefore, the four-point
function of the operators \ntwoform\ in the $\cn=2$ string computes
precisely the quantity of interest -- the coefficient of $t_8 F^4$ in
the effective Lagrangian of LST. This is why, by a slight abuse of
notation, we have denoted these operators in the same way as
\ampnorm.  Note that, using \reflprop\ and \ampo, 
the $\cn=2$ operators \ntwoform\
obey the simple reflection property
\eqn\refprop{\hO_{2j+1}^+=\hO_{k-2j-1}^-~.}

\noindent
In the next subsection we will compute the four-point
function of the operators \ntwoform, and in the next section we will
compare it with the coefficient of the $t_8 F^4$ term which we computed
in \S2.

\subsec{Four-point functions in $\cn=2$ string theory}

The analysis of symmetries performed in \S5.3 together with the
discussion of \S5.4 shows that there are several types of $\cn=2$ 
string correlators of the operators \ntwoform\ that should be 
non-zero:
\eqn\fourpt{\eqalign{
\langle &\hO^+_{2j_1+1}\hO^+_{2j_2+1}\hO^+_{2j_3+1}\hO^+_{2j_4+1}
\rangle;\;\;\;\;\;\;
j_1+j_2+j_3+j_4=k-2~,\cr
\langle &\hO^+_{2j_1+1}\hO^+_{2j_2+1}\hO^+_{2j_3+1}\hO^-_{2j_4+1}
\rangle;\;\;\;\;\;\;
j_1+j_2+j_3-j_4={k-2\over2}~,\cr
\langle &\hO^+_{2j_1+1}\hO^+_{2j_2+1}\hO^-_{2j_3+1}\hO^-_{2j_4+1}
\rangle;\;\;\;\;\;\;
j_1+j_2-j_3-j_4=0~,\cr
}}
and the complex conjugates of these. As we mentioned after equation \casesd,
it is enough to compute any one of the three classes of correlators
in \fourpt, since the others can then be obtained by using the reflection
property \refprop. 

In order to compute these correlators it is convenient to map some (or
all) of the RR operators in \fourpt\ to the NS-NS sector. This is
possible, since the spectral flow that relates these sectors is gauged
in the $\cn=2$ string. The operators that implement the spectral flow
transformation can be written as \BischoffBN\foot{We omitted a factor
$e^{\pm{1\over 2}c\tilde b}$ in the definition of the spectral flow
operators. This factor is needed for BRST invariance, but does not
influence correlation functions.}
\eqn\splusminus{S^\pm=e^{\pm\half(\varphi_2-\varphi_1)\pm{i\over2}(H+H')}~.}
These operators have the following properties \JunemannHI:
\item{(1)} $S^\pm$ are in the BRST cohomology of the $\cn=2$ string. They have
dimension zero.
\item{(2)} $\partial_z S^\pm$ is BRST exact, so in correlators that contain 
$S^\pm$ and other BRST-invariant operators, one can freely move the former 
around.
\item{(3)} $S^+$ and $S^-$ generate a ring. The product of $S^+$ and $S^-$ is 
one. 
The other independent OPE is
\eqn\ssplus{S^+(z)S^+(w)=e^{\varphi_2-\varphi_1+i(H+H')}~.}
\item{(4)} The OPE of $S^\pm$ with $\hO^\pm_{2j+1}$ is non-singular, so
we can multiply them (see below). Thus, the $\hO$'s form a module of the
ring generated by $S^+$ and $S^-$. 
\item{(5)} The algebraic structure associated with $S^\pm$ is very reminiscent
of the ground ring of two dimensional string theory \WittenZD. Both are  
useful in constraining the dynamics.  

\noindent
Due to the above properties, we can insert into any correlator of BRST
invariant operators the product $S^+ S^-=1$ (and their anti-holomorphic
counterparts) and change the positions of $S^+$ and $S^-$ freely
without changing the answer. The following results are useful for this
procedure\foot{In these formulae, and in the ones below which involve
$\Phi^{(su)}$, we write the form of the operators in the CHS background,
or in the deformed background for large $\phi$; however, the formulae
should be interpreted as exact formulae in the deformed background, when we 
replace the CHS operators by the corresponding operators in the deformed 
background, as described in \S4.3.}:
\eqn\sope{\eqalign{
S^+\bar S^+\hO^+_{2j+1}=&e^{-\varphi_1-\bar\varphi_1+i(H+\bar H)}
\psu_{j;j,j}e^{-Q(j+1)\phi}~,\cr
S^-\bar S^-\hO^+_{2j+1}=&e^{-\varphi_2-\bar\varphi_2-i(H'+\bar H')}
\psu_{j;j,j}e^{-Q(j+1)\phi}~,\cr
S^+\bar S^+\hO^-_{2j+1}=&e^{-\varphi_1-\bar\varphi_1+i(H'+\bar H')}
\psu_{j;-j,-j}e^{-Q(j+1)\phi}~,\cr
S^-\bar S^-\hO^-_{2j+1}=&e^{-\varphi_2-\bar\varphi_2-i(H+\bar H)}
\psu_{j;-j,-j}e^{-Q(j+1)\phi}~.\cr
}} 
Consider, for example, the four-point function on the third line of equation
\fourpt. By inserting $S^+\bar S^+ S^-\bar S^-=1$ and acting on the
$\hO^-$'s, using \sope, one finds (suppressing the worldsheet positions)
\eqn\ppmm{\eqalign{
\langle \hO^+_{2j_1+1}&\hO^+_{2j_2+1}\hO^-_{2j_3+1}\hO^-_{2j_4+1}\rangle=
\langle \hO^+_{2j_1+1}\hO^+_{2j_2+1}\cdot \cr
&e^{-\varphi_1-\bar\varphi_1+i(H'+\bar H')}\psu_{j_3;-j_3,-j_3}
e^{-Q(j_3+1)\phi} \cdot
e^{-\varphi_2-\bar\varphi_2-i(H+\bar H)}\psu_{j_4;-j_4,-j_4}e^{-Q(j_4+1)\phi}
\rangle~.\cr
}}
To simplify this expression further it is useful to note that
\eqn\gpmv{\eqalign{
G^-_{-\half} \psu_{j;j,j} e^{-Q(j+1)\phi}&=iQ(2j+1)e^{-iH'}
\psu_{j;j,j}e^{-Q(j+1)\phi}~,\cr
G^+_{-\half} \psu_{j;-j,-j} e^{-Q(j+1)\phi}&=iQ(2j+1)e^{iH'}
\psu_{j;-j,-j}e^{-Q(j+1)\phi}~.\cr
}}
Using this in \ppmm\ we find that
\eqn\pmone{\eqalign{
Q^2& (2j_3+1)^2\langle \hO^+_{2j_1+1}\hO^+_{2j_2+1}\hO^-_{2j_3+1}
\hO^-_{2j_4+1}\rangle=
\langle \hO^+_{2j_1+1}\hO^+_{2j_2+1}\cdot \cr
&e^{-\varphi_1-\bar\varphi_1-\varphi_2-\bar\varphi_2}\psu_{j_3;-j_3,-j_3}
e^{-Q(j_3+1)\phi}\cdot
G^+_{-\half}\bar G^+_{-\half}e^{-i(H+\bar H)}\psu_{j_4;-j_4,-j_4}
e^{-Q(j_4+1)\phi}
\rangle~,\cr
}}
where we used picture changing to move the operator $G^+_{-\half}\bar
G^+_{-\half}$ from the third to the fourth operator.  Recall
that our conventions are that $\varphi_1$ is the ghost conjugate to
$G^-$ and $\varphi_2$ the ghost conjugate to $G^+$ (see \brstcharge).

The fourth operator now has precisely the same form as the coupling
proportional to $\bar\lambda_{j_4}$ in the worldsheet Lagrangian
\pertB. Thus, taking this vertex operator to be the one which is integrated
over the worldsheet, we can write
\eqn\pmtwo{\eqalign{
& Q^2 (2j_3+1)^2 
\langle \hO^+_{2j_1+1}\hO^+_{2j_2+1}\hO^-_{2j_3+1}\hO^-_{2j_4+1}\rangle=\cr
&-{\partial\over\partial\bar\lambda_{j_4}} \langle \hO^+_{2j_1+1}
\hO^+_{2j_2+1}\cdot
e^{-\varphi_1-\bar\varphi_1-\varphi_2-\bar\varphi_2}\psu_{j_3;-j_3,-j_3}
e^{-Q(j_3+1)\phi}
\rangle~.\cr
}}
If we can compute the three-point function \pmtwo\ as a function
of the couplings $\lambda_j$, we can obtain from it the four-point
function we are after. Unfortunately, this involves solving the worldsheet
CFT \pertB\ at a generic point in moduli space, which seems difficult using
present techniques. In order to proceed, we will specialize to the case
\eqn\specialfour{j_4={k-2\over2}~,}
for which the coupling $\bar\lambda_{j_4}=\bar\mu$ and we can use known 
results for the amplitude on the right-hand side of equation \pmtwo\ in
the theory \slsutwo. In this case the sum rule in equation \fourpt\ takes 
the form
\eqn\sumrule{j_1+j_2=j_3+{k-2\over2}~.}
We can also ignore the ghosts $\varphi_1$ and $\varphi_2$ since their
correlation function is trivial. Thus, we arrive at:
\eqn\pmthree{\eqalign{
Q^2 (2j_3+1)^2 \langle
\hO^+_{2j_1+1}\hO^+_{2j_2+1}&\hO^-_{2j_3+1}\hO^-_{k-1}\rangle=
-{\partial\over\partial\bar\mu} \langle  e^{{i\over2}(H+\bar H- H'-\bar
H')}\psu_{j_1;j_1,j_1} e^{-Q(j_1+1)\phi}\cdot \cr & e^{{i\over2}(H+\bar H-
H'-\bar H')}\psu_{j_2;j_2,j_2}e^{-Q(j_2+1)\phi} \cdot
\psu_{j_3;-j_3,-j_3}e^{-Q(j_3+1)\phi}
\rangle~.\cr
}} 
The derivative with respect to $\bar \mu$ brings down a factor of $b/{\bar
\mu}$, where the the three-point function on the right hand side of equation 
\pmthree\ goes like $\mu^a\bar\mu^b$ \fourptscal. As in the discussion after 
\fourptscal, we can calculate $a$ and $b$ by imposing the sum
rules for $\phi$ charge and total $J_3$. In this case, using \sumrule, 
they give rise to the constraints
\eqn\phicharge{a+b+Q^2(2j_3+{k\over 2}+1)=0~,}
and
\eqn\ycharge{a-b+1=0~.}
Solving the two constraints one finds
\eqn\bbbb{b=-{Q^2\over 2}(2j_3+1)~.}
What remains is to compute the three-point function \pmthree\ in the
theory given by \slsutwo.

Consider first the $SU(2)/U(1)$ part of the calculation. The $SU(2)/U(1)$
component of the first operator
has dimension
\eqn\delone{\Delta_1={1\over8}+{j_1(j_1+1)\over k}-{(j_1+\half)^2\over k}
={1\over8}-{1\over 4k}~.}
So, it is a RR ground state. Its R-charge is (see \S4.3)
\eqn\Rone{R_1=\half(1-{2\over k})-{2\over k}j_1=\half-{2j_1+1\over k}~.}
The second operator has similar properties. For the $SU(2)/U(1)$ component
of the third operator, we have
\eqn\delRthree{\eqalign{
\Delta_3&={j_3(j_3+1)\over k}-{j_3^2\over k}={j_3\over k}~,\cr
R_3&={2j_3\over k}~,\cr }} 
so it is a chiral operator. As a check, the
sum of the three R-charges in \pmthree\ is:
\eqn\sumrch{\half-{2j_1+1\over k}+\half-{2j_2+1\over k}+{2j_3\over k}=0~,}
due to the sum rule \sumrule.

In order to compute the $SU(2)/U(1)$ correlator, we can use the decompositions
discussed in \S4.3, in particular equations \drjmslv, \vvjjmmgg. For the third
operator one uses the NS-NS sector analog of these equations, whose 
$SU(2)/U(1)$ parts are 
\eqn\pepepe{\psu_{j_3;-j_3,-j_3}=e^{-iQj_3Y}
V^{(su,susy)}_{j_3;-j_3,-j_3}~,}
and
\eqn\nsbbss{V_{j_3;-j_3,-j_3}^{(su,susy)}=V_{j_3;-j_3,-j_3}^{(su)}
\exp\left[{i(-2j_3)\over\sqrt{k(k-2)}}P\right]~.}
As a check, the $P$ charges (or $U(1)_R$ charges) of the three operators 
add up to zero, $2j_1+2j_2-(k-2)-2j_3=0$.

The $SU(2)/U(1)$ part of \pmthree\ now reduces to the 
three-point function of the bosonic parafermion operators
\eqn\threepara{
\langle V_{j_1;j_1,j_1}^{(su)}V_{j_2;j_2,j_2}^{(su)}
V_{j_3;-j_3,-j_3}^{(su)}\rangle~,}
which can be calculated using the
identity \paraplu\ and its complex conjugate
\eqn\paraid{V_{j;-m,-m}^{(su)}=
V_{{k-2\over2}-j;{k-2\over2}-m,{k-2\over2}-m}^{(su)}~.}
Applying these identities to all three operators in \threepara, we find
the three-point function
\eqn\reflthree{
\langle V_{{k-2\over2}-j_1;j_1-{k-2\over2},j_1-{k-2\over2}}^{(su)}
V_{{k-2\over2}-j_2;j_2-{k-2\over2},j_2-{k-2\over2}}^{(su)}
V_{{k-2\over2}-j_3;{k-2\over2}-j_3,{k-2\over2}-j_3}^{(su)}\rangle~.}
The sum rule \sumrule\ implies that
this $SU(2)/U(1)$ correlator can be calculated directly in $SU(2)$, since the
$J_3$ charge is conserved.
Furthermore, since this three-point function only involves
components with $|m|=j$, the $SU(2)$ three-point function
in question is just the structure constant that appears in
appendix B,
\eqn\cjjj{
\langle V_{{k-2\over2}-j_1;j_1-{k-2\over2}}^{(su)}
V_{{k-2\over2}-j_2;j_2-{k-2\over2}}^{(su)}
V_{{k-2\over2}-j_3;{k-2\over2}-j_3}^{(su)}\rangle
=C({k-2\over2}-j_1,{k-2\over2}-j_2,{k-2\over2}-j_3)~.
}
Using the results of \ZamolodchikovBD\ (reviewed in appendix B) 
and the sum rule \sumrule, one finds
\eqn\ccjjtt{
C({k-2\over2}-j_1,{k-2\over2}-j_2,{k-2\over2}-j_3)=
\left[{\gamma({1\over k})\gamma({2j_1+1\over k})
\gamma({2j_2+1\over k})\over
\gamma({2j_3+1\over k})}\right]^{\half}~,
}
where
\eqn\defgamma{\gamma(x) \equiv {\Gamma(x)\over \Gamma(1-x)}.}
This completes the $SU(2)/U(1)$ part of the calculation.
We next move on to the $SL(2)/U(1)$ part. As we did for the $SU(2)/U(1)$
three-point function, we will compute the $SL(2)/U(1)$ three-point
function by lifting it to a three-point function in $SL(2)$.

The $SL(2)/U(1)$ part of the first operator in \pmthree\ is (for large $\phi$)
\eqn\firstsl{e^{-{i\over2}(H'+{\bar H}')+iQ(j_1+\half)Y-Q(j_1+1)\phi}~.}
It can be lifted to a superconformal $SL(2)$ WZW vertex operator
as follows.
Denote by $\tilde \Phi$ bosonic $SL(2)$ vertex operators
(at level $k+2$, as necessary for the coset; these operators are discussed
in appendix B). The supersymmetric
$SL(2)$ model contains in addition three free fermions
$\chi^a$, with $a=3,+,-$. Denote by $H''$ the scalar field
that bosonizes $\chi^\pm$.
Now consider the vertex operator
\eqn\versltwo{e^{-{i\over2}(H''+{\bar H}'')}\tilde\Phi_{-j_1-1;j_1+1,j_1+1}~.}
Using the formula for the $U(1)_R$ current of $SL(2)/U(1)$ written
as a current in the full $SL(2)$ theory,
\eqn\rcurr{J^{\rm (sl)}=\left(1+{2\over k}\right)i\partial H''+
{2\over k} J_3~,}
one finds that the $SL(2)/U(1)$ R-charge of the vertex operator \versltwo\ is
\eqn\rrssll{R^{\rm (sl)}=-\half+{2j_1+1\over k}~.}
It is easy to check that this is the same as the 
R-charge obtained from equation (A.19), applied to
\firstsl. One can check that the dimensions also agree, as well
as the fact that \versltwo\ is a normalizable vertex operator in
$SL(2)$, in agreement with its expected properties in the coset theory.
Finally,  the sum of the $SU(2)/U(1)$ charge \Rone\ and the $SL(2)/U(1)$
charge \rrssll\ is zero, as expected from equation (A.21) applied to \pmthree.

The second operator in \pmthree\ can be lifted in the same way. For the third
operator, one can check that starting with the $SL(2)$ operator
\eqn\nssltwo{e^{i(H''+{\bar H}'')}
\tilde\Phi_{-j_3-{k\over2}-1;-j_3-{k\over2}-1,-j_3-{k\over2}-1}}
leads to an operator with the correct properties (dimension and R-charge) in
the $SL(2)/U(1)$ theory. 
Note that naively this operator does not have the correct asymptotic
behavior for large $\phi$ (see \expvjm), but the formulae (like
\expvjm) for the asymptotic behavior can only be trusted for small $j$
and large $k$, which is never the case for \nssltwo. In other cases one
can have operator mixings with operators having different asymptotic
behaviors (with the same global symmetry charges), so it seems
likely that the reduction of the operator \nssltwo\ to the $SL(2)/U(1)$
theory would have the asymptotic behavior that we
need here. 

Thus, after this lifting, we find that 
the $SL(2)/U(1)$ part of the correlator \pmthree\ is equal to the
$SL(2)$ correlator
\eqn\sltwocorr{\eqalign{
\langle e^{-{i\over2} (H''+{\bar H}'')}\tilde\Phi_{-j_1-1;j_1+1,j_1+1} \cdot
& \ e^{-{i\over2} (H''+{\bar H}'')}\tilde\Phi_{-j_2-1;j_2+1,j_2+1} \cdot \cr
& e^{i(H''+{\bar H}'')}
\tilde\Phi_{-j_3-{k\over2}-1;-j_3-{k\over2}-1,-j_3-{k\over2}-1}
\rangle~. \cr}}
Note that both the $H''$ charge conservation and the $J_3$ 
sum rule are satisfied in
this correlator, so there is no obstruction to calculating it directly in
the $SL(2)$ WZW theory.

Essentially this calculation is done in \gktwo, and is reviewed in appendix
B. Equation (4.19) of \gktwo\
contains the residue of the poles at $|m|=j+1$ for all the
external legs, which is what we need here. One has (see equation (B.25)) :
\eqn\sltwofin{\eqalign{
\langle \tilde\Phi_{-j_1-1;j_1+1,j_1+1}
\tilde\Phi_{-j_2-1;j_2+1,j_2+1}
&\tilde\Phi_{-j_3-{k\over2}-1;-j_3-{k\over2}-1,-j_3-{k\over2}-1}
\rangle=\cr
&\qquad\qquad{k\pi\over 2}\sqrt{\gamma\left(1-{2j_1+1\over k}\right)
\gamma\left(1-{2j_2+1\over k}\right)
\over \gamma\left({1\over k}\right)
\gamma\left(-{2j_3+1\over k}\right)}~.\cr}}
We are now ready to assemble all the pieces, to find the final answer for
the correlator on the third line of \fourpt\ 
for the special case \specialfour. 
We find
\eqn\ppmmfinal{\eqalign{
\langle \hO^+_{2j_1+1}\hO^+_{2j_2+1}&\hO^-_{2j_3+1}\hO^-_{k-1}\rangle \simeq
{1\over Q^2(2j_3+1)^2}\times {Q^2\over 2}(2j_3+1)\times\cr
&\left[{\gamma({1\over k})\gamma({2j_1+1\over k})
\gamma({2j_2+1\over k})\over
\gamma({2j_3+1\over k})}\right]^{\half}
\times {k\pi \over 2}
\left[\gamma\left(1-{2j_1+1\over k}\right)
\gamma\left(1-{2j_2+1\over k}\right)
\over \gamma\left({1\over k}\right)
\gamma\left(-{2j_3+1\over k}\right)\right]^{\half} ~.\cr}
}
Simplifying, we find:
\eqn\normform{
\langle \hO^+_{2j_1+1}\hO^+_{2j_2+1}\hO^-_{2j_3+1}\hO^-_{k-1}\rangle\simeq 
{{\rm const}}~,} 
where the constant appearing on the right-hand side 
is independent of the $j_i$ (and of $k$)\foot{This constant can be computed
using our techniques, but we will not attempt to fix it here.}.
Note that the result \normform\ is consistent with the reflection
property \refprop\ satisfied by the $\cn=2$ string operators: we can get
another amplitude of the same kind by taking, say, $j_2\to {k-2\over2}-j_2$
and $j_3\to {k-2\over2}-j_3$. Moreover, the reflection symmetry implies that 
whenever at least one of the four $j$'s in equation \fourpt\ is equal to
$(k-2)/2$ or to $0$, there is no need to do any further calculations, since by
using \refprop\ and the answer \normform\ above we can compute the amplitude 
for all of these cases. The computation for other values of the $j$'s
will be left for future work.

In the next section we will show that the gauge theory result \leffr\
indeed has the property predicted by the string calculation, that in the
appropriate normalization the analog of \normform\ for \fmu, \fmustar\ 
is indeed independent
of $j_1$, $j_2$ and $j_3$, and satisfies the selection rules \fourpt.

\newsec{Comparison between the type II and heterotic results}

In \S2 we wrote down the form of the $t_8 F^4$ coupling in the
heterotic theory \leffr\ in terms of the gauge fields $F^{(i)}$ of the
low-energy field theory. In order to compare with the type II computations 
of the previous section, it is useful to rewrite this coupling in terms of 
the operators $\co_n^\pm$ \fmu, \fmustar, which appear naturally in the type 
II string theory. For an $SU(k)$ gauge theory \leffr\ takes the form 
\eqn\coupl{{\cal L}_{F^4}=t_8\sum_{l>m=0}^{k-1}
{1\over M^2(\alpha_{lm})}(\vec F\cdot\vec\alpha_{lm})^4~,}
where $M(\alpha_{lm})$ is the mass of the W-boson corresponding to the
root $\alpha_{lm}$; in the $NS5$-brane realization this is a D-string stretched
between the $l$-th and the $m$-th $NS5$-branes. 
Using the form of the VEV of $B$
\vevab\ one can easily compute
this mass at the point in moduli space we are interested in :
\eqn\dbrane{
M(\alpha_{lm})=r_0 \left |\exp\left({2\pi i\over k}l\right)-
\exp\left({2\pi i\over k}m\right)
\right|=
2r_0 \left|\sin\left({\pi\over k}(l-m)\right)\right|~.}
In order to compare the result \coupl\ to the type II calculation
we should express $F$ in terms of the operators ${\cal O}_n^+$ (we
will generally suppress the space-time indices in this section).
Such a replacement makes sense at leading order in $M_s/M_W$, since at
this order we can simply
replace $B$ in \fmu\ by its expectation value, and then
solve for $F$ in terms of ${\cal O}_n^+$, obtaining (writing the
low-energy $SU(k)$ gauge field as $F_{\mu \nu} = {\rm diag}(F_{\mu \nu}^{(1)},
F_{\mu \nu}^{(2)}, \cdots, F_{\mu \nu}^{(k)})$)
\eqn\four{F^{(r)}={1\over k}\sum_{n=0}^{k-1}{{\cal O}_n^+\over r_0^n}
\exp\left(-{2\pi i\over k}rn\right)~.}
In this expression we included also an operator $\co_0^+$ (defined like
the other values of $n$ \fmu), which vanishes
due to tracelessness of $F_{\mu \nu}$,
so one can choose
the lower limit of the sum to be either $0$ or $1$.

Now, using \dbrane\ and \four\ one can express ${\cal L}_{F^4}$ of \coupl\ 
in the following form :
\eqn\couplo{{\cal L}_{F^4}=
t_8\sum_{n_i} {4t_{n_1,\,n_2,\,n_3,\,n_4}\over {k^4 r_0^{\sum_i n_i}}}
{\cal O}_{n_1}^+{\cal O}_{n_2}^+{\cal O}_{n_3}^+{\cal O}_{n_4}^+~,}
where
\eqn\couplO{t_{n_1,\,n_2,\,n_3,\,n_4}={1\over {r_0^2}}
\sum_{l>m=0}^{k-1}{{e^{-{\pi i\over k}(l+m)\sum n_i}
\over\sin^2\left[{\pi(l-m)\over k}\right]}}
\prod_{i=1}^4\sin\left[{\pi(l-m)\over k}n_i\right]~.}
We chose here to write the Lagrangian using the operators $\co_n^+$.
As discussed in \S5.2, when we replace the $B$'s in \fmu, \fmustar\ by their
VEVs, the operators $\co_n^+$ and $\co_{k-n}^-$ are identical (up to
a power of $r_0$). Thus, we can always replace (for any particular $n$) 
the operator $\co_n^+$ appearing in \couplo\ by $\co_{k-n}^-$.

The expression \couplO\ is quite complicated, so we would like to simplify it.
This simplification is discussed in appendix D. The result is that
\eqn\finaltn{t_{n_1,n_2,n_3,n_4} = {k^2\over {4r_0^2}} \min(n_i,k-n_i)~,}
if $\sum_{i=1}^4 n_i = 2k$, and it vanishes otherwise. 

Next, we would like to relate this result to the four-point functions
that we computed in the previous section. Using equations
\ttwwoonn\ and \relgastr\ we obtain that the contraction of an amputated
string theory operator $\hO_{2j+1}^-$ with a field theory operator
$\co_{2j+1}^+$ gives
\eqn\newcontract{\vev{\hO_{2j+1}^- \co_{2j+1}^+} = -r_0^{2j+1} \sqrt{{
{4k}\over{\pi}}},}
up to
factors of momentum which can be ignored following the discussion of the
previous section. The space-time Lagrangian contains in particular the 
vertex \couplo\ with all operators at zero momentum, and we can obtain a
non-zero four-point function of $\hO_{2j+1}^-$ operators 
by contracting the operators $\hO_{2j+1}^-$ with
the operators $\co_{2j+1}^+$ in the vertex \couplo. The low-energy field
theory result \couplo\ and equation \newcontract\ then imply that
the four-point functions computed in the previous section should be equal to 
\eqn\expect{\vev{\hO_{n_1}^- \hO_{n_2}^- \hO_{n_3}^- \hO_{n_4}^-} =
{64 \over {\pi^2 k^2}} t_{n_1,n_2,n_3,n_4}.}

Let us compare this with the results of the previous section.
First, if we look at a correlation function of the form \expect\ where
all $\alpha_i$ have the same sign, we see that the sum rule $\sum_i n_i = 2k$
required to get a non-vanishing $t_{n_1,n_2,n_3,n_4}$
is precisely the same as the sum rule we encountered in
the type II computation \casesd.  This is our first test of the
type II/heterotic duality for four-point functions.

Next, let us make a more precise comparison with the results of
\S5.5. In that section we computed the correlators for which at
least one of the $n_i$ was equal to either $1$ or $k-1$. In such cases,
\finaltn\ implies that $t_{n_1,n_2,n_3,n_4} = k^2 / 4 r_0^2$ with no
dependence on the other $n$'s, so we expect to find 
\eqn\expectn{
\vev{\hO_{n_1}^- \hO_{n_2}^- \hO_{n_3}^- \hO_{n_4}^-} =
\left({4\over \pi r_0 
}\right)^2~.}
The result \normform\ which we found in our type II computation (transformed
into a correlator of four $\hO^-$'s) has
precisely this form; 
the two expressions have precisely the same dependence on the $n_i$, 
and as we discussed after equation \sumphi\ they also have the same
dependence on $r_0$ if we reinstate the powers of $\mu$ into the
$SL(2)/U(1)$ correlators, so
they differ just by a constant multiplicative factor (which we did
not keep track of).
Thus, we find precise agreement between the four-point functions
we computed in type II string theory in \S5 and our heterotic (field
theory) expectations. This is a strong test of type II/heterotic 
duality, as well as of the techniques used in studying LST.

\newsec{Summary and discussion of future directions}

\subsec{Summary}

Since this paper is somewhat long, it is useful to briefly summarize 
our main results. The bulk of the paper was devoted to an analysis of a 
certain half-BPS $F^4$ term in the low-energy effective action of LST. 
We showed that this term can be efficiently computed in a topological 
version of LST, which is holographically dual to the ${\cal N}=2$ string 
on an asymptotically linear-dilaton background. Its coefficient is given 
by a tree level four-point function of certain normalizable operators in 
the ${\cal N}=2$ string.

At the same time, this term can be obtained by a one-loop
calculation in the heterotic string on $T^4$ near a point of enhanced
ADE gauge symmetry, or equivalently in the low-energy SYM theory corresponding
to the relevant ADE gauge group. The ${\cal N}=2$ string and heterotic 
computations of the $F^4$ term are apriori valid in different regions in 
moduli 
space. The expected agreement between them  
is a highly non-trivial consequence of heterotic/type II duality,
and of the non-renormalization theorem of this $F^4$ term in the effective
action. Therefore, computing it in the asymptotically 
linear dilaton background of the 
${\cal N}=2$
string provides a sensitive test of both S-duality and the non-renormalization 
theorem, and it also tests our understanding of LST at low energies.

While the heterotic one-loop calculation can be easily done for all values
of the moduli of LST, the ${\cal N}=2$ string calculation simplifies at
a specific point in moduli space, \vevab. We found that a large class of
four-point functions in the ${\cal N}=2$ string agree with the heterotic 
predictions at this point in moduli space. 

Our results passed several consistency checks:
\item{(1)} The holographic interpretation of physics in linear dilaton spaces
\abks\ suggests that while in the heterotic string one can consider
the gauge field strength $F$ directly, in the type II and ${\cal N}=2$ 
string calculations, the observables correspond to gauge invariant
operators in the full non-abelian gauge theory, such as $\co_n^\pm$,
\fmu, \fmustar. We indeed found that in order to compare the linear
dilaton results to those obtained in the heterotic string or low-energy
SYM theory, one has to perform the discrete Fourier transform \four. 
\item{(2)} The low-energy field theory analysis suggests that the 
normalizable versions of the two types of operators that we analyzed,
$\co_{n}^+$ and $\co_{k-n}^-$, should be the same (up to a constant), and
we found that this is indeed true (both in the type II string and in
the ${\cal N}=2$ string). In fact, the gauge theory provides a physical
interpretation of certain reflection symmetries of two dimensional CFTs
such as the parafermion theory and ${\cal N}=2$ minimal models, \plumin,
\paraplu, (and their $SL(2)$ analogs) that are known for many years. 
\item{(3)} Expressing the $t_8 F^4$ term (found from the heterotic
computation) in the basis of the operators $\co_n^{\pm}$,
we found that the resulting four-point functions should obey
specific selection rules. The same selection rules appeared in the 
type II computation in a completely different way (described in \S5.3).
\item{(4)} We explicitly computed using the ${\cal N}=2$ string 
specific four-point functions (related to $t_8 F^4$ terms) involving at
least one operator $\co_{2j+1}^{\pm}$ with $j=0$ or $j=(k-2)/2$, and 
found that they agreed (up to an overall constant that we did not
determine) with the heterotic computation of the same objects.

\noindent
In the process of computing the $t_8 F^4$ terms we clarified some additional 
issues in the study of LSTs. 
We showed that one could use normalizable 
vertex operators (which we carefully defined in \S5.2) to compute 
amputated correlation functions in LSTs (the same should be true 
for other holographic backgrounds). This implements the LSZ
reduction in the ``boundary theory'' directly in terms of the 
bulk variables and,
as we have seen, simplifies the computation of S-matrix elements and 
coefficients of terms in the effective Lagrangian in these theories. 
Also, in order to complete our computation we needed to carefully compute 
the two-point functions in string theory on $SL(2)/U(1)$. As in Liouville 
theory or $SL(2)$, this computation naively involves a ratio of infinities, 
but we showed (in appendix C) 
that the result can be uniquely determined
by relating the two-point functions to three-point functions. These results
are useful for many other computations in LST and in other holographic 
backgrounds.

We also showed that the correspondence between the asymptotically linear
dilaton background \slsutwo\ and LST sheds new light on the equivalence between
${\cal N}=2$ Liouville theory and the cigar ($SL(2)/U(1)$) CFT, conjectured in 
\gkone\ 
and further discussed in \refs{\HoriAX\newgk-\TongIK}. It suggests a picture, 
compatible with the worldsheet analysis, according to which the black hole 
metric
and the ${\cal N}=2$ superpotential \lntwo\ coexist in these backgrounds. 
In fact, our discussion of the normalizable operators in \S5.2,
applied to the moduli, shows that for any finite value of $\mu$, the
normalizable operators corresponding to the Liouville deformation
\lntwo\ and to the cigar deformation \bhpert\ should be identified (in
analogy with our identification \reflprop). There is only a single
deformation operator corresponding to moving along the flat direction
\vevab, though we can describe it in two (and, in fact, more)
different ways. This provides additional evidence for the equivalence
of the two theories.

One of our main results was the identification of some of the observables in
the topological LST (holographically described by the ${\cal N}=2$
string on the ``throat'' background). We found that these observables
correspond to normalizable versions of the vertex operators of the
LST. Their correlation functions compute amputated correlation
functions in the LST, which are related to protected terms in the 
effective action. Due to the spectral flow properties of ${\cal
N}=2$ strings, the same observables correspond in the type II string
both to RR operators and to NS-NS operators (the moduli of the
``throat'' background). We computed several three-point and four-point
functions of these observables, and found that both can be
non-zero. By differentiating with respect to the couplings $\lambda_j$
\lpertch\ one finds non-zero higher point functions as well. Note that 
this is different from the case of ${\cal N}=2$ strings in flat space, 
where all four and higher point functions vanish.

\subsec{Discussion of future directions}

There are several future directions that are suggested by our
results. The most obvious is to complete the analysis of 
$t_8 F^4$ terms. In this paper we only computed a subset of the
relevant four-point functions in the topological LST, at a particular
point in the moduli space \lamres. It would be nice to compute the 
rest of the four-point functions at this point in the moduli space,
and then understand the structure everywhere in moduli space. 

From the heterotic side we have predictions \finaltn, \expect\ for all 
four-point functions of the operators \ntwoform\ at the point \lamres\
in moduli space. We show in appendix D that the heterotic answer
can be naturally expressed using the fusion coefficients of the $SU(2)_{k-2}$ 
WZW theory. This is very suggestive, since at this point in moduli space, 
the relevant bulk background involves precisely this current algebra (see 
\slsutwo). It would be interesting to understand this relation to 
$SU(2)_{k-2}$ better. 

More generally, the heterotic result \leffr\ predicts the correlation
functions of the normalizable topological observables \ntwoform\ 
everywhere in moduli space. It would be interesting to verify these
predictions using ${\cal N}=2$ strings on the generic ``throat geometries''
\lpertch. The techniques developed in previous studies of topological 
string theories may be useful for this (see \DijkgraafQH\ for a review).

It should also be interesting to generalize our results to other
LSTs. We expect the generalization to $D_k$ and $E_k$ type LSTs to be
straightforward. It may be possible to generalize the results also to
six dimensional 
LSTs with ${\cal N}=(2,0)$ supersymmetry, which arise in type IIB
string theory near ALE singularities. In this case the low-energy effective
action contains two-form gauge potentials rather than one-forms, but
presumably there are still protected $H^4$ terms in the low-energy
effective action that may be computable using topological strings. It
is not clear if there is any useful dual description in this case.
Another interesting generalization is to the topological sectors
of the $3+1$ dimensional LSTs that are associated to generalized
conifold-type singularities (\eg\ \genconeq). The holographic dual 
of these topological LSTs involves ${\cal N}=2$ topological strings 
on ``throat'' backgrounds similar to those that we described in this 
paper. As we discussed in \S4.4, in this case the analog of the $t_8F^4$
term studied here is a tree level two-point function in the 
topological string theory. 

Our computations in this paper focused on three-point and four-point
functions at tree level in the ${\cal N}=2$ string on the ``throat''
background. Clearly, there are many other objects that can be computed
in this theory. Our analysis of \S5.5 showed that the four-point
functions in this theory were derivatives of three-point functions
with respect to the moduli, and one can similarly show that they are
second derivatives of two-point functions with respect to the moduli. So,
the full information about tree level correlation functions seems to
be contained in the two-point functions (as a function of the
moduli). One should be able to define a ``sphere partition
function'' such that the four-point functions would be its fourth
derivatives. Higher genus partition functions in this theory are 
related \berva\ to $R^4 F^{4g-4}$ terms in the effective action. We 
hope that such terms may again be computed using the duality to 
heterotic strings, and perhaps they can also be computed directly 
in the ${\cal N}=2$ string.

As we mentioned in the introduction, one of our motivations for
studying these topological LSTs is the hope that they can be a useful
toy model for studying dualities between open and closed strings, like
those that were found for ${\cal N}=2$ topological strings on conifold
backgrounds. We expect that these dualities should be similar to those of
$D\le 2$ dimensional bosonic and fermionic strings, where the closed string 
theory on the throat background is equivalent to an open string theory living 
on
D-branes localized inside the ``throat''\foot{The backgrounds discussed in 
this paper are very similar to the $c<1$ and $\hat c<1$ string theories. If one
replaces the ${\cal N}=0 (1)$ minimal model by an ${\cal N}=2$ minimal model,
and ${\cal N}=0 (1)$ worldsheet gravity by ${\cal N}=2$ worldsheet gravity,
one goes from the well understood examples related to the ``old matrix models''
to our system.}. For the case of the $A_1$ ALE space this conjecture was first 
made in \ovloops. We hope that our results, and in particular the 
identification 
of the observables of the topological LST, will be useful for understanding
this duality better; we hope to return to it in future
work. Recent results on closed and open ${\cal N}=2$ strings
\refs{\CheungYW\GluckWG-\GluckPA} should be useful for the further
study of ${\cal N}=2$ strings on ALE spaces.

\bigskip

\centerline{\bf Acknowledgements}

We would like to thank M. Berkooz, A. Giveon, E. Kiritsis, O. Lechtenfeld, 
Y. Oz, B. Pioline, 
O. Ruchayskiy, S. Sethi, S. Stieberger, P. Vanhove and E. Verlinde for 
useful discussions. 
OA would like to 
thank Stanford University, SLAC, Harvard University, the University of 
Chicago, the Aspen Center for Physics, and the University of British 
Columbia for hospitality during the work on this project. BF would like 
to thank the University of California at Santa Cruz, Stanford University, 
Rutgers University, and \'Ecole Normale Sup\'erieure (Paris) for hospitality. 
DK thanks the Weizmann Institute, Rutgers University, LPT at \'Ecole Normale 
Sup\'erieure (Paris), LPTHE at Universit\'e Paris VI, and the Aspen 
Center for Physics for hospitality. The work of OA and BF was supported 
in part by the Israel-U.S. Binational Science Foundation, by the ISF 
Centers of Excellence program, by the European network HPRN-CT-2000-00122, 
and by Minerva. OA is the incumbent of the Joseph and Celia Reskin career 
development chair. The work of BF is also supported by an European Community 
Marie Curie Fellowship. The work of DK and DS is supported in part by 
DOE grant \#DE-FG02-90ER40560.

\appendix{A}{Some results on CFT in the CHS background}

The CHS CFT contains a scalar $\phi$, a bosonic $SU(2)$ WZW model
at level $k-2$ ($k$ is the number of $NS5$-branes) with currents $J^i$,
$i=1,2,3$, and four fermions $\psi_\phi$, $\psi_1$, $\psi_2$, $\psi_3$.
The operator product expansions (OPEs) 
are (we take $\alpha'=2$ throughout this appendix):
\eqn\aa{\eqalign{
\phi(z)\phi(0)=&-\log|z|^2,\cr
\psi_a(z)\psi_b(0)=&{\delta_{ab}\over z},\cr
J^i(z)J^j(0)=&{\half(k-2)\delta^{ij}\over z^2}+
i\epsilon^{ijk} {J^k(0)\over z},\cr
}}
where $\epsilon^{123}=1$; $a,b=\phi,1,2,3$. Define
\eqn\bb{J^\pm=J^1\pm iJ^2.}
The OPE algebra on the last line of \aa\ is:
\eqn\cc{\eqalign{
J^3(z)J^3(0)=&{\half(k-2)\over z^2},\cr
J^3(z)J^\pm(0)=&{\pm J^\pm(0)\over z},\cr
J^+(z)J^-(0)=&{k-2\over z^2}+
{2J^3(0)\over z}.\cr
}}
Also define
\eqn\dd{\eqalign{
\psi^\pm=&{1\over\sqrt 2}(\psi_1\pm i\psi_2),\cr
\psi=&{1\over\sqrt 2}(\psi_\phi+i\psi_3),\cr
}}
which satisfy
\eqn\ee{\psi(z)\psi^*(0)=\psi^+(z)\psi^-(0)={1\over z}.}
Sometimes it is convenient to bosonize the fermions \dd\
and write them as
\eqn\bosferm{\psi^\pm=e^{\pm iH};\;\; \psi=e^{iH'}.}
As mentioned in \S4, the total $SU(2)$ currents of the supersymmetric
$SU(2)_k$ CFT, of level $k$, receive a
contribution also from the fermions. They are given by
\eqn\ggg{J_i^{\rm tot}=J_i-{i\over2}\epsilon_{ijk}\psi_j\psi_k,}
and in particular
\eqn\hh{J_3^{\rm tot}=J_3-i\psi_1\psi_2=J_3+\psi^+\psi^-.}
Note in particular that $\psi^\pm$ have charge $\pm1$ under $J_3^{\rm tot}$.

The stress tensor of the model is given by
\eqn\ss{
T=-\half\left(\partial \phi\right)^2-\half Q\partial^2\phi
+{1\over k+2}J_iJ_i-\half\psi^*\partial\psi-\half\psi\partial\psi^*
-\half\psi^+\partial\psi^--\half\psi^-\partial\psi^+,}
where for $\alpha'=2$ the slope of the linear dilaton is (see \qtwok)
the positive root of
\eqn\ff{Q^2={2\over k}.}
In particular, \ss\ implies that the dimension of $e^{\beta \phi}$ is
\eqn\tt{\Delta(e^{\beta \phi})=-\half\beta(\beta+Q).}
The CHS CFT has an $\cn=4$ superconformal symmetry. For our
discussion it is useful to exhibit an $\cn=2$ subalgebra of it.
The two superconformal generators are (see \KounnasUD) :
\eqn\ii{G=i\psi_\phi\partial \phi+iQ\partial\psi_0+Q(J_1\psi_1+J_2\psi_2+
J_3\psi_3-
i\psi_1\psi_2\psi_3),}
\eqn\jj{G_3=i\psi_3\partial \phi+iQ\partial\psi_3+Q(-J_3\psi_\phi+J_1\psi_2-
J_2\psi_1+
i\psi_1\psi_2\psi_\phi).}
One can define the generators
\eqn\ttffpp{G^\pm=G\pm i G_3,}
which are given by
\eqn\kk{\eqalign{
G^+=&\,i\psi(\partial \phi-QJ_3^{\rm tot})+iQ\partial\psi+QJ^-\psi^+,\cr
G^-=&\,i\psi^*(\partial \phi+QJ_3^{\rm tot})+iQ\partial\psi^*+QJ^+\psi^-.\cr
}}
These generators satisfy the OPE algebra
\eqn\ll{G^+(z)G^-(0)\simeq {2c\over 3z^3}+{2\over z^2}J(0)+
{1\over z}(2T(0)+\partial J(0)),}
from which one can find the form of the $U(1)_R$ current $J$.
In computing the OPE \ll, one notes that $G^+$ and $G^-$
split into two decoupled terms: the first two terms in each line
in \kk, and the last term.
Thus, in computing the OPE \ll\ we can separate:
\eqn\mm{\eqalign{
G^+(z)G^-(0)=-&\left[\psi(\partial \phi-Q J_3^{\rm tot})+
Q\partial\psi\right](z)\left[\psi^*(\partial \phi+Q J_3^{\rm tot})+
Q\partial\psi^*\right](0)+\cr
&Q^2J^-\psi^+(z)J^+\psi^-(0).\cr}}
The term that goes like $1/z^3$ is
\eqn\nn{(-1)(-1-Q^2{k\over2})-Q^2(-2)+Q^2(k-2)=2+{4\over k}+
{2\over k}(k-2)=4={2\over3}c,}
in agreement with \ll.

The coefficient of $1/z^2$ should be $2J$, the $U(1)_R$ current in the
$\cn=2$ superconformal algebra. From the square bracket term in \mm\ we get
\eqn\oo{2J_{sl}=2\psi\psi^*+2Q^2J_3^{\rm tot}.}
From the last term in \mm:
\eqn\pp{2J_{su}=Q^2(k-2)\psi^+\psi^-+Q^2(-2)J_3.}
After resolving the strong coupling singularity as discussed in section 4, \oo\
and \pp\ become the $U(1)_R$ currents of the $SL(2)/U(1)$ and $SU(2)/U(1)$
SCFTs, respectively. The sum of \oo\ and \pp\ is $2J$ with
\eqn\qq{J=\psi\psi^*+\psi^+\psi^-.}

\appendix{B}{Some facts about $SU(2)$ and $SL(2)$ WZW models}

In this appendix we review some properties of the $SU(2)$ and $SL(2)$
WZW models which are used in the main part of this paper.

The $SU(2)$ WZW model\foot{For more details and conventions see 
\ZamolodchikovBD.}
of level
\eqn\levl{k_{SU(2)}=k-2,}
has the central charge
\eqn\cntr{c_{SU(2)}={3(k-2)\over k}}
and 
contains an $SU(2)_L\times SU(2)_R$ affine Lie algebra of level $k_{SU(2)}$.
The operator algebra consists of primary fields of the current algebra
$\psu_j$ with $j=0,1/2,\cdots, (k-2)/2$ and their descendants.
The primaries  $\psu_j$ can be written in two equivalent bases
\ZamolodchikovBD, which are related by
\eqn\primr{\psu_j(y,\bar y;z,\bar z)=\sum_{m,\bar m=-j}^j
[C_{2j}^{m+j}C_{2j}^{\bar m+j}]^{1\over 2}
y^{j+m}{\bar y}^{j+\bar m}\psu_{j;m,\bar m}(z,\bar z),}
where $C_N^m={N!\over m!(N-m)!}$ are binomial coefficients and the
$\psu_{j;m,\bar m}$ are eigenfunctions of $J_3$ and ${\bar J}_3$ with
eigenvalues $m$ and $\bar m$.

The form of two- and three-point functions of the operators is fixed by
$SU(2)$ invariance.  In particular, the two-point functions of primaries
are determined up to an overall $j$-dependent factor, which can be
chosen to be one,
\eqn\twpt{\eqalign{&\langle\psu_{j_1}(y_1,\bar y_1)
\psu_{j_2}(y_2,\bar y_2)\rangle=
\delta_{j_1,j_2}|y_{12}|^{4j_1},\cr
&\langle\Phi^{(su)\dagger}_{j_1;m_1,\bar m_1}\psu_{j_2;m_2,\bar
m_2}\rangle=\delta_{j_1,j_2}\delta_{m_1,m_2}
\delta_{\bar m_1,\bar m_2},
}}
where $y_{ij}\equiv y_i-y_j$ and
\eqn\tdag{\Phi^{(su)\dagger}_{j,m,\bar m}=(-1)^{2j-m-\bar m}
\psu_{j,-m,-\bar m}.}
Here and below we suppress the $z$ dependence which is determined by
conformal invariance.
Similarly, the three-point functions of primary fields have the  form
\eqn\thpt{\eqalign{\langle\psu_{j_1}(y_1,\bar y_1)\psu_{j_2}(y_2,\bar y_2)
&\psu_{j_3}(y_3,\bar y_3)\rangle= \cr
&C(j_1,j_2,j_3)|y_{12}|^{2(j_1+j_2-j_3)}|y_{13}|^{2(j_1+j_3-j_2)}
|y_{23}|^{2(j_2+j_3-j_1)}~.\cr}}
$C(j_1,j_2,j_3)$ is uniquely determined  once we fix the normalization of the
operators using \twpt. One finds \ZamolodchikovBD
\eqn\stru{C^2(j_1,j_2,j_3)=\gamma\left({1\over k}\right)P^2(j_1+j_2+j_3+1)
\prod_{n=1}^3
\gamma\left(1-{2j_n+1\over k}\right){P^2(j_1+j_2+j_3-2j_n)\over P^2(2j_n)}, }
where $\gamma(x)$ and $P(j)$ are defined as follows :
\eqn\gamdef{\gamma(x)\equiv {\Gamma(x)\over \Gamma(1-x)},}
and
\eqn\pdef{P(j)\equiv \prod_{n=1}^j\gamma\left({n\over k}\right),\qquad
P(0)\equiv 1.}
One can translate \thpt\ into the $(j,m,\bar m)$ basis using equation \primr.
For the special case $j_3=j_1+j_2$, $m_2=\bar m_2=-j_2$, $m_3=\bar m_3=j_3$
that is used in \S5 we find, using \primr\ and \thpt,
\eqn\thpt{\langle\psu_{j_1;-j_1,-j_1}\psu_{j_2;-j_2,-j_2}
\psu_{j_1+j_2;j_1+j_2,j_1+j_2}\rangle=
C(j_1,j_2,j_1+j_2),}
where
\eqn\strusp{C^2(j_1,j_2,j_1+j_2)=\gamma\left({1\over k}\right)
\gamma\left(1-{2j_1+1\over k}\right)\gamma\left(1-{2j_2+1\over k}\right)
\gamma\left({2(j_1+j_2)+1\over k}\right).}

\noindent
We next  turn to a discussion of the $SL(2)$ WZW model of level
\eqn\levlsl{k_{SL(2)}=k+2,}
with central charge
\eqn\cntrsl{c_{SL(2)}={3(k+2)\over k}.}
The natural observables in the theory defined 
on the Euclidean version of $SL(2)$, 
$H_3^+~\equiv~SL(2,\IC)/SU(2)$, are primaries $\Phi_j^{(sl)}(x,\bar x)$
of the $SL(2)_L\times SL(2)_R$
current algebra \refs{\TeschnerFT,\TeschnerUG} with $j>-{1\over 2}$.
The worldsheet scaling dimension of $\Phi_j^{(sl)}(x,\bar x)$ is
\eqn\dimphij{\Delta(j)=-{j(j+1)\over k}~.}
In the papers \refs{\TeschnerFT,\TeschnerUG} the operators $\Phi_j^{(sl)}$
are normalized as follows:
\eqn\twptsl{\langle\Phi_{j_1}^{(sl)}(x_1,\bar x_1)
\Phi_{j_2}^{(sl)}(x_2,\bar x_2)\rangle=
\delta(j_1-j_2){k\over\pi}\left[{1\over k\pi}\gamma
\left({1\over k}\right)\right]^{2j_1+1}
\gamma\left(1-{2j_1+1\over k}\right)|x_{12}|^{-4(j_1+1)}.}
For our computations in section 5 it is more convenient to
choose a different normalization
\eqn\difnorm{\tilde\Phi_j(x,\bar x)\equiv {\Phi_j^{(sl)}(x,\bar x)\over
\sqrt{{k\over\pi}\left[{1\over k\pi}\gamma\left({1\over k}\right)\right]^{2j+1}
\gamma\left(1-{2j_1+1\over k}\right)}}.
}
In this normalization the two-point function is (compare to \twpt)
\eqn\twptfn{\langle\tilde\Phi_{j_1}(x_1,\bar x_1)
\tilde\Phi_{j_2}(x_2,\bar x_2)\rangle=
\delta(j_1-j_2)|x_{12}|^{-4(j_1+1)}.}
For discussing the coset $SL(2)/U(1)$ it is convenient to choose a 
different basis for 
the primaries $\tilde\Phi_j$ 
\eqn\difbas{\tilde\Phi_{j;m,\bar m}=\int d^2x \, x^{j+m}\bar x^{j+\bar m}
\tilde\Phi_j(x,\bar x),}
which is analogous to the $(j;m,\bar m)$ basis of $SU(2)$ WZW \primr.
The two-point function in this basis was computed in \refs{\FZZ,\gktwo} : 
\eqn\twptmm{\langle\tilde\Phi_{j;m,\bar m}\tilde\Phi_{j';-m,-\bar m}\rangle=
\pi \delta(j-j')
{\Gamma(-2j-1)\Gamma(j-m+1)\Gamma(1+j+\bar m)\over\Gamma(2j+2)\Gamma(-j-m)
\Gamma(\bar m-j)}.}

The three-point function takes the form
\eqn\thrptsl{\eqalign{\langle\tilde\Phi_{j_1}(x_1,\bar x_1)
& \tilde\Phi_{j_2}(x_2,\bar x_2)
\tilde\Phi_{j_3}(x_3,\bar x_3)\rangle=\cr
&\tilde
D(j_1,j_2,j_3)|x_{12}|^{2(j_3-j_1-j_2-1)}|x_{13}|^{2(j_2-j_1-j_3-1)}
|x_{23}|^{2(j_1-j_2-j_3-1)}, }} 
where the structure constants $\tilde
D(j_1,j_2,j_3)$ were computed in
\refs{\TeschnerFT,\TeschnerUG} :
\eqn\ddef{\eqalign{\tilde D&(j_1,j_2,j_3)={1\over 2\pi}
{1\over\sqrt{\gamma\left({1\over k}\right)\prod_{i=1}^3
\gamma\left(1-{2j_i+1\over k}\right)}}\times\cr
&{G(-j_1-j_2-j_3-2)G(j_3-j_1-j_2-1)G(j_2-j_1-j_3-1)G(j_1-j_2-j_3-1)\over 
G(-1)G(-2j_1-1)G(-2j_2-1)G(-2j_3-1)}.
}}
$G(j)$ is a special function which satisfies the following useful identities :
\eqn\propG{\eqalign{G(j)&=G(-j-1-k),\cr
G(j-1)&=\gamma(1+{j\over k})G(j),\cr
G(j-k)&=k^{-(2j+1)}\gamma(j+1)G(j).
}}

In the $j,m,\bar m$ basis the three-point function for $m={\bar m}$ is given by
\eqn\basthree{\eqalign{
\langle \tilde\Phi_{j_1;m_1, m_1} & \tilde\Phi_{j_2;m_2,m_2}
\tilde\Phi_{j_3;m_3,m_3}
\rangle=\tilde D(j_1,j_2,j_3)\times\cr
&F(j_1,m_1;j_2,m_2;j_3,m_3)\int d^2 x|x|^{2(m_1+m_2+m_3-1)}~,\cr}}
where
\eqn\jmjmjm{\eqalign{
F(j_1,m_1;j_2,m_2;&j_3,m_3)=\int d^2 x_1 d^2x_2|x_1|^{2(j_1+m_1)}
|x_2|^{2(j_2+m_2)}\times\cr
&|1-x_1|^{2(j_2-j_1-j_3-1)}|1-x_2|^{2(j_1-j_2-j_3-1)}
|x_1-x_2|^{2(j_3-j_1-j_2-1)}~.\cr}}
The integral over $x$ in \basthree\ ensures winding number conservation
$m_1+m_2+m_3=0$. The function $F$ \jmjmjm\ does not seem to be expressible
in terms of elementary functions. The same two-point functions and
three-point functions arise also in the coset $SL(2)/U(1)$ when we look
at correlation functions preserving the winding number; in the coset,
additional correlation functions are non-vanishing as well.

A special case that plays a role in \S5 is
\eqn\folfor{\langle\tilde\Phi_{-j_1-1;j_1+1,j_1+1}
\tilde\Phi_{-j_2-1;j_2+1,j_2+1}
\tilde\Phi_{-j_1-j_2-2;-j_1-j_2-2,-j_1-j_2-2}\rangle.}
As we argued in section 5 this correlator computes the residue of
the pole in the correlator of non-normalizable operators
$\tilde\Phi_{\tj_i;j_i+1,j_i+1}$ as $\tj_i$ approaches
$j_i$.  This residue is computed in \gktwo, and in the normalization
\twptfn\ it takes the following form :
\eqn\tfkp{\langle\tilde\Phi_{-j_1-1;j_1+1,j_1+1}\tilde\Phi_{-j_2-1;j_2+1,j_2+1}
\tilde\Phi_{-j_3-1;-j_3-1,-j_3-1}\rangle=
{k\pi\over 2}\sqrt{\gamma\left(1-{2j_1+1\over k}\right)
\gamma\left(1-{2j_2+1\over k}\right)
\over \gamma\left({1\over k}\right)
\gamma\left(1-{2j_3+1\over k}\right)},}
where $j_3=j_1+j_2+1$.

\appendix{C}{Normalization of two-point functions in Liouville and 
$SL(2)$ backgrounds of string theory}

In string theory in flat space-time $\IR^{d-1,1}$, it is well-known
that the two-point function of physical, on shell, operators on the
sphere (as well as the zero and one-point functions, which we will not
discuss here) vanishes.  From the worldsheet point of view this is
natural since the CFT two-point function is finite\foot{It is
typically proportional to the volume of space-time, which is infinite,
but, like in field theory (in momentum space), one is interested in the 
contribution to the correlation function that is proportional to the volume,
\ie\ preserves momentum.}, but one has to
divide by the volume of the conformal Killing group (CKG) of the sphere with
two punctures, which is infinite. Equivalently, the two-point function
vanishes since it does not saturate the zero modes of the
reparametrization ghosts $c$, $\bar c$ on the sphere. In space-time,
this is natural as well, since the two-point function corresponds to
the inverse propagator $p^2+m^2$, which indeed vanishes on-shell.

In backgrounds that involve $SL(2)$, such as Liouville theory, $AdS_3$, 
$SL(2)/U(1)$, or the Nappi-Witten spacetime \NappiKV,  it is similarly
well-known that the two-point function (as well as the zero and 
one-point function) does not vanish\foot{This is also true for backgrounds
involving other asymptotically $AdS$ spaces.}. From the worldsheet point of 
view this is due to the fact that while in string theory one still needs to 
divide by the volume of the conformal Killing group of the sphere with two 
punctures, the CFT correlator is typically infinite, due to a diverging 
integral over bosonic zero modes in the CFT. This infinity precisely cancels 
the volume of the conformal Killing group, and leaves behind a finite answer.

From the space-time point of view this is natural as well. Correlation
functions of non-normalizable operators in such backgrounds correspond
to off-shell Green's functions in a dual ``boundary theory''. In Liouville
theory and $SL(2)/U(1)$ the dual theory is in general a LST \abks,
and in particular low dimensional examples it can be described alternatively
by a large $N$ matrix model. In $AdS_3$, the dual theory is a two dimensional
CFT \adscft. In the Nappi-Witten model, the dual is not known but is
expected to exist \ElitzurRT. In all these cases, the two-point function
is expected to be non-zero in general. 

To compute the finite two-point function in string theory in the
backgrounds mentioned above, one has to evaluate the ratio of infinite
volumes of the bosonic zero modes in the CFT and the conformal Killing
group of the sphere with two punctures. Naively one might expect that
this just gives a constant, independent of the quantum numbers of the
operators whose two-point function is being computed, but this is
known to be incorrect. The purpose of this appendix is to compute the
two-point function in string theory for $SL(2)/U(1)$.  We start with a
discussion of the more familiar Liouville case, and then move on to
$SL(2)$. On general grounds, one expects that the ratio of
determinants in question should be the same in Liouville theory,
$SL(2)$, $SL(2)/U(1)$, and other related theories, 
since it has to do with the same divergence
in the underlying $SL(2)$ CFT.  We will see that this is indeed the
case.

We start with (bosonic) Liouville theory. We will use results from
\FateevIK\ but modify them to be consistent with our normalizations and
conventions\foot{Because of different choices for $\alpha'$ and for the sign
of $\phi$ in \FateevIK, this involves multiplying $Q$ in that paper by
$1/\sqrt{2}$ and multiplying $\phi$ by $(-1/\sqrt{2})$.} in appendix A.
The central charge of the Liouville theory is related to the
linear dilaton slope $Q$ as follows:
\eqn\centliouv{c=1+3Q^2~.}
The Lagrangian contains an interaction term
\eqn\lliinntt{{\cal L}_{\rm int}=\mu e^{-\sqrt{2}b\phi}~,}
where $b$ is defined by the relation $Q/\sqrt{2}=b+b^{-1}$.  
The operators of interest are
\eqn\opsliouv{V_\alpha=e^{-\sqrt{2}\alpha\phi}~.}
Their worldsheet scaling dimensions are \tt\ 
\eqn\dimliouv{\Delta(\alpha)=\alpha({Q\over \sqrt{2}}-\alpha)~.}
Note that the Liouville term in the Lagrangian \lliinntt\ is
$\mu V_b$; according to \dimliouv\ it is marginal.

The two-point function of the operators $V_\alpha$ is given by
\eqn\twoptliouv{\langle V_{\alpha_1}(z) V_{\alpha_2}(0)\rangle=
{\delta(\alpha_1-\alpha_2)D(\alpha_1)\over |z|^{4\Delta(\alpha_1)}}~,}
where
\eqn\ddaall{D(\alpha)\equiv 
\left(\pi\mu\gamma(b^2)\right)^{{1\over b}({Q\over \sqrt{2}}-2\alpha)}
{\gamma(2b\alpha-b^2)\over b^2\gamma(2-{2\alpha\over b}+{1\over b^2})}~,}
with $\gamma(x)$ defined in \gamdef.
As mentioned above, this CFT result is divergent when $\alpha_1=\alpha_2$. The
divergence arises from the integration over the bosonic zero mode of $\phi$ 
in the CFT, and is cancelled in string theory by the 
volume of the corresponding conformal Killing group.
The question we would like to address is what is the finite result obtained by
taking into account this ratio of infinite factors. 

A simple trick that allows one to resolve this issue is to differentiate
the two-point function \twoptliouv\ with respect to 
the cosmological constant $\mu$. 
This brings down from the action the operator $V_b$ (with a minus sign). 
In the CFT this operator is integrated over the worldsheet, and the
divergent factor $\delta(\alpha_1-\alpha_2)$ arises from the integral. 
In string theory, it is clear that the right prescription is to drop the
integral over $V_b$ and replace it by $c\bar c V_b$. This eliminates both
of the infinities mentioned above. The (unintegrated) 
Liouville three-point function
$\langle V_b V_\alpha V_\alpha\rangle$ is finite, and all the $c$, $\bar c$
zero modes on the sphere are soaked by the vertex operators, or equivalently,
the conformal Killing group of the sphere with three punctures is trivial. 
Thus, we conclude that the two-point function 
$\langle V_\alpha V_\alpha\rangle$
in string theory can be obtained by integrating the relation
\eqn\twothreeliouv{{\partial\over\partial\mu}
\langle V_\alpha V_\alpha\rangle_{\rm string}=-\langle V_b V_\alpha 
V_\alpha\rangle~.
}
The three-point function on the right-hand side of \twothreeliouv\ is a special
case of those calculated in \refs{\DornXN,\ZamolodchikovAA}.
Using the results of these papers and integrating \twothreeliouv, 
one finds that 
\eqn\twostringliouv{
\langle V_\alpha V_\alpha\rangle_{\rm string}={1\over\pi}({Q\over \sqrt{2}}
-2\alpha) 
D(\alpha)~.}
Comparing to \twoptliouv\ we see that the ratio of infinite volumes
produces in this case the finite factor $(Q/\sqrt{2}-2\alpha)/\pi$, 
which depends
on the particular operators whose correlation function is being computed.

We next move on to the case of $SL(2)$. This case was already analyzed
by slightly different methods in \MaldacenaKM, with results that agree
with the results that we will find below\foot{The results stated in 
\MaldacenaKM\ include only the $j$-dependent factors in the two-point functions
and not the additional $j$-independent factors. Reinstating these factors
the result precisely agrees with our result below.}. The two-point function in
the $SL(2)$ CFT is given in \twptsl. Restoring the $z$-dependence, 
we can write it as
\eqn\defdj{\langle\Phi_{j_1}^{(sl)}(x_1;z_1)
\Phi_{j_2}^{(sl)}(x_2;z_2)\rangle={\delta(j_1-j_2)
D(j_1)\over|x_{12}|^{4(j_1+1)}|z_{12}|^{4\Delta(j_1)}}~,}
where $\Delta(j)$ is given by \dimphij\ and  
\eqn\defdjtwo{D(j) = {k\over \pi} \left[ {1\over {k\pi}} \, \gamma\left(
{1\over k} \right)
\right]^{2j+1} \gamma\left(1 - {{2j+1}\over k}\right).}
Again, in order to compute the corresponding correlation function in string
theory we need to resolve the ratio of infinite factors coming from
the $\delta(j_1-j_2)$, and from the volume of the CKG. To do that we
will use a result from \newgk, where it is proven that (in the notations of
this paper) the following identity should hold in string theory 
(see equation (3.28) in \newgk):
\eqn\ipp{\langle I \Phi_{j}^{(sl)}(x_1,\bar x_1)
\Phi_{j}^{(sl)}(x_2,\bar x_2)\rangle={2j+1\over k}\langle 
\Phi_{j}^{(sl)}(x_1,\bar x_1)\Phi_{j}^{(sl)}(x_2,\bar x_2)\rangle~,
}
where
\eqn\idef{
I={1\over k^2}\int d^2z J(x;z)\bar J(\bar x;\bar z)
\Phi_{0}^{(sl)}(x,\bar x;z,\bar z)~.}
See \eg\ \newgk\ for the definition and properties of the $SL(2,\IR)$
current $J(x;z)$. Equation \ipp\ is a special case of a more general
relation that is proven in \newgk. 
Just like for the Liouville case, in order to make sense
of \ipp\ in string theory, we drop the integral in the definition
of $I$ \idef, compute the left-hand side of \ipp\ (which involves no
divergences) and take it to be the definition
of the right-hand side. 
Using the results of appendix B and the Ward identities
of $SL(2)$ currents summarized in \newgk, one finds that (suppressing 
the dependence on $x$ and $z$)
\eqn\iipphhii{\langle I\Phi_{j}^{(sl)}\Phi_{j}^{(sl)}\rangle
={1\over2\pi^2}\left(2j+1\over k\right)^2D(j)~.}
Substituting into \ipp\ we conclude that the string theory
two-point function is given by (again 
suppressing the dependence on $x$ and $z$)
\eqn\phistr{\langle\Phi_{j}^{(sl)}\Phi_{j}^{(sl)}\rangle_{\rm string}=
{1\over 2\pi^2}{2j+1\over k} D(j)~.}
The $SL(2)/U(1)$ result follows immediately from this.
Comparing to \defdj\ we see that the string two-point function is corrected
relative to the coefficient of $\delta(j_1-j_2)$ in the CFT two-point function 
by the
factor ${1\over 2\pi^2}{2j+1\over k}$. To compare to the Liouville result 
\twostringliouv\ we need to take into account 
the relation between the Liouville 
momentum $\alpha$ and $j$, $\alpha=-Qj/\sqrt{2}$. Thus, $Q/\sqrt{2}-2\alpha
=Q(1+2j)/\sqrt{2}$. 
This has to be multiplied further by $Q/\sqrt{2}$ 
to account for the difference between
$\delta(\alpha_1-\alpha_2)$ in \twoptliouv, and $\delta(j_1-j_2)$ in \defdj. 
Thus, the Liouville answer is in $SL(2)$ variables\foot{Recall that in our
conventions $Q^2=2/k$ \qtwok.} 
${1\over {2\pi}}Q^2(2j+1)={1\over\pi}
{2j+1\over k}$. The $SL(2)$ answer \phistr\ has the same dependence on $j$ 
and $k$;
it differs from the Liouville result by the factor $1/2\pi$. We do not know the
origin of this minor discrepancy.

\appendix{D}{Simplification of the formula for $t_{n_1,n_2,n_3,n_4}$}

In \S6.1 we found the formula
\eqn\couplOn{t_{n_1,\,n_2,\,n_3,\,n_4}={1\over {r_0^2}}
\sum_{l>m=0}^{k-1}{{e^{-{\pi i\over k}(l+m)\sum n_i}
\over\sin^2\left[{\pi(l-m)\over k}\right]}}
\prod_{i=1}^4\sin\left[{\pi(l-m)\over k}n_i\right]~,}
which we would like to simplify here.
Let us introduce a new summation variable
$p=l-m$.
In terms of this variable, the sum in \couplOn\ can be rewritten as
\eqn\newsum{\sum_{l>m=0}^{k-1}=\sum_{p=0}^{k-1}\sum_{m=0}^{k-1-p}~,}
where we added a $p=0$ term without changing the expression, since 
the summand vanishes for $p=0$. Using the relation
\eqn\rel{\sin n\alpha=\sin\alpha\sum_{l=0}^{n-1}e^{i(n-2l-1)\alpha}~,}
one can further simplify \couplOn\ to the form
\eqn\coupla{t_{n_1,\,n_2,\,n_3,\,n_4}={1\over {r_0^2}}\sum_{p=0}^{k-1}
\sum_{m=0}^{k-1-p}e^{-{2\pi i m\over k}\sum n_i}\sum_{l_i=0}^{n_i-1}
e^{-{2\pi i p\over k}(\sum l_i+2)}\sin^2\left({2\pi p\over k}\right)~.}
The behavior of the sum over $m$ in the expression above depends on 
the value of $\sum n_i$. Let us first consider the case $\sum n_i\notin k \IZ$.
Then, one can easily do the sum over $m$ and find
\eqn\couplb{\eqalign{ t_{n_1,\,n_2,\,n_3,\,n_4}=
&{k\over {4r_0^2}}\left(e^{-{2\pi i\over k}\sum n_i}-1\right)^{-1}\times\cr
\bigg[&\sum_{l_i=0}^{n_i-1}
\left(\delta_{\sum n_i-\sum l_i-3,k \IZ}+
\delta_{\sum n_i-\sum l_i-1,k \IZ}-
2\delta_{\sum n_i-\sum l_i-2,k \IZ}\right)-\cr
&\sum_{l_i=0}^{n_i-1}\left(\delta_{\sum l_i+3,k\IZ}+
\delta_{\sum l_i+1,k\IZ}-
2\delta_{\sum l_i+2,k\IZ}\right)\bigg]~.
}}
It is easy to see that this expression is actually zero.
Indeed, by a change of variables
\eqn\vanew{l_i\rightarrow\tilde l_i=n_i-l_i-1~,}
one can show that the third line of \couplb\ exactly cancels the second line.
This is precisely what we expect, since the theory at the specific point of
moduli space we are at has a $\IZ_k$ symmetry under which the $\co_n^+$ have
charge $n$, and this symmetry should not be broken by the $F^4$ vertex.

Now we turn to the case $\sum n_i \in k\IZ$.
Define $N$ via $\sum n_i=kN$. The coupling in this case takes
the form
\eqn\couplc{t_{n_1,\,n_2,\,n_3,\,n_4}={1\over {r_0^2}}\sum_{p=0}^{k-1}
(k-p){(-1)^{Np}\over \sin^2{p\pi\over k}}
\prod_{i=1}^4\sin{\pi p\over k} n_i\equiv
\sum_{p=0}^{k-1}(k-p)f(p)~.}
The function $f(p)$ introduced in \couplc\ has the following properties :
\eqn\prop{f(p)=f(-p)=f(p+k);\quad f(0)=0~.}
Using these properties one can easily show that
\eqn\propa{\sum_{p=0}^{k-1}(k-p)f(p)={k\over 2}\sum_{p=0}^{k-1}f(p)~,}
so we conclude that
\eqn\coupld{t_{n_1,\,n_2,\,n_3,\,n_4}={k\over {2r_0^2}}\sum_{p=0}^{k-1}
{(-1)^{Np}\over \sin^2{\pi p\over k}}
\prod_{i=1}^4\sin{\pi p\over k} n_i~.}
Without loss of generality we will assume that
$n_1\leq n_2\leq n_3\leq n_4$.
We can use the following identity to simplify \coupld\ :
\eqn\iden{\sum_{n=0}^{min(r,r^\prime)-1}
\sin{\pi p\over k}\sin{\pi p\over k}(r+r^\prime-2n-1)=
\sin\left({\pi p\over k}r\right)\sin\left({\pi p\over k}r^\prime\right)~.}
Then, \coupld\ takes the form
\eqn\coupldd{
\eqalign{t_{n_1,\,n_2,\,n_3,\,n_4}=
-{k\over {4r_0^2}}\sum_{p=0}^{k-1}(-1)^{Np}
\sum_{l_1=0}^{n_1-1} & \sum_{l_3=0}^{n_3-1}
\big[\cos{\pi p\over k}(Nk-2(l_1+l_3)-2)-\cr
&\cos{\pi p\over k}(n_1+n_2-n_3-n_4-2(l_1-l_3))
\big]~.}}
Simplifying, one arrives at the following expression
for the coupling
\eqn\couplfin{t_{n_1,\,n_2,\,n_3,\,n_4}=
-{k^2\over {4r_0^2}}\sum_{l_1=0}^{n_1-1}\sum_{l_3=0}^{n_3-1}
\left[\delta_{l_1+l_3+1,k\IZ}-\delta_{l_1+l_3+n_4+1,k\IZ} \right]~.}
Let us analyze this result. Since the $n_i$
lie between $1$ and $k-1$, the sum of the $n_i$'s is
\eqn\sumn{4\leq\sum_{i=1}^4 n_i\leq 4k-4~,}
or equivalently $1\leq N\leq 3$ (recall that $N={1\over k}\sum_i n_i$ 
must be integer to get a non-vanishing coupling), 
which means that we have three cases to consider. Let us consider them in turn. 

\item{(1)} For $N=1$ one can see that neither
of the two delta functions in \couplfin\ is saturated and hence
the coupling in this case is zero.

\item{(2)} For $N=2$ we see from the ordering of the $n_i$ that
\eqn\niin{\eqalign{&n_1+n_3\leq n_2+n_4 \Rightarrow n_1+n_3\leq k,\cr
&n_1+n_2\leq n_3+n_4 \Rightarrow n_1+n_2\leq k~,
}}
which means that the first delta function in \couplfin\
is never saturated, while the second gives the following
value for the coupling
\eqn\couplfina{t_{n_1,\,n_2,\,n_3,\,n_4}=
{k^2\over {4r_0^2}}\min(n_1,k-n_4)~.}

\item{(3)} For $N=3$ we conclude using $n_1 \leq n_2 \leq n_3 \leq
n_4 \leq k-1$ that
\eqn\niina{k+2\leq n_1+n_3\leq \left[{3k\over 2}\right];\quad
n_i+n_j\geq k+2,\,\forall i,j~, }
where $[3k/2]$ is the integer part of $3k/2$. From this we conclude
that the contribution from the first delta function is
\eqn\couplfinb{\sum_{l_1=0}^{n_1-1}
\sum_{l_3=0}^{n_3-1}\delta_{l_1+l_3+1,k\IZ}=\min(n_1,n_1+n_3-k)=n_1+n_3-k~,}
while the second delta function contributes
\eqn\couplfinc{-\sum_{l_1=0}^{n_1-1}
\sum_{l_3=0}^{n_3-1}\delta_{l_1+l_3+n_4+1,k\IZ}=
-(\min(n_1,k-n_4)+\min(n_1,k-n_2))=-(2k-n_2-n_4)~.}
We see that in this case the coupling is again vanishing.

\noindent
To summarize, using the symmetry of $t$ in its four indices,
we found that 
\eqn\finalt{t_{n_1,n_2,n_3,n_4} = {k^2\over {4r_0^2}} \min(n_i,k-n_i)}
if $\sum n_i = 2k$, and it vanishes otherwise.

It is interesting to note that \finalt\ can be written
in terms of the fusion coefficients of the $SU(2)$ WZW theory of level $k-2$,
\eqn\couplc{t_{n_1,\,n_2,\,n_3,\,n_4}={k^2\over 4r_0^2} \sum_{l=0}^{k-1}
N^l_{n_1-1,n_2-1}N^l_{n_3-1,n_4-1}\quad {\rm for}\,\,\sum n_i=2k~,}
where (see for example \HIV)
\eqn\fusion{N^{n_1-1}_{n_2-1, n_3-1}=
{2\over k}\sum_{p=1}^{k-1}{1\over \sin{\pi p\over k}}\sin{\pi pn_1\over k}
\sin{\pi pn_2\over k}
\sin{\pi pn_3\over k}}
are the fusion coefficients of $SU(2)_{k-2}$,
\eqn\fusone{N_{l_1,l_2}^{l_3}=\left\{\eqalign{&1\quad{\rm
if}\quad|l_1-l_2|\leq l_3\leq
{\rm min}(l_1+l_2,2(k-2)-l_1-l_2)
\cr&0\quad{\rm otherwise~.}}\right.}
The indices $l_i$ are related to the $SU(2)$ spins $j_i$, 
which label the primaries of
$SU(2)_{k-2}$,  as follows: $l_i=2j_i$. A useful identity for verifying
\couplc\ is:
\eqn\sinid{\sum_{l=0}^{k-1}\sin{\pi m l\over k}\sin{\pi n l\over k}=
{k\over 2}(\delta_{m-n,2k\IZ}-\delta_{m+n,2k\IZ})~.}

\listrefs
\end